\documentclass[aps,prl,reprint,superscriptaddress]{revtex4-2}
\usepackage[T1]{fontenc}
\usepackage[utf8]{inputenc}
\usepackage{amsmath,amssymb}
\usepackage[
pdffitwindow=true,
colorlinks=true,
frenchlinks=false,
linkcolor=blue,
anchorcolor=blue,
citecolor=blue,
filecolor=blue,
urlcolor=blue,
bookmarks=true,
bookmarksopen=true,
bookmarksnumbered=true,
bookmarksopenlevel=1,
plainpages=false,
pdfpagelayout=TwoPageLeft,
pdfpagelabels=true,
breaklinks=false]{hyperref}
\usepackage[main=english]{babel}
\usepackage[dvipsnames]{xcolor}
\usepackage[per-mode=symbol,separate-uncertainty]{siunitx}
\usepackage{graphicx}
\usepackage{isotope}
\usepackage{cleveref}
\usepackage{epstopdf}
\usepackage{physics}
\usepackage[normalem]{ulem}
\usepackage{float}
\usepackage{todonotes}
\usepackage{tabularx}

\begin{document}

\title{Practical quantum tokens: challenges and perspectives }

%%Quantum tokens: experimental challenges and perspectives
%% A perspective on practical implementation of quantum tokens 
%% The practical implementation of quantum tokens: architectures, challenges and perspectives. 
%% Older title: Underlying physics and practical implementation of quantum tokens
%% Project: QuaMToMe 

\author{Nadezhda~P.~Kukharchyk}
\email[]{Nadezhda.Kukharchyk@wmi.badw.de}
\affiliation{Walther-Mei{\ss}ner-Institut, Bayerische Akademie der Wissenschaften, 85748 Garching, Germany}
\affiliation{School of Natural Sciences, Technical University of Munich, 85748 Garching, Germany}
\affiliation{Munich Center for Quantum Science and Technology (MCQST), 80799 Munich, Germany}

\author{alphabetically: Holger~Boche}
\affiliation{Lehrstuhl für Theoretische Informationstechnik, School of Computation, Information and Technology, Technical University of Munich, 80333 Munich, Germany}
\affiliation{Munich Center for Quantum Science and Technology (MCQST), 80799 Munich, Germany}

\author{Christian~Deppe}
\affiliation{Institut für Nachrichtentechnik, 38106 Braunschweig, Germany}
\affiliation{Munich Center for Quantum Science and Technology (MCQST), 80799 Munich, Germany}

\author{Kirill~G.~Fedorov}
\affiliation{Walther-Mei{\ss}ner-Institut, Bayerische Akademie der Wissenschaften, 85748 Garching, Germany} 
\affiliation{School of Natural Sciences, Technical University of Munich, 85748 Garching, Germany}
\affiliation{Munich Center for Quantum Science and Technology (MCQST), 80799 Munich, Germany}

\author{Martin~E.~Garcia}
\affiliation{Institute of Physics, Center for Interdisciplinary Nanostructure Science and Technology (CINSaT), University of Kassel, 34132 Kassel, Germany}

\author{Ilja~Gerhardt}
\affiliation{Institute for Solid State Physics, Leibniz University Hannover, 30167 Hannover, Germany}

\author{Rudolf~Gross}
\affiliation{Walther-Mei{\ss}ner-Institut, Bayerische Akademie der Wissenschaften, 85748 Garching, Germany}
\affiliation{School of Natural Sciences, Technical University of Munich, 85748 Garching, Germany}
\affiliation{Munich Center for Quantum Science and Technology (MCQST), 80799 Munich, Germany}

\author{Thomas~Halfmann}
\affiliation{Institut für Angewandte Physik, Technische Universtität Darmstadt, 64289 Darmstadt, Germany} 

\author{Hans~Huebl}
\affiliation{Walther-Mei{\ss}ner-Institut, Bayerische Akademie der Wissenschaften, 85748 Garching, Germany}
\affiliation{School of Natural Sciences, Technical University of Munich, 85748 Garching, Germany}
\affiliation{Munich Center for Quantum Science and Technology (MCQST), 80799 Munich, Germany}

\author{David~Hunger}
\address{Karlsruher Institut f\"ur Technologie, Physikalisches Institut, 76131 Karlsruhe, Germany}
\address{Karlsruher Institut f\"ur Technologie, Institut f{\"u}r QuantenMaterialien und Technologien , 76344 Eggenstein-Leopoldshafen, Germany}

\author{Wolfgang~Kilian}
\affiliation{Physikalisch-Technische Bundesanstalt (PTB), 10587 Berlin, Germany}

\author{Roman~Kolesov}
\affiliation{3rd Institute of Physics, University of Stuttgart, 70569 Stuttgart, Germany}

\author{Juliane~Krämer}
\address{Faculty of Informatics and Data Science, University of Regensburg, %Bajuwarenstr. 4, 
93053 Regensburg, Germany}

\author{Alexander~Kubanek}
\affiliation{Institute for Quantum Optics, Ulm University, 89081 Ulm, Germany}
\affiliation{Center for Integrated Quantum Science and Technology (IQST), Ulm University, Albert-Einstein-Allee 11, 89081 Ulm, Germany}

\author{Kai~M\"uller}
\affiliation{Walter Schottky Institute, Technical University of Munich, 85748 Garching, Germany}

\author{Boris~Naydenov}
\affiliation{Department Spins in Energy Conversion and Quantum Information Science (ASPIN), Helmholtz-Zentrum Berlin für Materialien und Energie GmbH, 14109 Berlin, Germany}

\author{Janis~Nötzel}
\affiliation{Emmy-Noether Group Theoretical Quantum Systems Design, School of Computation, Information and Technology, Technical University of Munich, 80333 Munich, Germany}
\affiliation{Munich Center for Quantum Science and Technology (MCQST), 80799 Munich, Germany}

\author{Anna~P.~Ovvyan}
\affiliation{Kirchhoff-Institute for Physics, Heidelberg University, 69120 Heidelberg, Germany}

\author{Wolfram~H.~P.~Pernice}
\affiliation{Kirchhoff-Institute for Physics, Heidelberg University, 69120 Heidelberg, Germany}

\author{Gregor~Pieplow}
\affiliation{Department of Physics, Humboldt University of Berlin, 12489 Berlin, Germany}

\author{Cyril~Popov}
\affiliation{Institute of Nanostructure Technologies and Analytics (INA), Center for Interdisciplinary Nanostructure Science and Technology (CINSaT), University of Kassel, 34132 Kassel, Germany}

\author{Tim~Schr\"oder}
\affiliation{Department of Physics, Humboldt University of Berlin, 12489 Berlin, Germany}
\affiliation{Ferdinand-Braun-Institute, 12489 Berlin, Germany}

\author{Kilian~Singer}
\affiliation{Institute of Physics, Center for Interdisciplinary Nanostructure Science and Technology (CINSaT), University of Kassel, 34132 Kassel, Germany}

\author{Janik~Wolters}
%\email[]{Janik.Wolters@dlr.de}
\affiliation{Institute of Space Research, German Aerospace Center (DLR e.V.), 12489 Berlin, Germany}%
\affiliation{Institutes of Physics, Technische Universit\"at Berlin, 10623 Berlin, Germany}
\affiliation{Einstein Center Digital Future (ECDF)}
\affiliation{AQLS UG (haftungsbeschränkt), 10587 Berlin, Germany}

\date{\today}
%\pacs{}
%\keywords{} 

\begin{abstract}
The concept of quantum tokens dates back alongside quantum cryptography to Stephen Wiesner's seminal work in 1983~\cite{Wiesner.1983}. 
Already this initial work proposes society-relevant applications such as  secure quantum banknotes, which can be exchanged between a bank and a customer. This quantum currency is  based on various physical states that can be easily verified but is protected from being copied by the fundamental quantum laws. Four decades later, these ideas have flourished in the field of quantum information, and the concept of quantum banknotes has not only adopted many varying names, such as quantum money, quantum coins, quantum-digital payments, and quantum tokens, but also reached its first experimental demonstrations. In this perspective article, we discuss the current state-of-the-art of quantum tokens in the field of quantum information, as well as their future perspectives. We present a number of physical realizations of quantum tokens with integrated quantum memories and their applicability scenarios in detail. Finally, we discuss how quantum tokens fit into the information security ecosystem and consider their relationship to post-quantum cryptography.
\newpage
\end{abstract}

\maketitle
%______________________________________________________________________________

%\newpage
\section{Introduction}
Quantum states of light and their interaction with matter have been in the spotlight since the 1970s. Numerous experiments have confirmed quantum-mechanical properties of light and matter, including but not limited to the  observation of quantum entanglement and superposition in individual atomic systems and macroscopic condensates~\cite{EPR.1935,Bouwmeester.1997,Horodecki.2009,Kun.2025,Terhal.2025}. The underlying non-classical effects have catalyzed the development of numerous innovative protocols and technologies for information processing, sensing or quantum applications in general. In addition, quantum communication has been identified as highly relevant for secure information transmission and distributed quantum applications. Also, quantum banknotes were already conceptualized in 1983, which is associated with the advent of quantum tokens nowadays~\cite{Wiesner.1983}. This quantum currency is based on the idea of an unconditionally secure quantum banknote that can be exchanged between banks and their customers. This concept is intimately linked to an early version  of the no-cloning theorem~\cite{Wootters.1982,Dieks.1982} and an initial proposal for a quantum key distribution scheme~\cite{Bennett.1984}. Nowadays, this quantum-secured exchange is celebrated as the BB84 protocol. 

\begin{figure}[t!]
    \centering
    \includegraphics[width=0.5\textwidth]{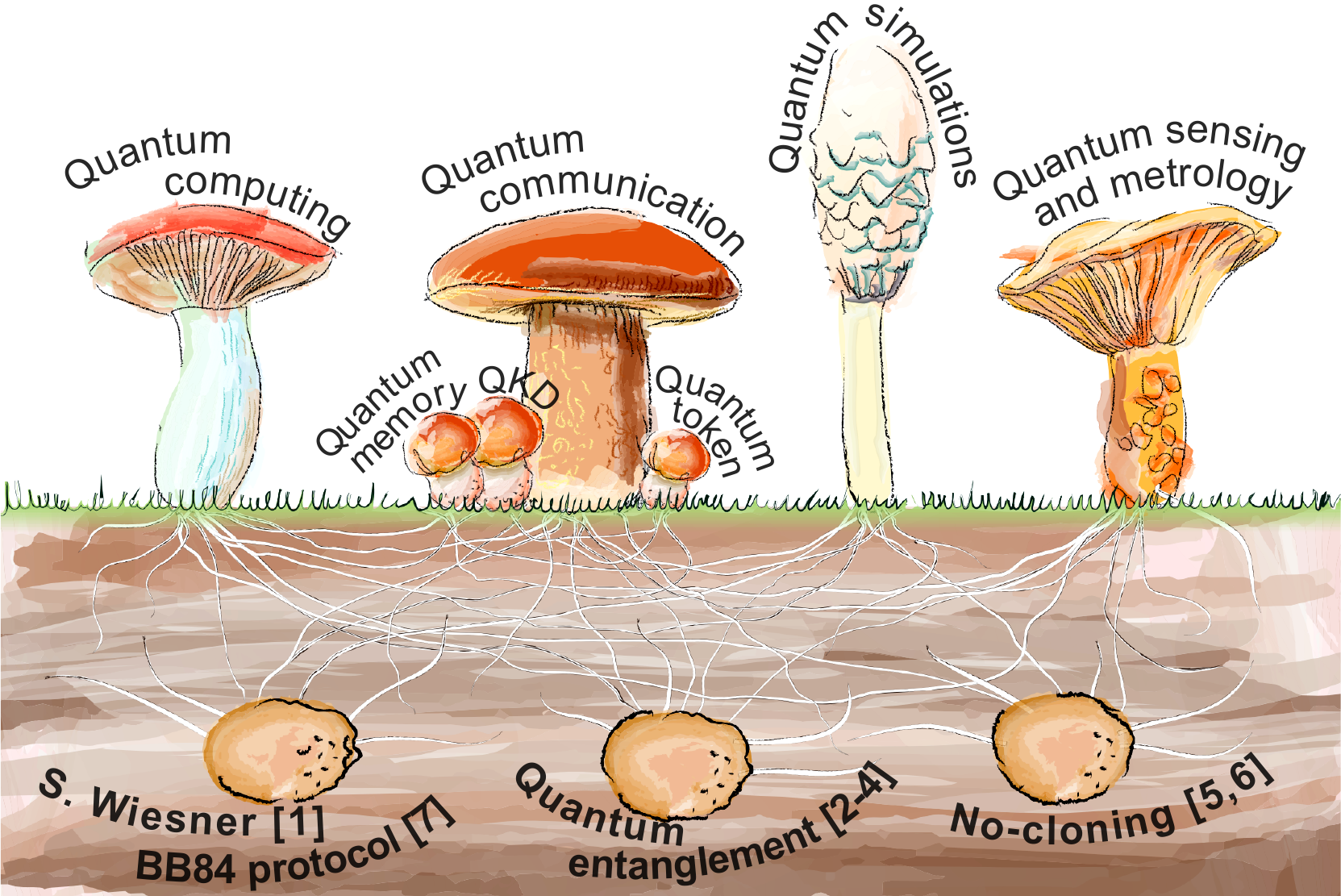}
    \caption{The loose structure of quantum information fields and sub-fields in relation to selected milestone ideas.
}
    \label{fig:intro_mushrooms}
\end{figure}

When originally conceived, the idea of quantum tokens was beyond the reach of experimental implementation. 
In the last decades, one of the main directions of experimental studies with quantum systems has been the controlled creation and manipulation of non-classical states for applications in various quantum technologies~\cite{Horodecki.2021}. The latter are often classified into the following key pillars: quantum communication, quantum computing, quantum simulation, quantum sensing, and quantum metrology (see Fig.\,\ref{fig:intro_mushrooms}). While these cornerstones of quantum science ask highly relevant intrinsic scientific questions and test quantum mechanics in itself, they also offer opportunities for present-day and future applications. In particular, the question of whether suitable quantum states, protocols, and algorithms can provide an advantage over classical communication, information processing, simulation or sensing schemes is part of active research. Presently, aspects such as task processing times, signal-to-noise ratios, provable security, among others, are investigated. Some quantum advantages have already been experimentally addressed and demonstrated, such as reaching the unconditional security of communication~\cite{Bennett1992,Chen2021},  performing classically impossible communication tasks~\cite{Bouwmeester.1997,Fedorov.2021}, or demonstrating the quantum advantage in computation~\cite{Arute2019,Liu2025}. The field of quantum communication, which is key to the idea of quantum tokens, has been at the forefront of these developments and has historically been pioneered by experiments in the optical regime. In the field of quantum computing, superconducting circuits operating at microwave frequencies are currently one of the leading candidates for future scalable quantum computers~\cite{Kjaergaard2020}. The combination of these two prominent, but experimentally very different, platforms gives rise to the concept of hybrid quantum communication and computation. This approach allows one to exploit the best of the two worlds' properties to ultimately achieve the goal of quantum advantage in practically relevant tasks.
Thanks to significant experimental progress in manipulating quantum states through their preparation, storage, and readout -- concepts surrounding quantum tokens can now be realized in proof-of-principle experiments.

The original quantum token ideas by S.\,Wiesner have received renewed attention over the last decade, in no small part because the rapid advances in quantum computing threaten to undermine the existing classical cryptography protocols which underpin the entire modern communication network. While post-quantum encryption approaches are considered to be a mid-term strategy~\cite{Singh.2025}, they may not guarantee the protection of sensitive information in the long term, because of their purely computational security and the corresponding vulnerability to the "harvest now, decode later" attack, as opposed to the information-theoretic security provided by quantum encryption schemes. Respectively, the field of research on quantum tokens can be defined as a combination of quantum cryptography protocols with quantum memories. Similar to classical keys, quantum tokens are typically split into two categories: public and private. Combined with their quantum features, these tokens promise to address several security issues, such as eavesdropping detection, privacy, unforgeability, and even contribute to the overarching problem of secure authentication preceding any remote communication between two or more parties. While authentication — the process of verifying one’s identity — has been a foundational element of human society for millennia, it has traditionally relied on physical objects that can be generally copied or forged. Quantum tokens follow these initial verification principles and add unconditional security to the authentication process by using quantum states, which are guaranteed by the laws of quantum physics, in particular, the no-cloning and measurement theorems. Naturally, this assumes that the original generation and distribution of tokens is implemented correctly. Quantum tokens are closely tied to the concepts of quantum key distribution, but can extend beyond them. Issued by one party to another, a quantum token state requires an on-demand verification routine that may occur at a specific point in time and space. It extends the applicability scenarios of quantum tokens beyond those of typical QKD protocols, making quantum tokens more demanding in realization, but more useful in applications. Most importantly, conventional quantum tokens must include a quantum memory element, in one form or another, which can be accessed by request and may also be recycled. Practically, a \textbf{quantum token can be understood as a device which stores a codebook of selected quantum states}. These stored quantum states can be retrieved and verified during an authentication or payment procedure. As long as the quantum nature of these states is preserved, it conceptually prevents a duplication of the quantum token, thereby protecting against forgery attacks and guaranteeing validation security. It has also been proposed that probing of an extended reservoir of quantum states can be done selectively in a non-destructive manner, so that this reservoir can be probed and verified again~\cite{Aaronson.2018}. This approach could potentially allow for multiple uses (recycling) of the single-issued quantum token, while still benefiting from the information-theoretic protection provided by quantum mechanics. 

In the existing literature, the concept of quantum tokens appears under different names, such as quantum money~\cite{Wiesner.1983,Gavinsky.2012,Bartkiewicz.2017}, quantum payment~\cite{Schiansky.2023}, quantum ticket~\cite{Pastawski.2012}, quantum tokenized signature~\cite{Chen.9520219102021}. All these names address the problem of unforgeability~\cite{Murray.2021,Brassard.2005,Chen.9520219102021}, which is the central property and requirement for the quantum token.

A somewhat alternative approach to the quantum token ideas is to use a quantum analogue of classical physically unclonable functions (PUFs)~\cite{Pappu.2002}. Here, one is supposed to use a classical physical system, the clonability of which is considered practically unrealistic. Then, one can employ quantum states to probe properties of this PUF and protect a corresponding communication channel by quantum laws~\cite{Arapinis.2021}. Such systems are sometimes referred to as the quantum-read PUFs (QR-PUFs)~\cite{Skoric.2010,Skoric.2012}. Their advantage is that they do not necessarily require a long-living quantum memory~\cite{Nikolopoulos.2017} and, therefore, are much easier to implement experimentally. The obvious drawback is that the ultimate security of such QR-PUFs relies on the ignorance of their intrinsic PUF structure, which, in theory, can be broken.

Another standalone type of quantum tokens is the so-called S-money protocol, which can be realized without any storage element. In general,  S-money relies on short-distance quantum communication and space-time restrictions for the verification of the corresponding classical keys distributed within a network of trusted agents~\cite{Kent.2019,Kent.2022}. The S-money thus relies on QKD protocols to exchange a classical key via a short-range quantum channel, which protects this information transfer step by the no-cloning theorem and leaves the task of secure local storage of classical keys to be implemented using classical or post-quantum protection routines. Finally, the unconditional protection of the classical key between agents stems from space-time restrictions on the classical communication channels between them~\cite{Jiang2024}. A recent realization of the S-money protocol highlights technological advantages of this approach~\cite{Jiang2024}. 

In this perspective article, we refer to the quantum token as to a device containing a quantum memory element storing a selected codebook of quantum states. Here, it is important to highlight the recent advances in the experimental realization of long-living quantum memories. For instance, it has been demonstrated that one can achieve coherence times of up to \SI{18}{\hour} by using an optical quantum memory based on rare-earth elements~\cite{wang_nuclear_2025} in combination with its unique temperature stability. Such memories could be made compatible with other experimental requirements in order to encode and validate the quantum token states at optical frequencies~\cite{Bozzio.2018,Guan.2018}. In the gigahertz frequency range -- relevant for superconducting quantum circuits and processors -- electronic spin ensembles have demonstrated record coherence times, with storage durations reaching up $\SI{2}{s}$ in nuclear spin denuded silicon \cite{Tyryshkin:2011fi}. 

At the same time, recent demonstrations of various quantum communication and cryptography protocols with propagating microwaves~\cite{Fedorov.2021,Fesquet2024} demonstrate the experimental feasibility of this approach. In this publication, we present selected experimental realizations and theoretical proposals for quantum tokens, comprising a quantum memory with the corresponding storage and readout routines. Finally, we present our vision for quantum tokens in terms of applications and quantum information science, as well as accounting for the competing field of post-quantum cryptography and its place in the overarching information security landscape.

\newpage

\section{Overview of possible physical realizations}

\begin{figure}[ht!]
    \centering
    \includegraphics[width=0.47\textwidth]{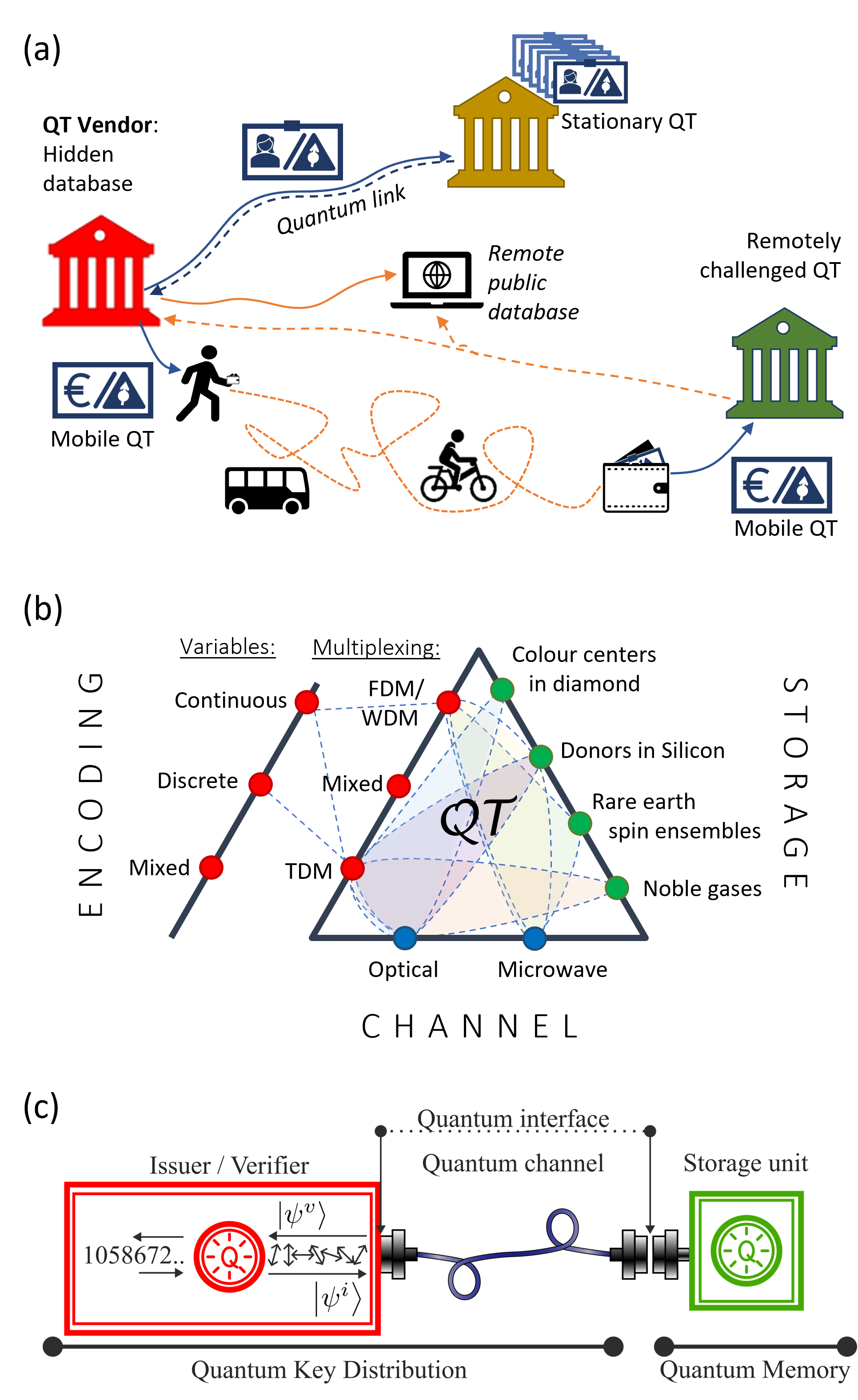}
    \caption{\\
    \textbf{(a)} Simplified implementation of quantum token. Here, a flying quantum token is an authentication or transaction key that does not require any intermediate storage. The on-submission quantum token, on the other hand, requires a storage element. The on-submission quantum token can be verified by the holder at the latest by the time of expiry of the quantum storage element.\\     
    \textbf{(b)} Structural representation of a quantum token, which in a general case requires an encoding protocol, a transmission channel, and a storage element. The encoding protocol is defined by the choice of type of variables (continuous, discrete or mixed) and multiplexing method: TDM (time-division multiplexing), FDM/WDM (frequency- or wavelength-division multiplexing) and mixed. The transmission channel defines the frequency of wavelength at which the respective quantum token is being generated and transmitted; it is bound to the available infrastructure.\\
    \textbf{(c)} Key elements of physical realization of quantum token, which include issuing and verifying parties, transmission channel, and storage element.\\
    \\
    }
    \label{fig:sec2:QT_overview}
\end{figure}

Practical realizations of quantum tokens are driven by their use-cases, which embrace their means of transport, storage, and verification routines (see \Cref{fig:sec2:QT_overview}(a)). One cornerstone is a dedicated vendor that provides quantum tokens. Those  can then be exchanged either via a quantum link followed by storing the token in a state-of-the-art long-lived quantum memory, or emitted directly in the form of a physically transportable entity. At a later stage, these tokens can be locally or remotely challenged to ensure their validity, e.g., for authentication purposes. This concept dictates three main ingredients: a storage unit or \textbf{\textit{quantum memory}}, a \textbf{\textit{transmission channel}}, and an \textbf{\textit{encoding and verification protocol}}, as schematically shown in \Cref{fig:sec2:QT_overview}(b). A unique combination of these three parts forms a specific realization of quantum tokens.

One key and unsolved challenge in realizing quantum tokens is establishing the \textbf{\textit{quantum memory}}. After identifying a potential candidate, its properties determine the subsequent choice of the transmission channel and the encoding protocol. 
Quantum memories are physical systems capable of coherently storing a quantum state for a sufficiently long time and returning it on demand. Potential candidates span from natural atoms or ions to artificial atoms and cavities. They come in the form of ensembles of atomic gases, color centers in diamonds, donors in silicon, and rare earth spin ensembles. In applications, which do not require ultra-long storage times, other systems are no less useful, such as 3D microwave cavities~\cite{Romanenko.2020,Xie.2018,Nojiri.2024}, superconducting artificial atoms~\cite{Wang.2022,Somoroff.2023,Place.2021}, quantum dots~\cite{Yoneda.2018}, trapped atom and ions~\cite{Wang2021,Ruster.2016,Levine.2019,Dudin.2013}, and even phononic modes~\cite{MacCabe.2020,Seis.2022}. 

In this article, we focus on a selection of ultra-long coherent spin memories that combine electronic and nuclear spins in the context of quantum tokens. We explore their potential for information encoding, possible transport and communication protocols, and application scenarios. Specifically, we detail spin systems in Cesium-Xenon gas mixtures~\cite{esg23,jut24,mes23}, (Si,V, Sn, N)-V centers in diamonds~\cite{OrphalKobin2023,torun_optical_2023,Karapatzakis2024,Delgado_2025a}, Phosphorus donors in silicon~\cite{morello2010,Weichselbaumer:2020hn}, and rare earth ions~\cite{Kukharchyk2018coh,casabone_dynamic_2021,deshmukh_detection_2023}, all of which have already demonstrated outstanding coherence results and are therefore considered to be the key-systems envisioning accessible quantum memories. 

Due to semi-shielding of their $4f$-electrons, the rare earth ions are an excellent example of spin ensembles exhibiting outstanding quantum coherence of optical and spin degrees of freedom \cite{goldner_chapter_2015}, culminating in a recent demonstration of record-long nuclear spin coherence lifetime of up to 18 hours \cite{wang_nuclear_2025}. 
The color centers in diamonds are a large group of defects, which can couple to the long-coherent nuclear spins of \isotope[13]{C} and guarantee fast photon read-out rates~\cite{Klotz.2025,Maurer.2012,Beukers2025,Afzal2024,Trusheim2020}. 
For the prototypical system of electron spins, localized phosphorus donors in isotopically pure, nuclear spin-free silicon have demonstrated coherence times up to seconds~\cite{Steger:2012ev} and, if transferred into a coupled nuclear spin system~\cite{Davies:1974gu,Hoehne:2011ii}, up to hours~\cite{Saeedi.2013}. 
Gas mixtures of alkali metals and noble gases allow for storing optical signals in an electron state in the alkali metal ground state and to coherently transfer this state further to more coherent states of nuclear spins of noble gas atoms~\cite{wol17, bus22,nam17}. 

In these systems, the quantum state is often stored in an excitation of  an ensemble of electron and nuclear spins in the form of so-called "spin-wave excitation"~\cite{Afzelius2013} or a collective Dicke state~\cite{Dicke.1954}: 
$\ket{\psi} = \sum^N_j c_j e^{i\delta_i t} e^{-jkz_j} \ket{r_1 \dots e_j \dots g_N} $, which implies deployment of such storage protocols as atomic frequency comb (AFC)~\cite{Businger.2020, Ma.2021,Seri.2017} or off-resonant Raman-type~\cite{Saunders.2016,Kalachev.2019}. 
With techniques initially developed in pulsed spin resonance, the quantum state can be further transferred from the electronic spins into coupled nuclear spins~\cite{Davies:1974gu,Hoehne:2011ii}, which offers enhanced storage times in the range of hours~\cite{Saeedi.2013}. To extract information from these systems, refocusing pulses are applied, as implemented in the Hahn echo sequence, and are followed by read-out optical pulses. 
Other approaches explore the possibility of selectively addressing single ions or atoms to store single quantum states~\cite{deshmukh_detection_2023}. This would allow the addition of spatial and frequency dimensions to quantum memory multiplexing~\cite{Lan.2009,Kunz.2019,VernazGris.2018}.

The properties of the selected quantum memory impose requirements regarding the implementation of the \textbf{\textit{channel}} used for the transmission, which we attribute to either the optical or the microwave domain. Some of the specific systems discussed later can also be addressed via both optical and microwave excitations, thus enabling control via either channel.  However, motivated by optical quantum communication technologies, many of them have been primarily researched for the storage of optical photons. Particular optical interest lies in the compatibility with the existing infrastructure of fiber networks, which could be used for quantum communication, thereby reducing construction efforts.

While communication between very distant quantum nodes is best ensured by optical fiber technologies, local direct (wireless) communication can alternatively be implemented using microwave and GHz technologies. In particular, when considering the integration with superconducting microwave quantum circuits, this approach allows for circumventing the necessity of transducing the quantum signal between the gigahertz and optical frequencies.~\cite{Fesquet.2023,GonzalezRaya.2022}.  
The development of quantum microwaves and the associated technologies began more than ten years ago as a pioneering scientific achievement~\cite{Blais.2004,Wallraff.2004,Chiorescu.2004}. 
Demonstration that quantum properties (such as path entanglement) can be observed in propagating microwave signals is one of the crucial achievements~\cite{Menzel.2012,Fedorov2018,Renger.2022}. Due to the low signal energy of quantum microwaves, suitable measurement methods had to be developed first, with which propagating squeezed states could then be detected~\cite{Fedorov.2021,Pogorzalek2019, Menzel2010, Eichler:2011kt}. 
%At the same time, important discoveries regarding signal conversion and dynamics were made in the area of spin systems~\cite{Weichselbaumer:2020hn,Zollitsch:2015ut}.
Combining spin systems with quantum microwaves allows one to close the technology gap in the generation and storage of purely microwave quantum tokens, as these play a central role in the realization of quantum networks in the GHz frequency range.

The third pillar of basic quantum token representation is the \textbf{\textit{encoding}} routine, which is also researched in the QKD field.
One quantum token is a combination of generated single quantum states, which can be distributed in time (time domain multiplexing or TDM) or in frequency/ wavelength (frequency/wavelength domain multiplexing or TDM/WDM), the choice of which is limited by the chosen quantum memory system. As a basic example, microwave quantum memory based on ensembles of donors in silicon has a very narrow available frequency bandwidth and thus facilitates only TDM~\cite{Weichselbaumer:2020hn}. Optical quantum memories based on rare earth spin ensembles, on the other hand, are nicely compatible with either TDM, or FDM, or a combination of both due to their inhomogeneously broadened optical transitions and developed storage routines~\cite{Afzelius:2009gc}. %more references
%proper transition

Additionally, the encoding relies on the type of variables used to generate the key, with two standard types: discrete and continuous variables. The discrete-variable (DV) protocols use individual photons and their properties, such as polarization, to encode information, whereas continuous-variable (CV) protocols rely on the quadratures of light, like amplitude and phase. The aforementioned BB84 protocol~\cite{Bennett.1984,Bennett.1983} is a textbook example of a DV protocol, which uses strictly defined combinations of photon polarizations. It typically achieves longer transmission distances than CV schemes but requires expensive single-photon detectors. There are other DV protocols, such as E91~\cite{Ekert.1991}, BBM92~\cite{Bennett1992} or SARG04~\cite{Scarani.2004}, which aim to mitigate the weaknesses and pitfalls of the BB84 protocol.
The CV protocols can operate with standard telecommunication components and offer higher key rates over short distances. However, CV systems are more sensitive to noise and loss, making them more challenging over long distances~\cite{Jain.2022,Zhang.2024b,Hajomer.2024}. Both approaches have unique advantages and are actively developed for different use cases in quantum communication~\cite{Hajomer.2025}.

The basic representation of the quantum token outlined in \Cref{fig:sec2:QT_overview}~(b) does not capture key \textbf{performance metrics} that go well beyond the coherence time of the chosen quantum memory system. These metrics, such as \textit{fidelity, write-read efficiency, qubit count parameter, bit-rate, etc.}, are essential for evaluating the practical applicability of quantum token solutions~\cite{Liu.2023}. Some of these parameters are highly dependent on the quality of individual devices, while others are intrinsically linked to the internal properties of the selected spin systems and the encoding protocols employed. 
Nearly the most important quantity characterizing the applicability of the QT device is the \textbf{\textit{fidelity}} $F$, which measures how close the verified quantum state $\ket{\psi^v}$ is to the issued quantum state $\ket{\psi^i}$, in the general case:
$$F(\rho(\ket{\psi^v}),\rho(\ket{\psi^i}) = \left( \mathrm{tr}\sqrt{ \sqrt{\rho(\ket{\psi^v})} \rho(\ket{\psi^i}) \sqrt{\rho(\ket{\psi^v})}} \right)^2.$$ Here, $\rho(\ket{\psi^{v,i}})$ are the density matrices of the verified and issued quantum states, respectively. The fidelity can be specified for separate elements of quantum tokens: quantum channel, quantum memory, and the interface between them (cf.~\Cref{fig:sec2:QT_overview}(c)). However, the overall fidelity of the quantum token device is of utmost importance, as it quantifies the total loss of quantum information. 
The \textbf{\textit{write-read efficiency }}$\eta$ is the quantitative characteristic of the quantum memory. It is defined as the ratio of retrieved (verified) photons to the issued ones. However, it lacks information on whether the quantum state is preserved. 
The complexity of the token key depends on its size, which is described by the \textbf{\textit{qubit count parameter}}. This parameter is limited by the storage and control capabilities of the quantum memory and by the \textbf{\textit{bit-rate}} of the channel-memory combination. The additional metrics for quantum tokens are related to the previously mentioned main ones. They can be derived first with respect to the QKD and memory parts independently, and then combined to achieve the overall device performance. 

Most of the quantum memories discussed here are solid-state-based, thus suffering significantly from interactions with lattice vibrations (phonons). To reduce the impact of phonons, one has to freeze them out by cooling down such memories to liquid helium temperatures or even below. Moreover, one has to decouple the remaining phononic excitations from the frequencies of the utilized spin states. This leads to a significant overhead of operational infrastructure, consisting of a cryostat, often requiring the supply of liquid $^4\mathrm{He}$ or a $^4\mathrm{He}/^3\mathrm{He}$ mixture. For spin-based systems, in addition, a magnet system is required, providing magnetic fields from hundreds to thousands of \si{\milli\tesla}. Quantum tokens integrating such memories can be demonstrated soon. However, they require substantial running costs and maintenance resources, which large institutions can afford but are out of reach for the average private user. Nevertheless, such tokens could be deployed by larger stakeholders, such as banks or governments, as stationary units, see \Cref{fig:sec2:QT_overview}~(a).

One of the most fascinating aspects is certainly the idea of a mobile quantum token, which can be carried by any single user as easily as in a pocket and can be used to securely validate any transaction or prove identity. This is the holy grail of quantum memories, requiring the preservation of a quantum state in ambient conditions when being exposed to mechanical shocks, electromagnetic radiation, and other noise sources. Thermalized to room temperatures and about the size of a finger, quantum memories based on mixtures of alkali metal and noble gas atoms in atomic vapor cells address the problem of storing photonic quantum states for several minutes, hours, or even up to days under ambient conditions. 
Once ensuring the required robustness and technological simplicity of the overall system, they are expected to outperform solid-state quantum memories and pioneer the mobile version of quantum tokens.
% see details in \Cref{sec:Q-ToRX}.
On the other hand, solid-state quantum memories are making rapid progress and are meanwhile addressing the possibility of preserving a quantum state at natural thermal conditions. This has been experimentally demonstrated recently with an ensemble of donors in isotopically engineered silicon, where a quantum bit has been stored for 39 minutes~\cite{Saeedi.2013}. The possibility of room-temperature storage is also being actively pursued for devices based on nitrogen-vacancy centers in silicon, where the quantum state is transferred into a nuclear spin state with long coherence time.

As discussed in more detail in the following sections, the key emphasis of this publication is to review the current status and perspectives of a variety of experimentally explored quantum memory systems, including their potential to enable a wide range of promising applications and use cases. In the following sections, we discuss each system in detail, along with the appropriate protocols and implementation scenarios.

\newpage
%% -----------------------------
%% Project part: 
%% -----------------------------

\section{Microwave quantum tokens for superconducting quantum networks}
%\subsection*{N. Kukharchyk, H. Huebl, K. G. Fedorov}
\begin{center}
{N. Kukharchyk, H. Huebl, K. G. Fedorov}
\end{center}
\label{sec:QuaMToMe}

\begin{figure*}
    \includegraphics[width=0.95\linewidth]{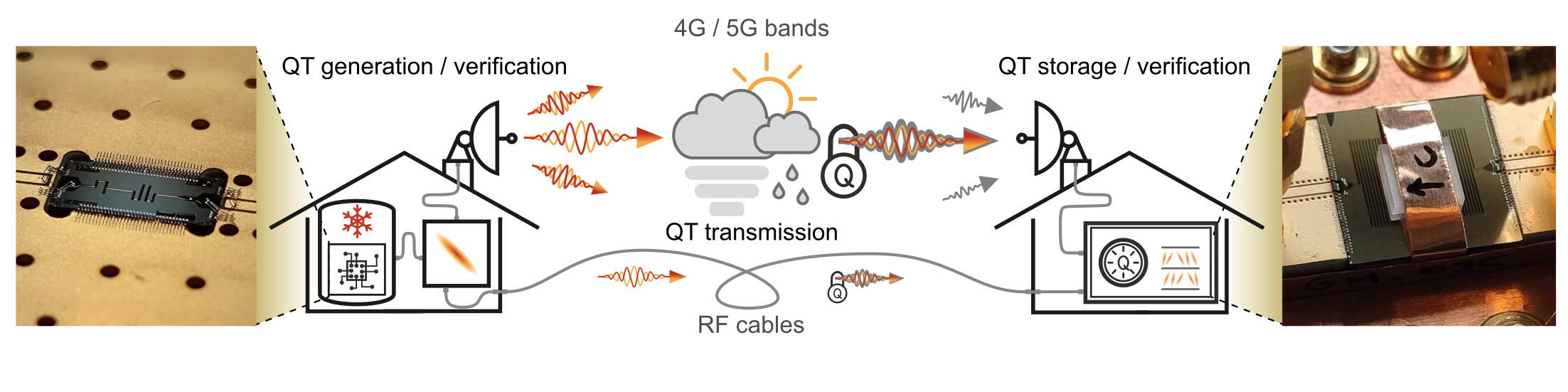}
    \caption{Microwave quantum tokens show strong potential for the advancement of secure short-range quantum communication. Beyond traditional cable-based connections, wireless links in the \SI{2}{\giga\hertz} to \SI{12}{\giga\hertz} frequency range are feasible due to the exceptionally low absorption losses in open air. These links also exhibit strong resilience to weather-related disturbances, such as fog and rain. Recent studies \cite{Fesquet2024} report high fidelities for the wireless transmission of microwave quantum states. Taken together, these findings indicate that secure information transmission via quantum tokens can be seamlessly integrated into existing classical microwave communication infrastructures.}
    \label{fig:QuaMTome:fig1}
\end{figure*}
Recent advances in generating quantum states in the microwave domain, coupled with the progress in superconducting circuit–based quantum information processing and the persistent challenge of efficient microwave-to-optical transduction, underscore the importance of developing quantum tokens and their storage directly within the gigahertz frequency range. In this section, we follow a holistic experimental approach operating in the microwave frequency domain. We address both the generation of quantum tokens as well as strategies for their storage and retrieval in solid-state spin ensembles. The key generation is based on continuous variable (CV) quantum communication representing an established concept in quantum key distribution (QKD)~\cite{Scarani2009}, quantum teleportation~\cite{Pirandola2015}, dense coding~\cite{Braunstein2000}, and Q-tokens~\cite{Pastawski.2012}, which has been successfully translated to the field of gigahertz communications using superconducting quantum circuits~\cite{Fedorov.2021,Pogorzalek2019,Axline2018,Leung2019,Fesquet2024}. Notably, the latter concepts are directly compatible with the superconducting quantum processors. In particular, they discuss the generation of Q-tokens assembled from displaced squeezed states corresponding to the continuous-variable (CV) regime of quantum communication. Especially for small-scale local networks with characteristic ranges of up to about \SI{100}{\meter}, direct communication between quantum nodes using GHz signals is expected to offer clear advantages over approaches based on the frequency conversion to optical frequency bands~\cite{Magnard2020,Yam2025}. Such microwave quantum networks can be complemented further by using ultra-long-lived quantum memories based on spin ensembles, which can be directly interfaced with microwave quantum states. Specifically, we discuss here phosphorus dopants in isotopically engineered, nuclear-spin-diluted, silicon host crystals as well as erbium-based spin ensembles in LiYF\textsubscript{4} and CaWO\textsubscript{4} matrices. Both systems have demonstrated electron spin coherence times exceeding seconds~\cite{Tyryshkin:2011fi,Saeedi.2013} and are considered as prime candidates for quantum memory applications \cite{OSullivan:2022}.

We implement the generation of a microwave Q-token in the gigahertz frequency band using concepts inspired by a well-known CV quantum key distribution protocol~\cite{Cerf2001}. This scheme is based on two central operations: (i) squeezing and (ii) displacement of propagating microwave states. The latter imprints the secret token information and encodes the classical key as a quantum state, while the squeezing operation ensures the unconditional security of the protocol. A corresponding microwave CV-QKD protocol with finite-size codebooks of displaced squeezed microwaves has recently been implemented with superconducting circuits and demonstrated the information-theoretic security~\cite{Fesquet2024}. Respectively, such codebooks or ensembles of displaced squeezed states represent the Q-token, which can be stored in a quantum memory. This type of Q-token guarantees the information security against copying attempts via the no-cloning theorem. Evaluating the fidelity of retrieved quantum states enables verification of whether a duplicate has been made. For instance, if the state fidelity $F_{\mathrm{nc}}$  exceeds the threshold of $2/3$, assuming an infinite Gaussian Q-token codebook, one can assert that the state has not been copied, thereby confirming the authenticity of information therein. In real systems, one has to rely on finite-energy codebooks which lead to somewhat increased no-cloning threshold values, $F_{\mathrm{nc}} > 2/3$~\cite{Owari2008,Jensen2011}.

\begin{figure}[b!]
    \centering
    \includegraphics[width=0.75\linewidth]{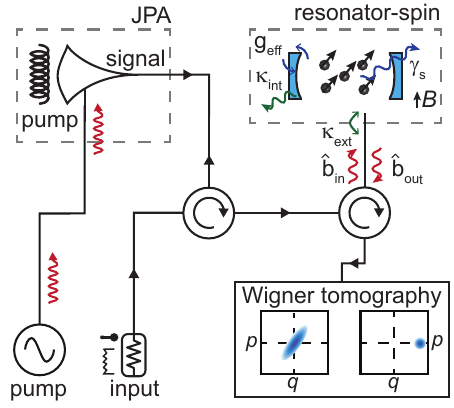}
    \caption{Conceptual experimental setup for the coupling of propagating squeezed microwaves and quantum tokens to a spin ensemble. A Josephson parametric amplifier (JPA), controlled by a microwave pump signal, generates squeezed microwave states, which are subsequently displaced to form a quantum token. This signal is then transferred to a spin-resonator system engineered to transduce microwave excitations into spin excitations. The quantum properties of the emitted signal from the spin-resonator are assessed using Wigner tomography.}
    \label{fig:QuaMTome:fig2}
\end{figure}

In our particular experiments, we generate microwave squeezed states using superconducting Josephson parametric amplifiers (JPAs)~\cite{Honasoge2025}, which typically consist of a $\lambda/4$ coplanar waveguide resonator shorted to ground via a direct-current superconducting quantum interference device (dc-SQUID). The dc-SQUID provides a nonlinear inductance, enabling a flux-tunable JPA resonance frequency. The periodic modulation of the magnetic flux through the dc-SQUID with a strong coherent tone at twice the JPA resonance frequency enables a parametric amplification process, which leads to squeezing of the JPA's input signals. These squeezed states are subsequently emitted by the JPA and can be guided to spin ensembles, as sketched in \Cref{fig:QuaMTome:fig2}.

One approach to realizing a quantum memory relies on resonantly converting microwave signals into spin excitations via an intermediate microwave resonator inductively coupled to the spin ensemble~\cite{Afzelius2013}. The initial information from the input signal can be stored in the spin memory for the duration of its coherence time and recovered by using a spin echo sequence~\cite{Julsgaard2013,OSullivan:2022}. Donor states in nuclear spin-depleted $^{28}$Si crystals have been extensively studied in this context and possess long relaxation and coherence times at cryogenic temperatures~\cite{Weichselbaumer:2020hn, Tyryshkin:2011fi, feher59b, OSullivan:2022}. One of the spin systems considered here is an electron spin ensemble constituted by phosphorus donors. This spin ensemble is hosted by an isotopically purified $^{28}$Si host crystal, and has a donor concentration of [P] = $10^{17}\mathrm{cm}^{-3}$. It is inductively coupled to a planar lumped element superconducting microwave resonator~\cite{Weichselbaumer:2019hi, Weichselbaumer:2020hn}. When the transition frequency of one of the electron spin resonance transitions becomes resonant with the microwave resonator, the corresponding modes can hybridize, resulting in a hybrid resonator-spin system. The Hamiltonian of this system is typically described using the Tavis-Cummings Hamiltonian \cite{Wesenberg:2009es, Zollitsch:2015ut, Weichselbaumer:2020hn}.
% \begin{align} 
% \label{eq:Fesquet-Hamiltonian_system}
% 	&\frac{\hat{H}}{\hbar} = \hspace{1cm}\\ 
%     &\sum_{j}^{N} \left[ g_j (\hat{\sigma}_{-}^j \hat{a}^{\dagger} + \hat{\sigma}_{+}^j \hat{a}) + \frac{\Delta_j}{2} \hat{\sigma}_{\mathrm{z}}^j 
%     + \sqrt{2 \gamma_{\mathrm{s}}}(\hat{\sigma}_{-}^j \hat{f}_j^{\dagger} + \hat{\sigma}_{+}^j \hat{f}_j)\right] \notag \\
%    &+  \left[(\sqrt{\kappa_{\mathrm{int}}}\hat{b}_{l} + \sqrt{\kappa_{\mathrm{ext}}}\hat{b}_{\mathrm{in}})\hat{a}^{\dagger} + H.c.\right] \mathrm{,} \notag
% \end{align}
% where $g_j = g_\mathrm{eff} / \sqrt{N}$ is the coupling strength of a single spin to the fundamental mode of the microwave resonator, which is described with the bosonic operator $\hat{a}$. In addition,  $\Delta_j$ denotes the detuning of the $j$-th spin and the resonator frequency $\omega_\mathrm{r}$ and $\gamma_s$ is the dephasing rate modeling spin decoherence using an associated bosonic noise mode $\hat{f}_j$. Moreover, we account for internal cavity losses via the  bosonic mode $\hat{b}_{l}$ and a coupling rate $\kappa_{\mathrm{int}}$. Also, the resonator-spin system is coupled to  microwave input modes $\hat{b}_{\mathrm{in}}$, with a coupling rate $\kappa_{\mathrm{ext}}$. 

Since a Q-token is supposed to consist of multiple individual quantum states drawn from the corresponding codebook, it is important to realize the storage of multiple of those states in the quantum memory. In particular,  time-domain multiplexing for the sequential storage of Q-token states using a single spin transition is a suitable approach for its implementation.

Alternatively,  spin systems, such as \isotope[167]{Er}, offer a rich structure of transitions in the microwave range available already at and close to zero magnetic fields~\cite{Kukharchyk2020,Strinic2025}. Often these systems show inhomogeneous broadening, which can be used for storage protocols based on a frequency division multiplexing approach. This scheme allows for the simultaneous storage of individual quantum states and  Q-tokens. While the protocol naturally differs from the time-domain storage concept above, the availability of multiple transitions offers access to multi-frequency storage concepts, which  necessitate the development of Q-token generation at multiple frequencies. 

Recently, we have demonstrated a high-precision microwave spectroscopy of such ensembles~\cite{Strinic2025}, proving the possibility of simultaneous access to several spin transitions. First experiments of coherent storage in this broadband regime have already demonstrated a retrieved echo efficiency on the order of $0.1\%$. While this number is insufficient for practical Q-token storage, it demonstrates the feasibility of this approach.

The successful verification of quantum token storage requires the realization of state estimation or tomography protocols. We perform  
%To verify fidelities of microwave Q-tokens, we perform 
Wigner tomography of the output microwave signals using a heterodyne detection setup, which is based on the calculation of statistical quadrature moments up to the fourth order and their application for the Gaussian state reconstruction~\cite{Fedorov2018,Fedorov2016}. For signal calibration, we perform Planck spectroscopy~\cite{Menzel2010,Gandorfer2025}. Analyzing the signals with respect to their moments then allow to estimate the output signal variance $\sigma_{\mathrm{s,out}}^2$, which can be translated to a squeezing level of the emitted output signal via $S = -10 \log_{10}(\sigma_{\mathrm{s,out}}^2/0.25)$ referenced to the vacuum level. %Using Eq.\,\Cref{eq:Fesquet-Hamiltonian_system}
For the Tavis-Cummings system, we can derive that the output variance $\sigma_{\mathrm{s,out}}^2$ is related to the variance $\sigma_{\mathrm{s,in}}^2$ at the input of the resonator-spin system by
\begin{equation} \tag{2}
	\label{eq:Fesquet-input_output}
	\sigma_{\mathrm{s,out}}^2 = r^2\,\sigma_{\mathrm{s,in}}^2 + l^2\, \sigma_{\mathrm{l}}^2 + t^2\,\sigma_{\mathrm{spin}}^2 \mathrm{,}
\end{equation}
where the complex coefficients $r$, $l$, and $t$ describe terms of the Langevin noise operator associated with the reflection from the cavity, the resonator internal losses, and coupling to the spin ensemble, respectively. Here, the variance $\sigma_{\mathrm{l}}^2$ originates from a thermal bath coupled to the resonator via its internal losses. Similarly, $\sigma_{\mathrm{spin}}^2$ is associated with a thermal bath formed by the spin ensemble. Initial experiments show coupling rates up to $t=0.6$, indicating that quantum signals prevail in spin excitations.

%When storing the microwave quantum token states into the ensemble of donors in silicon, both thermal baths are assumed to be at a temperature of $T = 142\, \mathrm{mK}$, according to characterization measurements. The squeezing levels of retrieved states are thus characterized in three different cases: first, we measure squeezing for the completely detuned case, where both the resonator and spins are detuned from the JPA frequency. In a second step, we tune the resonator into resonance with the JPA at the common frequency of $ 5.6169\, \mathrm{GHz}$. And finally, the spins are also tuned in resonance at the joint frequency of $5.61724\, \mathrm{GHz}$ leading to a significant decrease in the measured squeezing level, which we interpret as a successful coupling of the incoming squeezed states to the spin ensemble~\cite{Oehrl.2025}. Based on our model, we estimate a coupling to the spin ensemble of $t^2 \simeq 36\%$. We observe a good agreement between the model and the measurements. We note that for large pump powers above $ -20\, \mathrm{dBm}$, the model starts to diverge from the measurements. We attribute this deviation to higher-order nonlinearities in the JPA. Further analysis and measurements are ongoing and more details can be found in Oehrl et al.~\cite{Oehrl.2025}. 

The storage of microwave quantum tokens in solid-state spin ensembles will represents an important milestone in quantum technologies in general and  for quantum communication and Q-tokens in particular. The combination of tunable Zeeman spin transitions with very long coherence times at GHz frequencies makes these systems  a very promising choice for future microwave quantum memories. Experimentally, we have already shown a successful coupling of propagating microwave squeezed states to an ensemble of phosphorus donor atoms embedded in nuclear spin-depleted $^{28}$Si crystals~\cite{Oehrl.2025}. Additionally, we have demonstrated the broadband spectroscopy of rare earth spin ensembles which paves the way toward multi frequency multi-mode Q-token storage using frequency domain multiplexing protocols~\cite{Strinic2025}.

%Our preliminary experimentss  further indicate that squeezed microwave  states combined with  quantum memories  have a wide  potential for applications in quantum communication and, in particular, to Q-token protocols. 

It is worth to note that the generation of microwave Q-tokens as well as operation of spin quantum memories typically demands millikelvin temperatures. While this puts boundaries on consumer applications, the use in quantum information processing or the implementation of quantum-LANs connecting multiple superconducting quantum processors is evident due to its technical compatibility. The distribution and storage of unconditional secure Q-tokens  is essential for many  important concepts such as  distributed information processing and for the blind computing paradigm.

%These spin ensemble have been heavily studied for realization of optical quantum memories and repeaters~\cite{Rancic2018,Longdell2019}, developing numerous storage and control routines, such as atomic frequency comb~\cite{Usmani2010}, electromagnetically induced transparency~\cite{Hetet2008}, controlled inhomogeneous line broadening~\cite{Lauritzen2010}, and Raman storage protocols\cite{Moiseev2013}.

\newpage
\section{Long-lived quantum tokens based on rare earth ions} 
\begin{center}
{D. Hunger, T. Halfmann, R. Kolesov}
\end{center}

\label{sec_NEQSIS}
%\paragraph*{Network-compatible quantum memories with rare earth ions} %will be removed later
Rare-earth ions (REI) doped in dielectric solids constitute optically addressable spins systems with exceptionally long coherence times for both optical and spin transitions~\cite{goldner_chapter_2015}. The spin states can be directly addressed and controlled via narrow optical transitions, enabling deterministic initialization, manipulation, and readout of qubit states. Integration of REI into photonic devices such as cavities has enabled the enhancement of light-matter interactions and gives access also to single ions~\cite{dibos_atomic_2018,kindem_control_2020,ulanowski_spectral_2022,deshmukh_detection_2023} as qubits. Consequently,  REI are promising candidates for the core tasks of quantum token protocols: Efficient generation of memory-compatible photonic qubits, their long-lived storage, and efficient state retrieval~\cite{tittel_quantum_2025}.

%\Cref{fig:NEQSIS:fig1} a) schematically illustrates REI incorporated in optical cavities, acting as efficient light–matter interfaces. \Cref{fig:NEQSIS:fig1} b) shows the corresponding level scheme, and \Cref{fig:NEQSIS:fig1} c) depicts the inhomogeneous broadening that enables multiplexing and spectral selection of single ions.
%\textit{Ensemble memories.} 
REI have been widely used to store photonic qubits in macroscopic ensembles of ions with demonstrations of the longest storage times to date. Storage protocols for ensembles are mostly based on electromagnetically induced transparency (EIT)~\cite{heinze_stopped_2013,hain_light_2025} and atomic frequency combs (AFC)~\cite{afzelius_multimode_2009}. With this, weak coherent pulses have been stored for up to 1 h~\cite{ma_one-hour_2021}. One of the key challenges is the combination of long storage time with high storage efficiency and low noise. In this respect, using EIT storage with narrow-band spectral filtering, single photons could be stored for up to 1.3 s~\cite{hain_light_2025} with an efficiency of 12.5\%. To increase storage capacity, spectral, temporal, and spatial multiplexing has been implemented, and demonstrations have achieved storage in 250 spatio-temporal modes~\cite{teller_solid-state_2025}, and in 1650 spectro-temporal modes~\cite{wei_quantum_2024}.
\Cref{fig:NEQSIS:fig2} (a) depicts a bulk‑doped crystal that typically serves as the platform for such experiments.

\begin{figure}
    %\centering
    \includegraphics[width=0.5\textwidth]{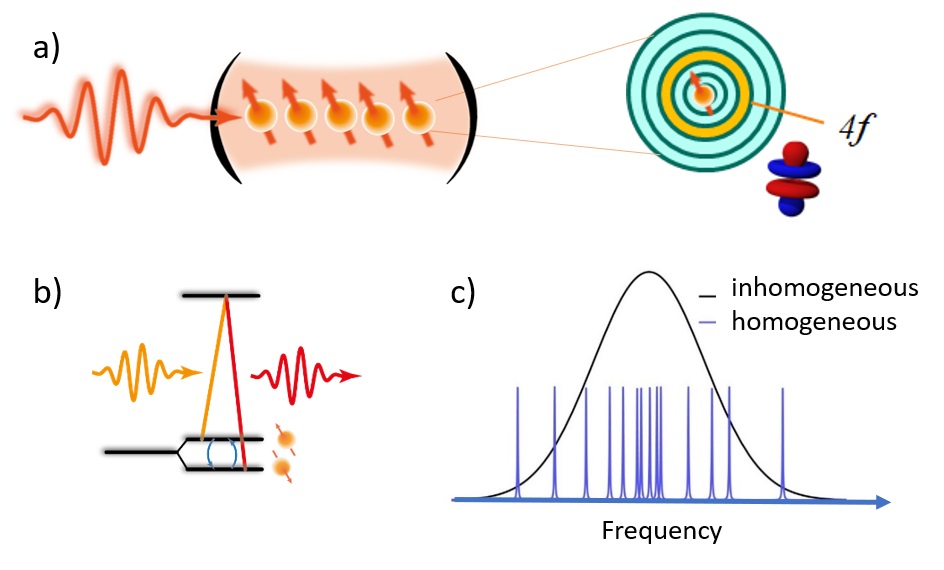}
    
    \caption{(a)~A photonic qubit is stored in a single REI qubit by cavity-enhanced absorption or cavity reflection. (b)~Schematic level scheme for the spin-photon interface. (c)~Inhomogeneously broadened optical transition (black) with narrow homogeneous linewidths (blue), enabling single- and multi-qubit addressing within a single cavity.}
    \label{fig:NEQSIS:fig1}
\end{figure}

%\textit{Single ion memories.} 
Scaling to deterministic, multi-qubit token generation benefits from addressing individual ions. Due to the dipole-forbidden character of the coherent $4f-4f$ transitions in REI, a key requirement is a significant enhancement of light-matter interactions with an optical microcavity. The enhancement is quantified by the coherent cooperativity  
\begin{equation}
C=F_P\frac{\gamma_0}{2\gamma},
\end{equation}
where $F_P=3/(4\pi^2)\times Q/V \times \lambda^3$ is the Purcell factor determined by the quality factor $Q$ and the mode volume $V$ of the cavity, $\gamma_0$ is the natural excited‑state decay rate, and $\gamma$ the homogeneous linewidth. Nanophotonic and open‑access microcavities have achieved Purcell factors up to $F_P=700$ ~\cite{dibos_atomic_2018,zhong_optically_2018}, enabling single‑ion detection, state read‑out and coherent control~\cite{kindem_control_2020,ulanowski_spectral_2022,deshmukh_detection_2023}. \Cref{fig:NEQSIS:fig1}~(a) shows the cavity‑enhanced absorption (or reflection) used to map a photonic qubit onto a single REI, while \Cref{fig:NEQSIS:fig1}~(b) details the three‑level $\Lambda$‑system employed for storage and retrieval. Spectral addressing of a chosen ion within an inhomogeneously broadened ensemble is illustrated in \Cref{fig:NEQSIS:fig1}~(c).
A range of experiments has achieved efficient single ion detection, state readout, and coherent control~\cite{dibos_atomic_2018,kindem_control_2020,ulanowski_spectral_2022,deshmukh_detection_2023}. Single-ion quantum coherence has reached up to 30\,ms lifetime~\cite{kindem_control_2020}, and also multi-qubit registers based on spectral multiplexing have been demonstrated~\cite{chen_parallel_2020,ulanowski_spectral_2022}. Furthermore, single REI can serve as single photon sources to produce a photon-based quantum token in a memory-compatible way~\cite{dibos_atomic_2018,xia_tunable_2022,ulanowski_spectral_2022}. Quantum tokens may require transmission over large distance, and, for example, Erbium ions offer suitable telecom transitions coupled to coherent spins~\cite{gritsch_optical_2025,chen_parallel_2020,gupta_dual_2023}.

%\textit{Methods for long coherence.} 
For quantum token storage, it is advantageous to choose long-lived nuclear spin states as qubit levels. Non-Kramers ions with zero effective electron spin have so far shown the longest coherence times. For example, with Pr$^{3+}$:YSO, coherent storage of up to 1\,min was demonstrated~\cite{heinze_stopped_2013}, and the longest nuclear spin coherence was observed for Eu$^{3+}$:YSO to date, yielding a hyperfine coherence of up to 18 hours~\cite{wang_nuclear_2025}, and coherent storage for up to one hour~\cite{ma_one-hour_2021}. For these examples, specific magnetic fields are used that lead to a zero first order Zeeman transition (ZEFOZ), which improves coherence by $\sim 10^3\mathrm{-fold}$. Additionally, carefully optimized dynamical decoupling sequences are harnessed~\cite{genov_arbitrarily_2017}, which can prolong coherence by another factor of $\sim 10^4$. Recent developments on Er$^{3+}$:CaWO$_4$~\cite{le_dantec_twenty-threemillisecond_2021} and $^{171}$Yb$^{3+}$:CaWO$_4$~\cite{tiranov_sub-second_2025} and earlier work on $^{167}$Er$^{3+}$:YSO~\cite{Rancic2018} show that Kramers ions can achieve second-long coherence.

%\textit{Photonic integration.} 
Efficient coupling of REI to guided or resonant optical modes is essential for both ensemble and single‑ion approaches.
Nanophotonic waveguides provide strong transverse confinement, and the on-chip geometry offers more possibilities for spatial multiplexing, increase of optical depth, and combination with additional structures such as RF electrodes for spin control. \Cref{fig:NEQSIS:fig2}~(b) shows an example geometry designed by R. Kolesov (unpublished) that optimizes light-matter interaction by an extended waveguide length through a meandering shape. Earlier work has used laser-written waveguides in Y$_2$SiO$_5$ crystals~\cite{corrielli_integrated_2016} or titanium-doped lithium niobate waveguides~\cite{saglamyurek_broadband_2011} to demonstrate efficient quantum storage.
Integration of REI into micro- and nanophotonic cavities and use of the Purcell effect has enabled enhanced light-matter interactions for small ensembles and single ions, which offer better scaling potential. \Cref{fig:NEQSIS:fig2}~(c) shows an open-access Fabry-Perot microcavity with a REI-doped membrane, which achieves full tunability together with high quality factors and moderately small mode volumes. This approach has enabled Purcell-enhanced spectroscopy of few ions in nanoparticles~\cite{casabone_cavity-enhanced_2018,casabone_dynamic_2021,eichhorn_multimodal_2025} and single ion detection in nanoparticles and membranes~\cite{deshmukh_detection_2023,ulanowski_spectral_2022}. Even stronger light-matter coupling can be achieved with nanophotonic cavities. Harnessing nanobeam geometries, smallest mode volumes with high quality factors have led to exceptionally large Purcell factors mentioned above~\cite{dibos_atomic_2018,kindem_control_2020}. \Cref{fig:NEQSIS:fig2} d) shows a design by R. Kolesov \cite{sardi_photonic_2024} that harnesses nanopatterned lithium niobate in a nanobeam photonic crystal geometry with side-wall modulation to realize a cavity. The additional electrodes allow one to tune the cavity resonance frequency, similarly to earlier work on microdisc cavities~\cite{xia_tunable_2022}, which allowed single ion detection and fast dynamic resonance frequency tuning.

Coupling REI to cavities is a key aspect for achieving efficient and thus near-deterministic single photon generation in a memory compatible way. The Purcell effect leads to a dominant emission into a cavity mode that can be collected with $>90\%$ efficiency. Enhancement of the emission furthermore leads to a broadening of the optical transition and can mask spectral diffusion. This has enabled the generation of Fourier-limited photons~\cite{yang_toward_2023,ourari_indistinguishable_2023} and the entanglement of quantum memories~\cite{ruskuc_multiplexed_2025}. Cavity-coupled REI are thus promising to achieve efficient quantum token generation, efficient writing into the memory, and efficient readout.

%\textit{Quantum token protocol}. 
The quantum token protocol suitable for REI is largely following the original proposal of quantum money~\cite{Wiesner.1983}, and one variant is analyzed in detail in~\cite{strocka_secure_2025}: The issuer of the quantum token produces random photonic quantum states, e.g., time-bin or polarization qubits, and sends the photons - possibly via an optical fiber link - to the customer. The customer receives the photon qubits and stores them in a multi-mode REI quantum memory or a single-ion quantum register coupled to a cavity. For verification, the stored photons are released and measured, and the issuer compares the measurement outcomes with the expected result.

\begin{figure}
    %\centering
    \includegraphics[width=0.45\textwidth]{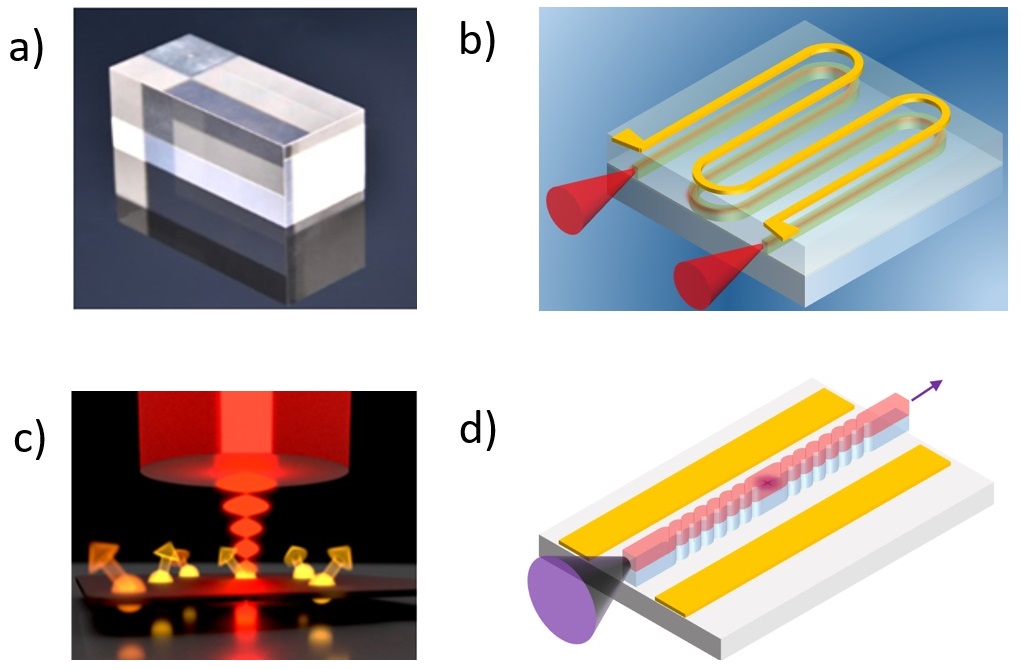}
        \caption{Examples of photonic structures for REI quantum token storage. (a)~A doped bulk crystal enabling multiplexed and long-lived storage. (b)~Waveguide with integrated RF waveguide (yellow) for spin control. (c)~Fiber-based Fabry-Perot microcavity with integrated REI-doped membrane. 
    (d)~Nanophotonic cavity based on a Lithium Niobate nanobeam with tuning electrodes and RF waveguide (yellow).}
    \label{fig:NEQSIS:fig2}
\end{figure}

Although REI experiments currently require cryogenic environments ($\sim$1.5\,K), highly stabilized lasers, vector magnets and RF control, recent engineering advances have yielded compact rack‑mounted platforms that integrate cryogenics, laser locking and microwave electronics. Combined with the photonic integration strategies outlined above, these systems open the way for field‑deployable quantum‑token nodes that can operate autonomously over long periods.

The confluence of record‑breaking nuclear‑spin coherence, near‑deterministic single‑photon generation via Purcell‑enhanced cavities, and versatile photonic integration positions REI as a uniquely powerful platform for deterministic, long‑lived, and efficiently transmittable quantum tokens. Future work will focus on achieving cooperativities well above unity for single‑ion memories, integrating fast electro‑optic tuning of cavity and emitter resonances, and demonstrating full end‑to‑end token protocols -- including long‑distance verification over deployed fiber networks.

\newpage

\section{Diamond based quantum tokens}

\begin{center}
{B. Naydenov, M. E. Garcia, C. Popov,  K. Singer}
\end{center}
\label{sec:DIQTOK}

\begin{figure}[!ht]
	\centering \includegraphics[width=0.95\linewidth]{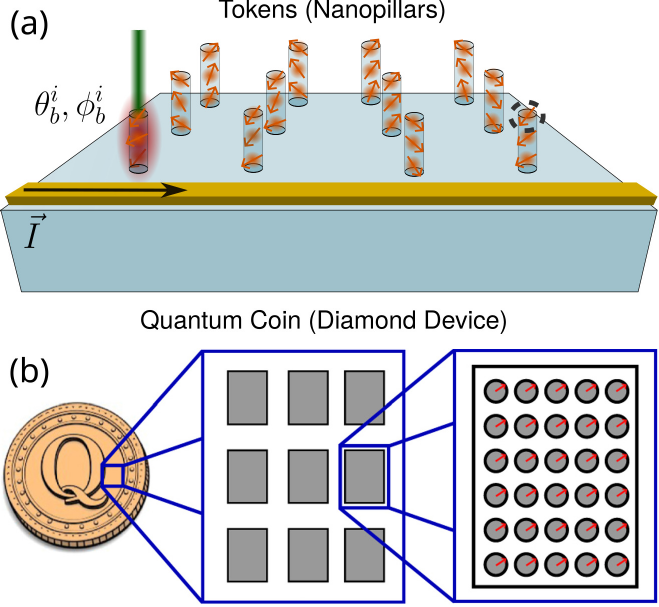}
	\caption{(a)~Schematic representation of the diamond-based quantum token. Small ensembles of NVs represent the token and are fabricated in the nano-pillars in order to increase the optical readout efficiency (green line). The NVs electron spins are used for initialization and readout of the quantum information presented by the angles $\theta_b^i$ and $\phi_b^i$ of the spin magnetization on the Bloch sphere. This is realized using radio and microwave pulses and the information is then stored in the nitrogen nuclear spins. The wire is used to apply RF and MW   for spin control. (b) The final quantum coin (left) consists of many quantum tokens, where each one has a certain number of nanopillars. \\
    }\label{fig:token}
    \end{figure}

%\paragraph*{Brief general description of your token realization}
The realization of quantum tokens pursued here is an innovative approach based on quantum projection noise of ensembles of nitrogen-vacancy (NV) color centers in diamond \cite{Diqtok_patent}. The envisioned device is schematically shown in \Cref{fig:token}(a). The developed in this work ensemble-based quantum protocol enables the implementation of non-copyable tokens containing an ensemble of identically prepared qubits.  It can replace, especially in case of ensembles, the conventional copy protection guaranteed by the quantum no-cloning theorem, which can be applied to single NVs. In addition, the realization of a quantum token based on single NVs is technologically rather challenging, as high precision in magnetic field alignment is required. In the ensemble approach, the quantum projection noise can reveal an undesirable copy operation, since it increases if a hacker tries to make an illegal measurement of the token on a basis different than the one prepared and issued by a bank \cite{Tsunaki_2025}. The overall security of the protocol has been verified by parallelized brute-force simulations in order to make accurate predictions about the number of ensembles and qubits required on a quantum coin. Advanced attacks on the protocol can assume that measurements can be performed on partial ensembles and that even individual qubits can be measured. Even though such an attack could be considered technically unfeasible, the security of the ensemble-based protocol under these advanced attacks has been proven~\cite{Bauerhenne_2025}. Following this approach, we have developed a protocol for designing a specific quantum coin with desired security requirements, as sketched in \Cref{fig:token}(b).

%\paragraph*{Quantum token protocol}
An ensemble-based quantum coin can be realized by fabrication of nanopillars in diamond which will correspond to the individual quantum tokens, each containing an ensemble of NV centers with the same alignment within it, see \Cref{fig:token}(a)~\cite{Tsunaki_2025,Bauerhenne_2025} .

The electronic spins of the NVs are used as auxiliary qubits, while the nitrogen nuclear spins serve as memory qubits. The number of NV centers in each ensemble should be small enough such that the quantum projection noise is still measurable~\cite{Tsunaki_2025} and can be determined by the diameter of the nanopillars as well as by the density of the created NVs. Ion implantation or delta-doping during diamond growth can be implemented for the creation of NVs with desired density and electron beam lithography followed by inductively coupled plasma reactive ion etching for fabrication of nanopillars with various diameters~\cite{Delgado_2025a, Delgado_2025b}. Such photonic nanostructures enhance photon collection efficiency. The NV ensemble in each nanopillar is initialized by a confocal microscope and laser illumination, and the electron spins manipulated by microwave fields in conjunction with static electric fields for addressing and for controlling the charge state of the NV centers. On the other hand, the number of quantum tokens (i.e. of the nanopillars) within a quantum coin should be chosen high enough such that the probability of a hacker to generate accidentally a copy of the quantum coin is arbitrarily small~\cite{Bauerhenne_2025}, as captured in \Cref{fig:token}(b). 
\begin{figure}[!b]
	\centering \includegraphics[width=0.95\linewidth]{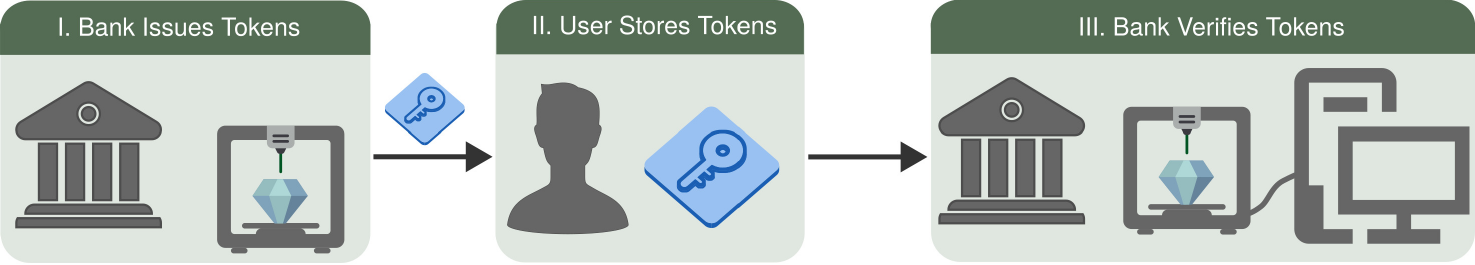}
	\caption{
		Scheme of the ensemble-based quantum token protocol. A bank generates many quantum tokens with angles $\theta_b^i$ and $\phi_b^i$ in a quantum coin device. The latter is then physically handed to the user, who stores it until authentication by the bank is required, which consists in measuring the tokens in the coin and comparing them with the previously generated angles. The coin is accepted if a minimum number of tokens in the coin above a certain security threshold is measured in the right state. See~\cite{Tsunaki_2025,Bauerhenne_2025} for more details.
	}\label{fig:protocol}
\end{figure}

The proposed protocol is schematically presented in \Cref{fig:protocol}.
First, the bank optically initializes the electronic spins of a token and prepares a quantum state by applying resonant microwave pulses. The state is then transferred to the nuclear spins, which have a longer coherence time that can be further extended by dynamic decoupling and charge state control of the NVs. To authenticate the token angles, the bank performs a single-shot readout of the nuclear spin, using the electronic spin as an ancillary qubit. The same protocol, but with different angles, is then applied to all tokens in the coin.
This protocol is benchmarked on five IBM quantum processors and a general attack scenario is analyzed, in which the hacker attempts to read the bank token and forge a fake one, based on the information gained from this measurement. It has been experimentally demonstrated that the probability for the bank to erroneously accept a forged coin composed of multiple tokens can reach values below 10$^{-22}$ , while the probability that the bank accepts its own coin is above $0.999$~\cite{Tsunaki_2025}.

%\paragraph*{Quantum memory approach}
The main idea behind this practical realization of QT is to store quantum coherence for a certain period of time and then reactivate it so that it can be used in a verification routine. First, coherence is generated on the electron spins in NV centers because they can be easily optically polarized and read out. There the electron spin coherence T$_{2e}$ in this system has a maximum lifetime of $2.4$\,ms~\cite{Herbschleb_2019} (in isotopically pure $^{13}$C diamond) and cannot be extended due to the strong coupling of the electron spins to magnetic and electrical environmental noise. However, it is possible to transfer the coherent state to neighboring nuclear spins, where it can be stored for several seconds or even minutes. The reason for this is that the magnetic moment of the nuclear spins is three orders of magnitude smaller than that of the electron spins. This quantum memory can be realized in diamond with the intrinsic nuclear spins to NV centers $^{14}$N ($I = 1$) or $^{15}$N ($I = 1/2$) or using carbon nuclei with $^{13}$C ($I = 1/2$). The main challenge here is to develop a method that allows for fast coherence transfer, compared to the electron spin coherence time and without losses to the quantum memory. Coherence can also be generated directly on the nuclear spins using radio frequency pulses and transferred to the electron spins via a (back) swap gate and later read out optically. For this purpose, the nuclear spins are first initialized (polarized) via the NV centers.

The nuclear spin coherence time T$_{2n}$ is limited by magnetic interaction with spins of the same type and also by coupling to different electron spins (NV centers, P1 centers, other paramagnetic defects) in the diamond crystal. In order to extend the lifetime of the stored coherence state, which would lead to a longer storage time for the quantum tokens, the nuclear spins must be decoupled from the environment. The goal of decoupling protocols (also known as dynamical decoupling~\cite{Munuera-Javaloy_2021}) is to suppress certain terms in the magnetic dipole-dipole interaction and, depending on the application, they can suppress the coupling between the same spins (homogeneous decoupling, e.g., MREV~\cite{Maurer_2012}) or the coupling to other spins (heterogeneous decoupling, e.g., CPMG~\cite{Naydenov_2011} or XY8~\cite{Munuera-Javaloy_2021}). 

%\paragraph*{Limitations and application scenarios}

The application scenario we are envisioning is a diamond-based portable quantum token that can be miniaturized and equipped with integrated photonic structures and microwave electronics. This can generate dynamic decoupling pulses during transport, which allows nuclear spins to be optimally decoupled from electron spins. The current state of the art with integrated photonic chips in combination with the microwave electronics found in mobile phones would allow in the near future an ensemble-based quantum token to be integrated into an ID card, for example. The bank would then be a system at border control. This would allow transit at the airport to be checked with a built-in expiration date. By transmitting secretly the alignment parameters of the quantum tokens (the angles $\theta_b^i$ and $\phi_b^i$), it would also be possible to check the token at another local position.

The main limitation currently are the coherence times of the nuclear spins in diamond. Although they can reach values of several seconds even at room temperature~\cite{Maurer_2012}, this is still orders of magnitude shorter compared to what would be required for a practical implementation. However, since the ensemble based quantum token protocol is hardware agnostic, it could be easily adapted to other promising physical systems like trapped ions and cold atoms. The latter possess long coherence times, but the minituarization of the physical devices is still a challenge due to the required complex experimental setups, like vacuum chambers.

\newpage
%% -----------------------------
%% Project part: 
%% -----------------------------

\section{Quantum token architecture based on hybrid quantum photonics} 
\begin{center}
{A. Kubanek, W. H. P. Pernice, A. P. Ovvyan}
\end{center}

\label{sec:HybridQToken}

%\paragraph*{HybridQToken} %will be removed later

\begin{figure*}[ht!]
  \centering
    \includegraphics[width=.95\textwidth]{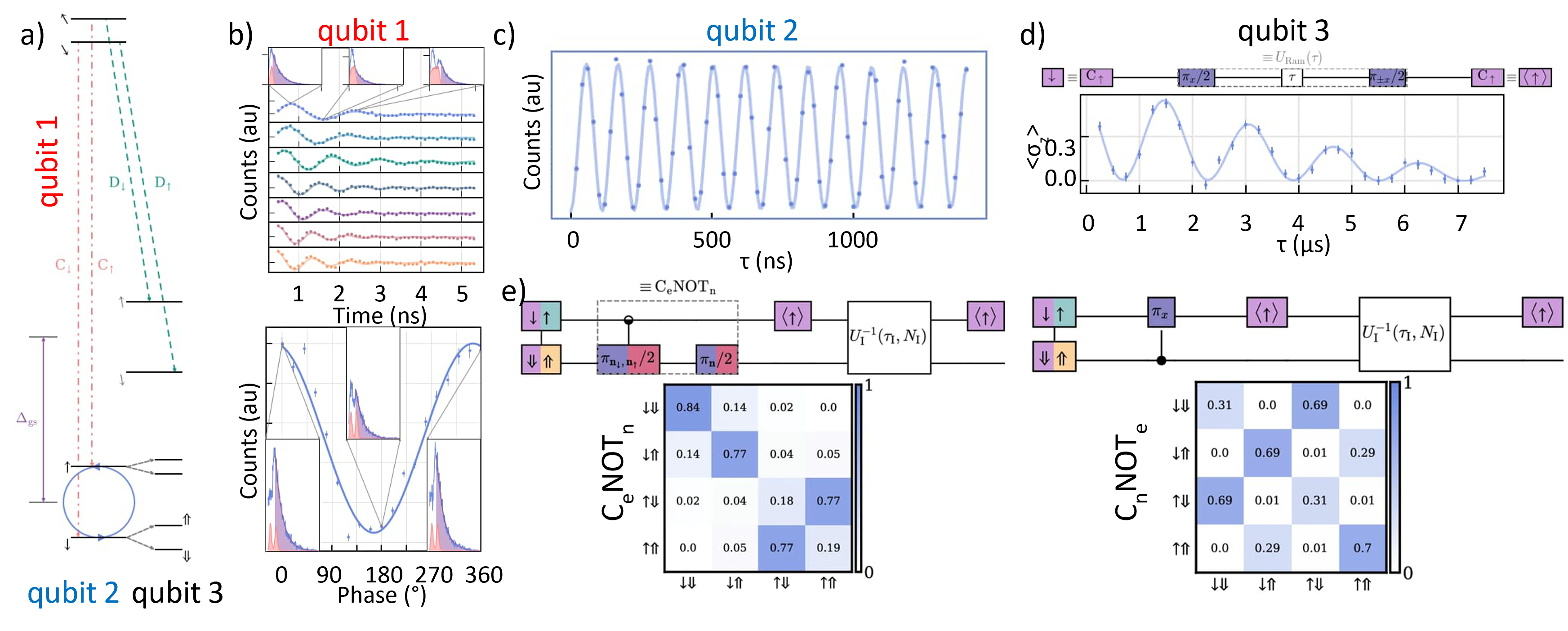}
  \caption{\textbf{Coherent control of the three qubits\textsuperscript{*}.}
  (a) Schematic of the level scheme. The optical dipole of transition C is connected to the SiV's electron spin state. Hyperfine-coupling to nearby nuclear spins splits the electron spin level. (b) Coherent amplitude and phase control of the optical dipole with Rabi-frequencies of about \SI{1}{\giga\hertz}. (c) Coherent, microwave-based electron spin control is demonstrated with a Rabi frequency of about 10 MHz. (d) The coupling to nearby nuclear spins is revealed by electron Ramsey measurement, here with a coupling strength of $A_\parallel /2 \pi = 621.8$ kHz. (e) Electron-mediated nuclear spin control. Amplitude transfer matrix of a $C_eNOT_n$ and $C_nNOT_e$ gate with the corresponding fidelities.  
  \\
  \textsuperscript{*} \Cref{fig:HybridQToken:Fig1} (a) - (e) Reprinted with permission from \cite{klotz_2025}. Copyright 2025, Nature Publishing Group under the terms of the Open Access Publishing Agreement. DOI: /10.1038/s41534-025-01049-2. 
}
  \label{fig:HybridQToken:Fig1}
\end{figure*}

$^{13}$C nuclear spins in diamond are excellent candidates for quantum memories due to their good isolation from the environment. The week interaction with the environment results in long storage times, known to be capable of storing information on the timescale of seconds \cite{Maurer_2012}. At the same time, the quantum memory needs to be accessed and, ideally, interfaced with photons to enable the information exchange over long distance. Strongly coupled spins, including $^{13}$C nuclear spins, have been detected with sub-megahertz resolution in nanodiamonds (ND) \cite{Klotz_2024}. The control of strongly-coupled $^{13}$C has been demonstrated via a nearby SiV center \cite{klotz_2025}, with further improved sensitivity due to modified electron-phonon coupling in NDs \cite{Klotz_2022}, as well as due to large strain \cite{klotz_2025}. Thereby, three coherent processes are established, which enable the exchange of quantum information between long-lived quantum memories and information transmission over long distances. As depicted in \Cref{fig:HybridQToken:Fig1} the electronic spin of the SiV center connects to photons via the optical dipole moment (here transition C). Hyperfine coupling between the SiV centers electron spin and $^{13}$C nuclear spin states enables to mediate coherent interactions to the memory unit. All three qubits, the optical dipole (qubit 1), the electron spin (qubit 2) and the nuclear spin (qubit 3) are coherently controlled. The coherent control rates of the optical dipole was realized with a Rabi-frequency of about \SI{1}{\giga\hertz}, see \Cref{fig:HybridQToken:Fig1}(b). The coherent control of the SiV's electron spin was realized at microwave frequencies with a Rabi frequency of about 10 MHz, see \Cref{fig:HybridQToken:Fig1}(c). The hyperfine coupling to the nuclear spin environment was investigated for a large variation of coupling strengths depending on the relative position and orientation of the spins \cite{Klotz_2024}. For the three qubit system discussed above, the hyperfine coupling strength of $A_\parallel /2 \pi = \SI{621.8}{\kilo\hertz}$ was resolved by means of electron Ramsey measurements, see \Cref{fig:HybridQToken:Fig1}d) to a nearby $^{13}$C nuclear spin. For the exchange of quantum information respective gatter-operations need to be constructed, between the different qubit systems. As an example, the amplitudes of the gatter's transfer matrix for the two CNOT-gates between electron spin and nuclear spin, $C_eNOT_n$ and $C_nNOT_e$, with the respective fidelities are depicted in \Cref{fig:HybridQToken:Fig1} (e). 

\begin{figure}
  \centering
  \includegraphics[width=0.8\linewidth]{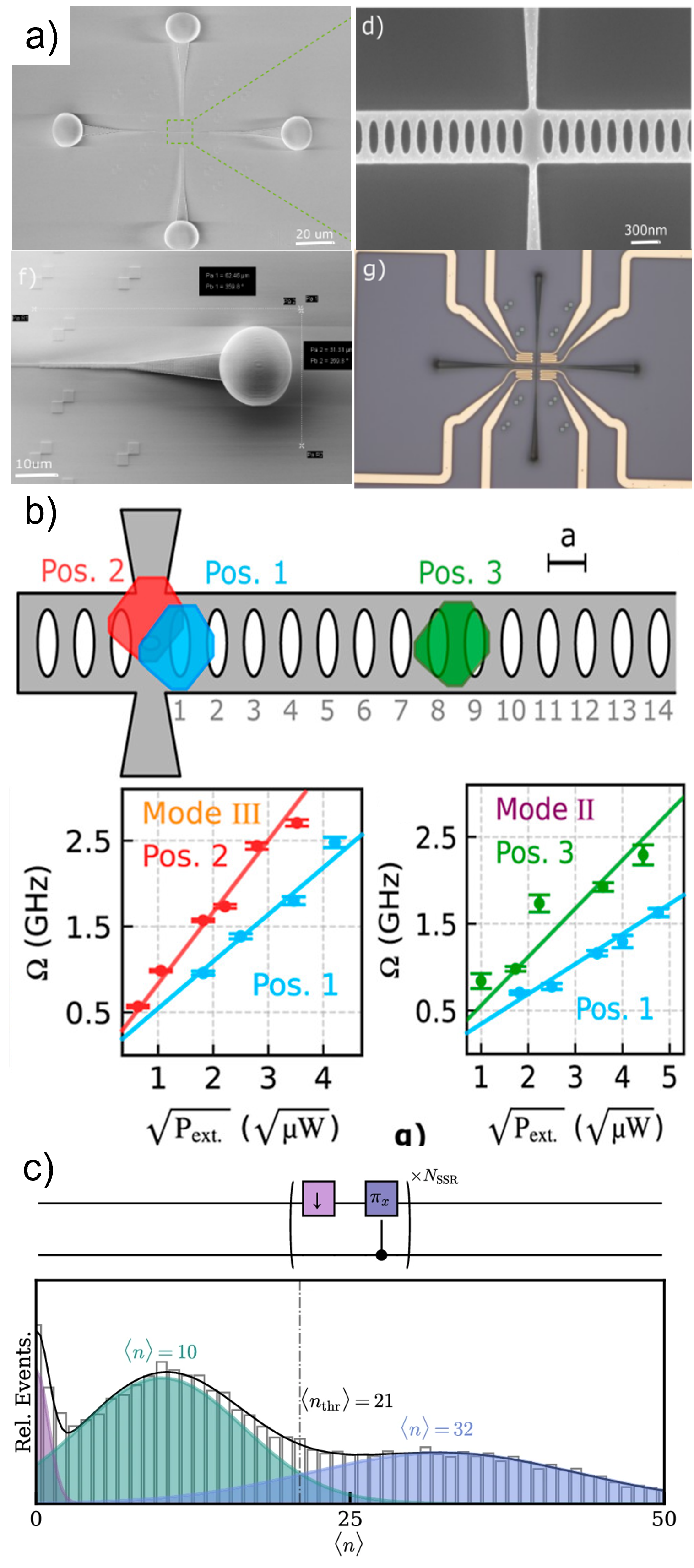}
  \caption{\textbf{Hybrid Quantum Photonics.} (a) $Si_3N_4$-photonics was optimized for its use in the quantum regime including waveguides, photonic crystal cavities with predefined interaction zones, 3D couplers and electronic infrastructure. 
  (b)\textsuperscript{*} AFM-based quantum postprocessing enables the optimization of all degrees of freedom of the optical coupling while in-situ monitoring the hybrid systems' performance by means of maximizing the optical Rabi frequency  (here shown depending on the ND's position within the photonic crystal cavities mode).
  (c)\textsuperscript{**}  
  Photon statistics obtained from a 3 ms long single-shot readout sequence with a separation in the mean photon number between the bright state n=32 and dark state n=10, enables to compensate limitations in the initialization process.
  \\
  \textsuperscript{*}  \Cref{fig:HybridQToken:Fig2} 
  (b) Reprinted with permission from \cite{Lettner_2024}. Copyright 2024, American Chemical Society under the terms of the Open Access Publishing Agreement, licensed under CC-BY 4.0 DOI: /10.1021/acsphotonics.3c01559.
  \\
  \textsuperscript{**} 
  \Cref{fig:HybridQToken:Fig2} 
  (c)  Reprinted with permission from \cite{klotz_2025}. Copyright 2025, Nature Publishing Group under the terms of the Open Access Publishing Agreement. DOI: /10.1038/s41534-025-01049-2
  }
    \label{fig:HybridQToken:Fig2}
    
\end{figure}

The three-qubit unit is realized in a ND with sizes typically below 100 nm. Since the spatial expansion is small compared to the optical wavelength the NDs can be integrated into photonic platforms by means of AFM-based nanomanipulation, realizing efficient coupling between the optical dipole and an optical mode, which is defined by the photonic structure. The quantum unit can be incorporated in a variety of photonic platforms. As a promising material platform, we developed $Si_3N_4$ chip technology with predefined interaction zones and optimized it for its use in the quantum regime \cite{Kubanek_2022}. An example is the strong background fluorescence in $Si_3N_4$, originating from green laser illumination, which was a known limitation for the use of quantum signals in $Si_3N_4$ devices. The developed pump-probe design in a crossed-waveguide architecture enabled a suppression of background signal by 20 dB \cite{Fehler_2020}. Integrated, 1D photonic crystal cavities (PCC) enable an efficient interface between the optical dipole of the SiV center and a single waveguide mode by means of Purcell enhancement \cite{Fehler_2020, Fehler_2021}. 3D-printed coupler build an efficient optical interface to Gaussian, free-space optics and electrodes, microawave- and RF-structures as well as microheaters enable the complete on-chip control of the spin system, as illustrated in \Cref{fig:HybridQToken:Fig2}(a).\\
The quantum post-processing is based on AFM-nanomanipulation and, besides the positioning with nanometer accuracy, enables the control of all degrees of freedom of the optical coupling to individual PCC modes \cite{Lettner_2024}. 
As depicted in \Cref{fig:HybridQToken:Fig2} (b), the quantum post-processing enables the optimization of the hybrid device directly by evaluating the optical Rabi oscillations, while the position and the dipole orientation of selected SiV centers is repeatedly optimized with respect to the overlap with the optical mode. Here, the optical Rabi frequency was increased to about 2.7 GHz. \\
As a result of the enhanced light-matter interaction the corresponding optical transition spectrally broadens. For such spectrally broadened transition, optical access to the electron spin states becomes more efficient \cite{Antoniuk_2024} and the integrated quantum memory units can be accessed on-chip. The resulting hybrid devices combine advantages of an industry-relevant photonics platform, such as $Si_3N_4$, with individually optimized quantum systems enabling full on-chip control under relaxed experimental conditions. \\

Quantum token protocols involve entanglement distribution based on time- or frequency-binned spin-photon entangled states \cite{Bernien_2013}. The optical dipole of a single SiV center is the interface to single photons at a wavelength of about \SI{737}{\nano\meter}. These single photons have shown to be indistinguishable, even for SiV center in seperate NDs \cite{Waltrich_2023}. A low-noise yet highly efficient quantum frequency conversion into the telecom frequency band, such as the telecom C-band, can be adapted in a modular way \cite{Schäfer_2025} and is required for long-distance information exchange via established fiber networks \cite{Bersin_2024}. The electron spin coherence needs to be long enough to exchange quantum information with the $^{13}$C nuclear spin quantum memory, which has been demonstrated in above mentioned references. The demonstrated realization established nuclear spin initialization and readout as well as relevant gatter operations via the SiV's electron spin in a ND. The electron spin coherence time is extended to 273 $\mu s$ at 4K by applying dynamical decoupling sequences \cite{klotz_2025}. A large ground-state splitting reduces phonon-induced electron spin decoherence. The prolonged electron spin coherence enables to address individual $^{13}$C  nuclear spins via indirect electron spin control. The protocols are implemented starting with the initialization of the electron spin state. Using rotations on the electron spin in conjunction with electron-mediated nuclear rotations enables to construct initialization gates, which transfers the electron spin’s population onto the target nuclear spin. A target nuclear spin initialization between 0.647 and 0.682 was achieved, which is limited by the imperfections of the electron spin initialization fidelity. Based on the spin initialization together with conditional and unconditional rotations $C_eNOT_n$ gates as well as $C_nNOT_e$ gates were implemented, as depicted in \Cref{fig:HybridQToken:Fig1}(e). Again, the electron initialization fidelity of about 0.84 limits the gate fidelity which itself is well above 0.9. Imperfect nuclear spin initialization can be overcome by implementing single shot readout of the nuclear spin state. A clear separation of the mean photon number between both nuclear spin state of n=32 for the bright state and n=10 for the dark state, as shown in \Cref{fig:HybridQToken:Fig2} (c), results in a discrimination between both nuclear spin states with fidelities of $F_{bright}=0.935$ and $F_{dark}=0.824$.

Since the ND host can be integrated in conventional photonics, a potentially higher photon-collection efficiency enables single-shot readout on the electron spin, which circumvents remaining limitations of the electron spin initialization fidelity. Increasing the cooperativity of the hybrid quantum photonics systems makes spin-dependent reflectivity accessible which can be utilized for further protocol development \cite{bhaskar_2020} . Additionally, combining an ancillary nuclear spin with single-shot readout and a subsequent SWAP gatter operation can also improve electron spin initialization. An extension to quantum register consisting of multiple $^{13}$C  nuclear spins \cite{bradley_2019} enables fault-tolerance in the storage of quantum information. Extending the system to more than one nuclear spin further increase the complexity of applicable protocols. Finally, it is worth mentioning that these results neither required a vector magnet nor a dilution refrigerator, which enables to realize a simpler version table-top experiment with lower technical overhead and thus positively affects scalability concerns.
 
\newpage
%% -----------------------------
%% Project part: 
%% -----------------------------

\newcommand{\gp}[1]{{\color{blue} #1}}

\section{Quantum tokens based on color centers in diamonds}
\begin{center}
{T. Schröder, K. M\"uller, G. Pieplow}
\end{center}
\label{sec:QPIS}

\begin{figure*}[ht!]
  \centering
    \includegraphics[width=.95\textwidth]{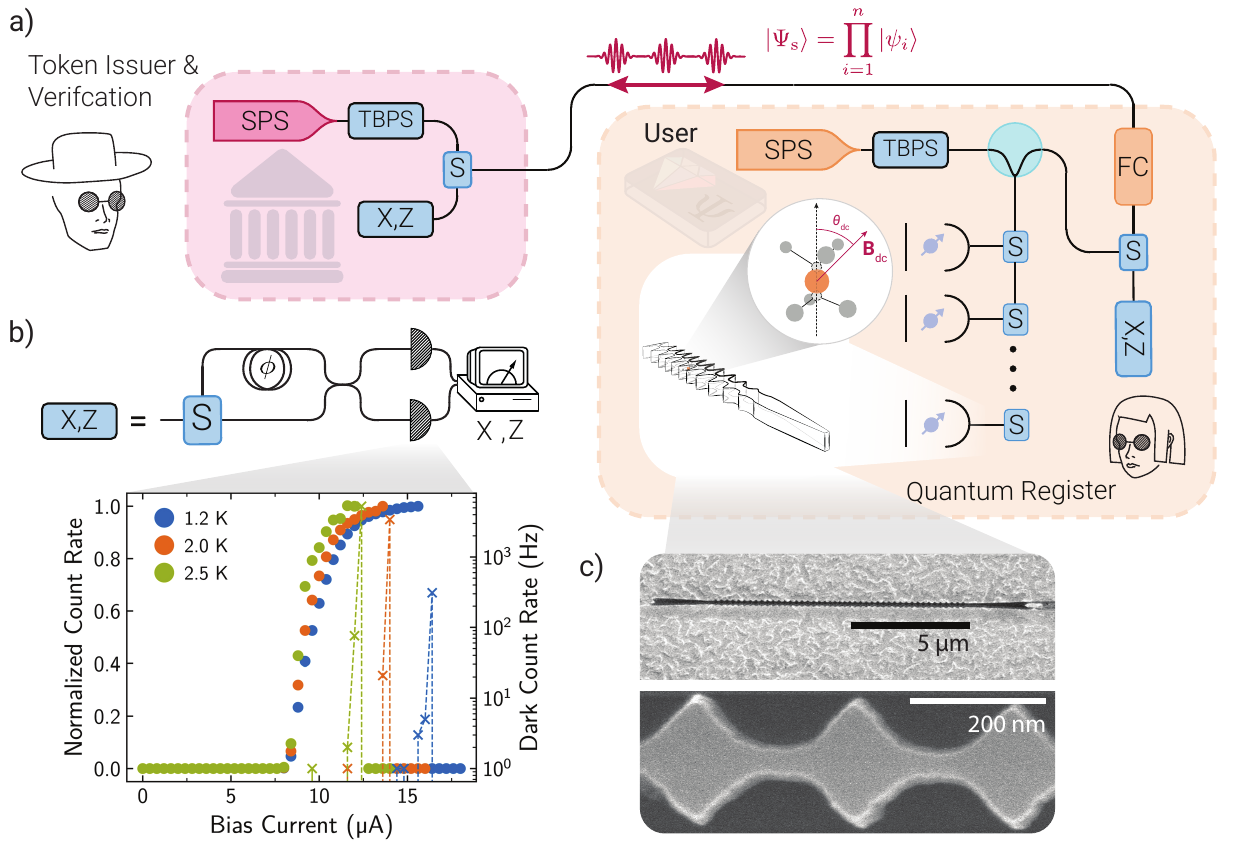}
    \caption{a) The SnV enabled quantum token scheme encompasses token creation, storage, retrieval, and verification. The issuer generates a quantum token $\ket{\psi_{\rm s}}$ with a unique serial number $s$, composed of photonic qubits $\ket{\psi_i}$ prepared via a single-photon source (SPS) and time-bin qubit preparation stage (TBPS) \cite{Lee2019,Bouchard2022,Yu2025}. Photons are encoded in the desired states and transmitted in the infrared, requiring frequency conversion (FC) only at the user end for G4V interaction. Users store the token in a high-efficiency sawfish spin-photon interface array \cite{bopp_sawfish_2024, pregnolato_fabrication_2024}, using electronic or nuclear $^{13}$C spin. Fast switches (S) enable efficient routing and $X$/$Z$-basis measurements, with the $X$-basis implemented via a switch-integrated imbalanced Mach-Zehnder interferometer with controllable phase $\phi$. Storage completes after sequential $X$-basis measurements. For retrieval, the spin state is entangled with an SPS photon sent to the verifier; a $Z$-basis spin measurement follows. The figure was adapted from \cite{strocka_secure_2025}. b) Normalized count rates of a SNSPD illuminated with 780 nm laser light operated at different temperatures. The count rate is saturating towards unity with the bias current approaching the critical current of 
    for lower operating temperatures (figure taken from \cite{grotowski_optimizing_2025}). c) Fabricated sawfish resonator in diamond (figure taken from \cite{pregnolato_fabrication_2024}).} 
    \label{fig:qpis-overview}
\end{figure*}

The token implementation we pursue is based on Wiesner’s seminal quantum money protocol \cite{Wiesner.1983}. Specifically, we studied the generation, storage, and retrieval of Wiesner quantum money states, where storage and retrieval are accomplished through an integrated and highly efficient spin-photon interface \cite{bopp_sawfish_2024, pregnolato_fabrication_2024}, as well as integrated superconducting nanowire single-photon detectors (SNSPDs) \cite{grotowski_optimizing_2025, strohauer_current_2025, Majety2023, zugliani_tailoring_2023, strohauer_site-selective_2023,flaschmann_optimizing_2023}. The token scheme was explained in detail by Strocka \textit{et al.}~\cite{strocka_secure_2025}, from which we adapted \Cref{fig:qpis-overview}. To date, a full implementation of such a message and retrieve token scheme has not yet been achieved with solid state quantum memories, with a recent demonstration \cite{Mamann.31.03.2025} implemented with cold atoms. In this work, we consider the SnV as a quantum memory, which is a group-IV vacancy center (G4V) in diamond. G4Vs include the silicon vacancy (SiV), germanium vacancy (GeV), tin vacancy (SnV), and lead vacancy (PbV). G4Vs in diamond form inversion-symmetric defects with ${\rm D_{3d}}$ symmetry, where the dopant atom occupies an interstitial site between two vacancies. This high symmetry strongly suppresses sensitivity to electric-field noise compared to the NV center, making G4Vs ideal for integration into nanostructures where surface charge fluctuations can be significant 
\cite{OrphalKobin2023,pieplow_quantum_2025,DeSantis2021}.

\begin{figure}[hb!]
    \centering
    \includegraphics[width=\linewidth]{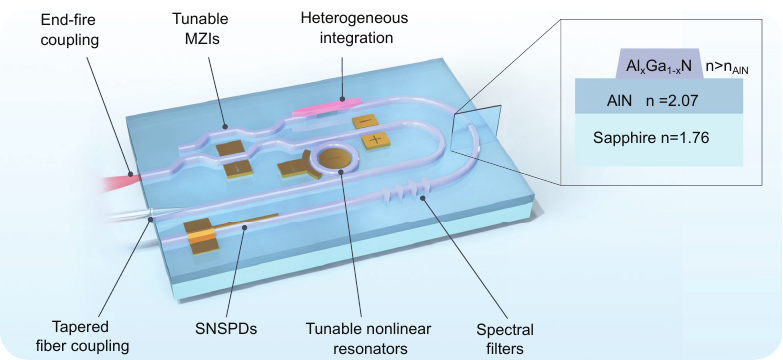}
    \caption{AlGaN/AlN heterostructures for all the functionalities required for routing and switching photons (figure taken from \cite{gundogdu_alganaln_2025}).}
    \label{fig:qpis-2}
\end{figure}

The token protocol we adapted begins with the creation of a Wiesner quantum money state by a token issuer, as depicted in \Cref{fig:qpis-overview}. With the aid of a single-photon source (SPS) and a time-bin qubit preparation stage \cite{Yu2025, Bouchard2022, Lee2019}, the token state $\ket{\Psi_{\rm s}} = \prod_{i=1}^n \ket{\psi_i}$ is produced through the successive emission of single photons, thus utilizing time-division multiplexing (TDM). Here, $n$ denotes the number of photonic qubits, and each $\ket{\psi_i}$ is selected uniformly at random from the set of time-bin qubits ${\ket{E}, \ket{L}, (\ket{E} + \ket{L})/\sqrt{2}, (\ket{E} - \ket{L})/\sqrt{2}}$, where $\ket{E}$ ($\ket{L}$) represents the early (late) time-bin state. Each token state is uniquely identified by a serial number $s$. Time-bin encoding is chosen for this project because it is particularly well-suited to the storage and retrieval protocols developed for G4Vs \cite{Wei2025, Omlor2025,Knaut2024,Bhaskar2020}. When single photons are generated directly in the telecom band \cite{costa_telecom_2025, kaupp_purcell-enhanced_2023, birowosuto_fast_2012}, only a single frequency conversion (FC) stage is needed at the quantum memory node to translate them to a frequency compatible with the G4Vs. Conversion from the visible and near infrared to the telecom spectral range can be readily achieved with quantum frequency conversion systems \cite{schafer_two-stage_2025, chamier_low-noise_2025, stolk_metropolitan-scale_2024,mann_low-noise_2023, dreau_quantum_2018, zaske_visible-telecom_2012}. 
Through a series of routing operations, spin-photon cavity reflections, control operations, and photon detection events, the photons of the token state that survive transmission are stored in the memory register with high fidelity \cite{strocka_secure_2025}. During storage, the token can be processed using the noise-tolerant hidden-matching quantum retrieval game approach \cite{amiri_quantum_2017,Pastawski.2012} to verify and spend it. Alternatively, the state can be returned to the issuer or sent to a trusted verifier. In Strocka \textit{et al.}~\cite{strocka_secure_2025}, we investigated the latter to demonstrate the versatility of the quantum register. The retrieval of the state is performed using an inverse protocol that requires another SPS, which creates single photons that are reflected off the spin-cavity systems.
Notably, through heralded storage, more complex quantum states \cite{ghosh2024existential} can, in principle, be both stored and generated \cite{borregaard_one-way_2020}, such that this memory architecture is not limited to product states.

We focus on the SnV as a quantum memory, exploiting both its electronic spin and nearby $^{13}$C nuclear spins. More generally, the negatively charged G4Vs are optically active, effective spin-1/2 systems with zero-phonon-line (ZPL) transitions at 520 nm (PbV) \cite{trusheim_lead-related_2019}, 619 nm (SnV) \cite{iwasaki_tin-vacancy_2017}, 602 nm (GeV) \cite{neu_single_2011}, and 737 nm (SiV) \cite{neu_single_2011}. When coupled to a nanophotonic crystal cavity, their optically accessible spins provide an efficient and integrable quantum-memory platform \cite{bopp_sawfish_2024, pregnolato_fabrication_2024}. Photons can be stored in such a memory using two main approaches. The first is approach uses amplitude gates \cite{Wei2025, Knaut2024, Bhaskar2020}, which generate intermittent entanglement through spin-dependent reflection of incident time-bin photons combined with a spin control operation. The second approach involves controlled-phase gates \cite{Omlor2025, borregaard_one-way_2020}, which rely on a spin coupled to a one-sided open cavity, a single control operation, and a spin-dependent phase shift upon reflection. In the QPIS project we target the second method because it is not limited to $50$\% efficiency. Such a limit would break the token schemes security threshold for noise tolerance \cite{amiri_quantum_2017, Pastawski.2012}. For both methods the storage of a state is heralded by detecting the reflected time-bin qubits with a measurement device. Such a device also has to operate at near unity efficiency to guarantee a successful implementation of the protocol. Once the storage of a qubit is heralded, the electronic spin state can be swapped onto a nearby nuclear $^{13}$C spin \cite{Beukers2025, Afzal2024} for storage exceeding 17\,ms  \cite{Beukers2025}. The color centers that had the highest priority for us were SnVs \cite{Trusheim2020, Thiering2018}, which have the additional benefit of being sufficiently spin-coherent in the Kelvin range, due to their comparatively large spin-orbit splitting \cite{Harris2024, torun_optical_2023}.

For control of the memory and the creation of time-bin qubits, all-optical approaches have been analyzed both experimentally and theoretically \cite{strocka_secure_2025, pieplow_deterministic_2023, torun2023, Debroux2021}. In particular, a highly efficient microwave control scheme was developed and analyzed, exploiting the orientation of the magnetic control field \cite{Pieplow2024}. Theoretical analysis revealed that, for both microwave and optical control, the fidelity of the spin gates can exceed 99\% \cite{strocka_secure_2025, Pieplow2024}.

For the integrated routing of photons, which is required for scaling the G4V platform to multiple spin-coupled cavities, a new platform was developed \cite{gundogdu_alganaln_2025} (\Cref{fig:qpis-2}c). The detection that concludes the heralding requires highly efficient detectors, which in QPIS are based on integratable SNSPDs \cite{grotowski_optimizing_2025, strohauer_current_2025, Majety2023, zugliani_tailoring_2023, strohauer_site-selective_2023,flaschmann_optimizing_2023}(\Cref{fig:qpis-2}a).

Finally, the most central component of the scheme is the spin-photon interface. In QPIS, we base this interface on the sawfish nanophotonic crystal resonator, which, through its amplitude-modulated external corrugations, overcomes a fabrication challenge related to the adiabatic coupling of a resonator mode to a waveguide \cite{bopp_sawfish_2024, pregnolato_fabrication_2024} (\Cref{fig:qpis-2}b,c).

All these components are critical for the functioning of the scheme and were specifically developed and chosen for the QPIS project to aim at the most efficient nanophotonic operation of the targeted quantum memory.

As of now, G4Vs have demonstrated storage times ranging from 20\,ms for the electronic spin of a germanium vacancy (GeV) \cite{senkalla_germanium_2024} to up to 18\,s for a nuclear $^{13}$C spin coupled to the defect \cite{grimm_coherent_2025}. Achieving long coherence times with the GeV required millikelvin temperatures, which is a constraint expected to be less stringent for heavier G4Vs, such as tin and lead vacancies \cite{Harris2024}. 
Although these are record storage times for optically accessible solid-state spins, such storage times are too short for a realistic quantum money scheme, which would ideally hold token states indefinitely. However, signature schemes that allow an issuer to send a limited number of tokens, which an autonomous agent can use to sign a restricted set of instructions, for example steering commands in a satellite between base stations, could still be enabled by such short storage times \cite{BenDavid.2023}.

\newpage
%% -----------------------------
%% Project part: 
%% -----------------------------

\section{Room-temperature quantum token based on alkali metals and noble gases}  \label{sec:Q-ToRX}
\begin{center}
    {J. Wolters, W. Kilian, I. Gerhardt}
\end{center}

%\npkb{1. Update the fig 13 (your fig 1) following our discussion. Namely, change font/text sizes, line thickness. Put the cell dimensions to the cell image.
%\newline
%2. Do you have another good alternative for your second figure? The drawing does not really fit into here. }

Storing photonic quantum states for several minutes, hours, or even up to days under ambient conditions is hardly addressed in science, but a central challenge for the realization of practical quantum tokens. Quantum memories based on mixtures of alkali metal and noble gas atoms in atomic vapor cells can address this challenge with the aim of ideally extending the storage time into the hour range. At the same time, this approach implements a robust and technological simple overall system. 
The solution discussed here consists of storing optical signals into an electron spin wave in the alkali metal ground state and the subsequent coherent transfer of the spin wave to nuclear spins of noble gas atoms with very long coherence times. This allows us to profit from the efficient quantum optical interface of alkali metal atoms and at the same time the exceptionally long coherence time of noble gas spins, as they are well protected from the environment.

Using the optical interface of alkali metal atoms, single photons or weak coherent states can be stored in a phase coherent fashion~\cite{wol17, bus22}. Naturally, this can be extended to the storage of polarization states in a quantum superposition~\cite{nam17}, which renders these memory systems very flexibly in terms of encoding basis and continuous as well as discrete variable protocols can potentially be implemented. The acceptance band of alkaline memories is naturally limited to the optical transitions of alkali metal atoms. Most prominent are the D$_1$-lines of cesium and rubidium at 894~nm and 795~nm, respectively. Alternative options comprise the D$_2$ lines, or transitions to higher orbital angular momentum states such that a variety of near-infrared to the telecommunication C-band can be used~\cite{tho24, maa24, fin18}. Suited signal bandwidths range from few MHz up to the GHz regime, which allow for fast writing of quantum information into the memory. As sources of quantum token bits, weak coherent states or single photons can be used. While the former can be conveniently generated by attenuated lasers, the latter can be generated by semiconductor quantum dots, molecules, cavity-enhanced spontaneous parametric conversion~\cite{mot20} or four-wave-mixing in the same alkali metal species~\cite{cra23}. In both, recent experimental progress allowed first proof of concept experiments~\cite{maa25, tho24, bus22, wan25}.

The storage inside a ground state spin-wave is commonly limited by experimental factors, such as the spin-decoherence due to atomic collisions with the walls of the cell. Another disadvantageous process is the diffusion of atoms out of the optical probing region. Subsequently, storage times on the single photons level have been limited to the millisecond range but have been extended in the single-digit second range with coherent pulses and other measurement schemes. The direct storage of quantum information in nuclear spins of a noble gas would therefore be an interesting alternative. The research on the nuclear magnetic properties has been extensively performed on ideal spin systems such as $^{3}$He, $^{129}$Xe (both spins of $1/2$), or $^{21}$Ne (spin $3/2$). Unfortunately, a direct optical access to these systems is extraordinary hard~\cite{altiere_pra_2018}. Therefore, mediation by another species is needed for the polarization and optical read-out. Polarization of noble-gas spins is performed by spin-exchange optical pumping~\cite{Gentile2017_RMP_45004} (SEOP) for all rare gases.
%or performed by the pumping of the spin ensemble via a metastable state (metastablility-exchange optical pumping, MEOP) \cite{Gentile2017_RMP_45004}.
As the quantum information is stored in the form of single photons into the alkali part of the memory, an effective transition from the alkali ensemble towards the noble-gas ensemble is required. Such spin-exchange experiments have been carried out in the past decades~\cite{Kornack2002_PRL_253002}. Novel developments, which enhance the exchange towards strong coupling, have been performed recently~\cite{Shaham2022_NP}, and enable an effective interface for noble-gas atoms towards the storage of quantum information into the memory.

Photon storage in alkali metal vapors at near room temperature is particularly appealing, as it requires neither complex cooling mechanisms nor large magnetic fields. 
In alkaline $\Lambda$-type memories, a single photon is mapped to a spin wave $ |s\rangle =\sum_j e^{i\Delta k r_j} |\downarrow_1,..,\uparrow_j,\downarrow_N\rangle$, i.e.\ a coherent macroscopic superposition of all $N$ atoms 
\begin{figure}
    \centering
    \includegraphics[width=0.45\textwidth]{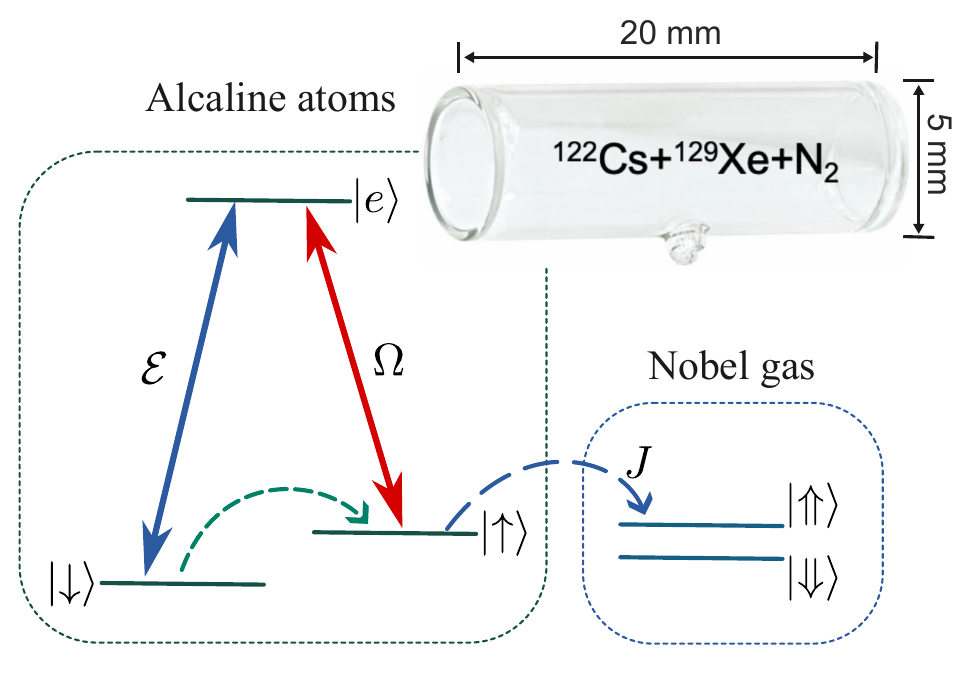}
    \caption{Visualization of the spin-exchange quantum memory scheme and an atomic vapor cell with filling of $^{133}$Cs, $^{129}$Xe with nuclear spin 1/2 and N$_2$ for fluorescence quenching. Initially, all alkali metal and noble atoms are prepared into the states $|\downarrow \rangle$ and $|\Downarrow \rangle$, respectively. A Photonic quantum state $\mathcal{E}$ and strong classical control field $\Omega$ induce a two photon transition between the alkaline ground states $|\downarrow \rangle$ and  $|\uparrow \rangle$. When the noble gas spin is tuned in resonance by a magnetic field, the spin exchange collisions will lead to the strong coupling $J$, such that a coherent spin transfer to the noble gas can be achieved. For readout the process is reversed and the photonic quantum state is retrieved.}
    \label{fig:Q-ToRXSpin:Exchange}
    %\end{wrapfigure}
\end{figure}
The maturity of such memories has been continuously improved since the first demonstrations of memories based on electromagnetically induced transparency (EIT) in the early 2000s~\cite{Phillips2001}. In the past decade this development of quantum memory implementations in alkaline vapor gained remarkable momentum; reviews can be found in~\cite{shi23, hes16, san11}. Here we will summarize a few highlights: the efficiency of a room-temperature EIT-like memory was pushed beyond 80\%, by control pulse optimization~\cite{guo19, rob24}; EIT-like quantum memories reached $\approx$ 1 GHz bandwidths~\cite{wol17}; storage of polarization states with more than 90\% fidelity has been shown~\cite{wan25,nam17}. 
%Recently, single photons were interfaced with such a quantum memory. 
We reduced the read-out noise of the memory considerably~\cite{esg23}, which allowed single photon storage and retrieval for the first time in a ground-state vapor cell memory, with an photon-photon correlation function of $g^{(2)}(0) = 0.177(23)$. This testifies the quantum character of the light recovered~\cite{bus22}. We demonstrated a multiplexed random-access optical memory that allowed for storage and retrieval of several independent signals~\cite{mes23}, which shows the potential to store large amounts of photonic quantum states. Moreover, these memories were readily integrated into a standalone mobile system~\cite{jut24, wan22}, and we developed simulation tools for accurate performance estimation~\cite{Rob25}.
Despite being a macroscopic superposition state, the spin wave exhibits a comparatively long coherence time, ranging from few \si{\micro\second} to several \si{\milli\second}~\cite{kat18}. To extend that into the regime of several minutes to hours, it was proposed by~\cite{kat22} to use strong coupling to noble gas spins during spin exchange collisions~\cite{Kornack2002_PRL_253002}, alike sketched in~\Cref{fig:Q-ToRXSpin:Exchange}. The ability to transfer the spin state from the alkali spin ensemble back and forth to the noble gas spins was also demonstrated just recently~\cite{Shaham2022_NP}, where hour long coherence times in the $^3$He noble gas ensemble were observed, also attainable in $^{129}$Xe ~\cite{Kilian2007_EPJD_197}.
Such a room-temperature quantum memory, based on alkalies and noble gases, could be implemented in small spectroscopy cells (cf.~\Cref{fig:Q-ToRXSpin:Exchange}) and would have the advantages of (i) continuous room-temperature operation, (ii) up to hour-long storage times, (iii) optical interface at the single-photon level, and (iv) ability of miniaturization \& transportability. 

%\begin{figure}
%    \centering
%    \includegraphics[width=\columnwidth]{Figures/Q-ToRX/qtorx_03.pdf}
%    \caption{Geometry of the alkali vapor gas memory. The cell is heated up to 100~$^\circ$C.}
%    \label{fig:Q-ToRX:03}
%    %\end{wrapfigure}
%\end{figure}

A quantum token based on alkali atoms and noble gases will be suitable for numerous applications. 
The quantum money protocol from~\cite{Bozzio.2018} can be implemented. For this purpose, the legitimates issuer (bank) generates a large scale of randomly selected quantum states and transfers them to the customer as a quantum token, which corresponds to a certain amount of money, via a fiber-optic interface. The customer stores these in a transportable, multimodal quantum memory. The customer can then hand over the stored quantum token to a legitimate receiver (retailer) for payment. The retailer carries out measurements on the quantum token (challenges) and sends the results to the bank for authentication. If authentication is successful, the bank authorizes the quantum token as genuine and transfers the corresponding amount of money to the merchant's account.
Another exciting use case is client-server identification using quantum physically uncloneable function (PUF). This application requires a secure identification protocol between the communicating parties to ensure a secure data exchange as outlined in Ref.~\cite{Doo21}.
Potential limitations to these application scenarios are that the process is so far limited to the usual alkali metal wavelengths, bulky magnetic shielding is required to suppress external field fluctuations. The number of required stored states for a token could increase due to noise/low fidelity, the robustness of the transmission of the spin wave from the alkali to the noble gas ensemble still needs to be experimentally demonstrated as well as the scalability. As the quantum superposition is linked to the internal states of freedom and a holding magnetic field is likely required, the quantum memory needs to be shielded within a homogeneous magnetic field from the environment. Magnetic disturbances and magnetic inhomogeneities would likely limit the stored quantum information. Moreover, a cross-talk to rotational sensing (gyroscopy) might be of interest. In effect, the memory can also act as an atomic gyroscope, and the relative orientation of the memory against the storage frame of reference will be required.

In summary, when these technical hurdles are overcome, atomic  ensemble memories offer a unique potential for real-world applications due to there long coherence times and robust room-temperature operation.  
%\npkb{While you have openly discussed all limitations, I would suggest to highlight/remind with another strong sentence that this your memory does nor require heavy cryogenic equipment and highly promising for room-temperature storage and beyond ( the gyroscope operation.)}
\newpage
\section{Quantum Physical Unclonable Functions and Quantum Tokens}
\begin{center}
    {J. Nötzel, C. Deppe, H. Boche}
\end{center}
\label{sec:Q-TOK}
Authentication, the act of proving one’s identity, has been a cornerstone of human society over centuries. It is expressed via objects such as keys and seals, via capabilities like signatures or via secrets, like passwords. Over the past decades, it has become a field of intense research in cryptography and is present in everyday-life via passwords, fingerprint scanners, signatures and personal identification numbers (PINs). 
Taking the way in which an ordinary key works as an example, identity may be proven by a lock in a simple way via the possession of a key: If a lock accepts only one key, then the person holding that key proves its identity by being able to open the lock. The lock then serves as mechanism that probes the key and, based on the response, either opens or stays closed. Such a mechanism which is able to verify the identity of a key is called a verifier.
Due to its immense importance, cryptography has made strong attempts of quantifying the level of security provided by an authentication method. While from technological perspective, the concept of identity may even be associated with a certain physical state of the human brain, in which the password or PIN is stored, the security of an authentication method is hard to quantify at this level. Rather, the security of such methods is analyzed in terms of the mathematical complexity of breaking them and of the physical security of the involved actions or devices.
The second approach replicates more closely the mechanism underlying present-day physical unclonable functions \cite{Pappu.2002}. 

Physical Unclonable Functions (PUFs) are hardware security primitives that exploit inherent, uncontrollable random physical variations introduced during the manufacturing process. These variations are practically impossible to replicate -- even by the original manufacturer -- making PUFs a powerful tool for authentication and key generation \cite{che2015puf, delvaux2014helper}. The behavior of a PUF is commonly characterized by a set of Challenge-Response Pairs (CRPs): for a given input challenge, the PUF produces a unique and unpredictable output response. Security relies on the assumption that, due to the underlying randomness, it is computationally infeasible to predict the correct response for an unseen challenge, even when a large number of CRPs are known \cite{shamsoshoara2020survey}. Therefore a PUF can be authenticated via the pre-stored CRPs.

Despite their promise, classical PUFs (cPUFs) remain vulnerable to side-channel and machine-learning attacks, which can compromise security by learning response behavior from observed data. 
To precisely understand the ultimate security limits of the PUF concept, recent work started to formalize it from a quantum theory perspective. Since the no-cloning property of carefully designed quantum states indicates that the \emph{unclonability} of a PUF may actually be turned into a proven feature resting on physical theories rather than an empirically justified assumption, first implementations are being attempted. 

Currently, two principal research directions can be clearly distinguished in the literature:
The first one contains quantum tokens and Quantum-Read Physical Unclonable Functions (QR-PUFs), the second one the so-called Quantum Physical Unclonable Functions (qPUFs). The concept of a quantum token is based on Wiesner's conjugate coding scheme \cite{Wiesner.1983}, while the so-called QR-PUFs were introduced in \cite{skoric2009qrpuf}. The work \cite{skoric2009qrpuf} made no reference to \cite{Wiesner.1983}, but from a theoretical perspective both approaches share important construction principles. One of these is the underlying idea of utilizing, as a means of authentication of the prover, a physical system that is \emph{unclonable}. While this property follows from the no-cloning property in the case of quantum tokens, it is an assumption in \cite{skoric2009qrpuf}. Further, both QR-PUFs and quantum tokens utilize classical information or lookup tables (also called \emph{trusted enrollment data}) at the verifier. This information can in principle be copied, as it is not protected by the no-cloning principle. 

An authentication scheme following the concept of conjugate coding is depicted in \Cref{fig:qtok:qr-puf}. The alternative realization following the exposition in \cite{skoric2009qrpuf} is captured therein by assuming the measurement is realized based on the injection of quantum states, which are themselves generated based on classical information stored in the lookup table. 

In contrast, qPUFs (that were introduced in \cite{Arapinis.2021}), can be operated solely on the basis of quantum information at the verifier's side -- no such memory is a priori needed at the prover. However, for qPUFs only the hardness of cloning the hardware in possession of the prover is proven within a formal hardware model, thereby enabling information-theoretic security statements. We thus observe two design approaches with different security assumptions or properties, and pronounced differences in the design challenges.

\begin{figure}[t!]
	\includegraphics[width=.5\textwidth]{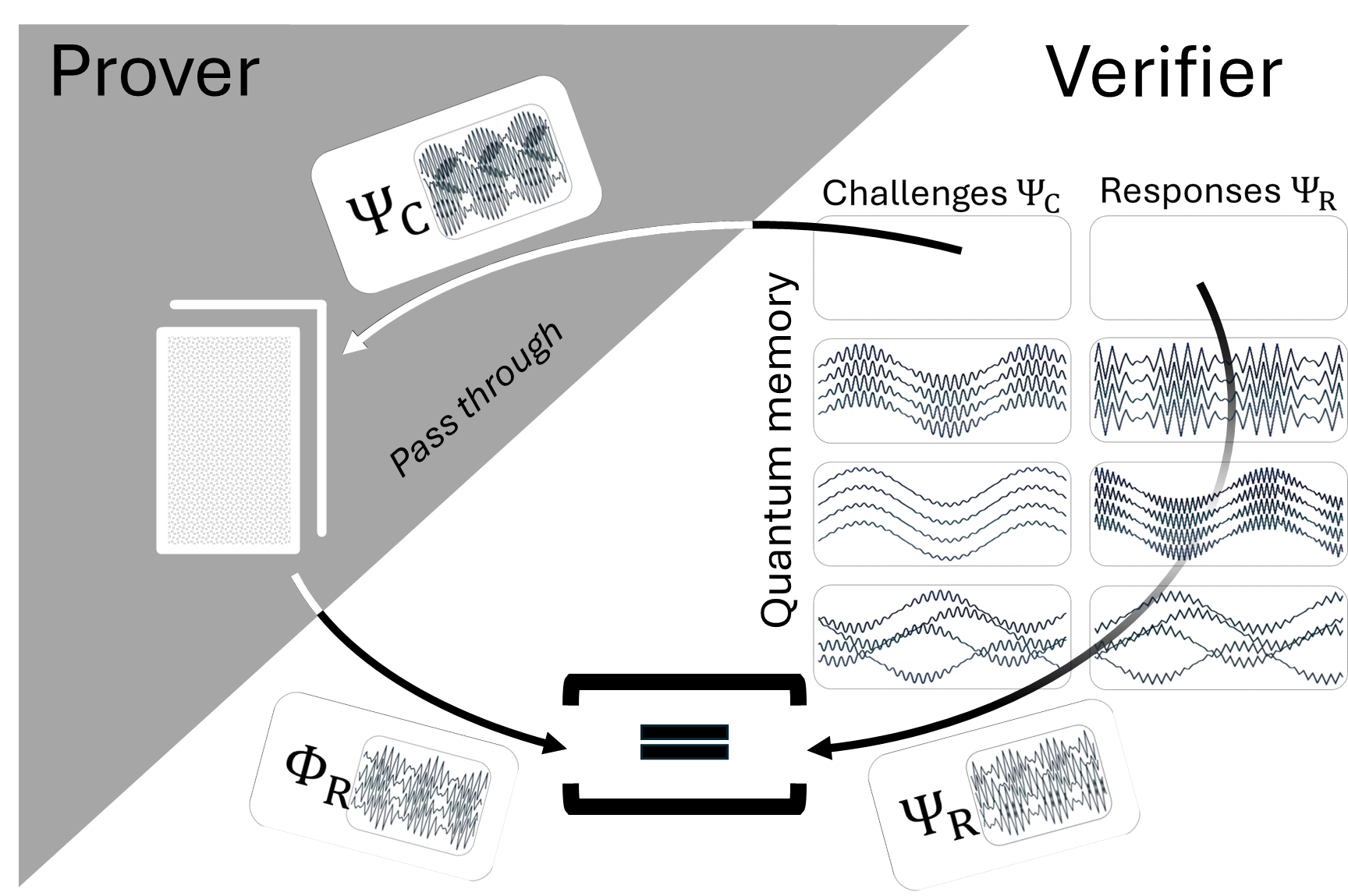}
	\caption{\label{fig:qtok:qPUF} Authentication scheme based on qPUFs \cite{ghosh2024existential}. Both $\Psi_C$ and $\Psi_R$ are generated at the time where the authentication takes place based on pre-stored entangled states.
	}
\end{figure}

\begin{figure}[b!]
	\includegraphics[width=.5\textwidth]{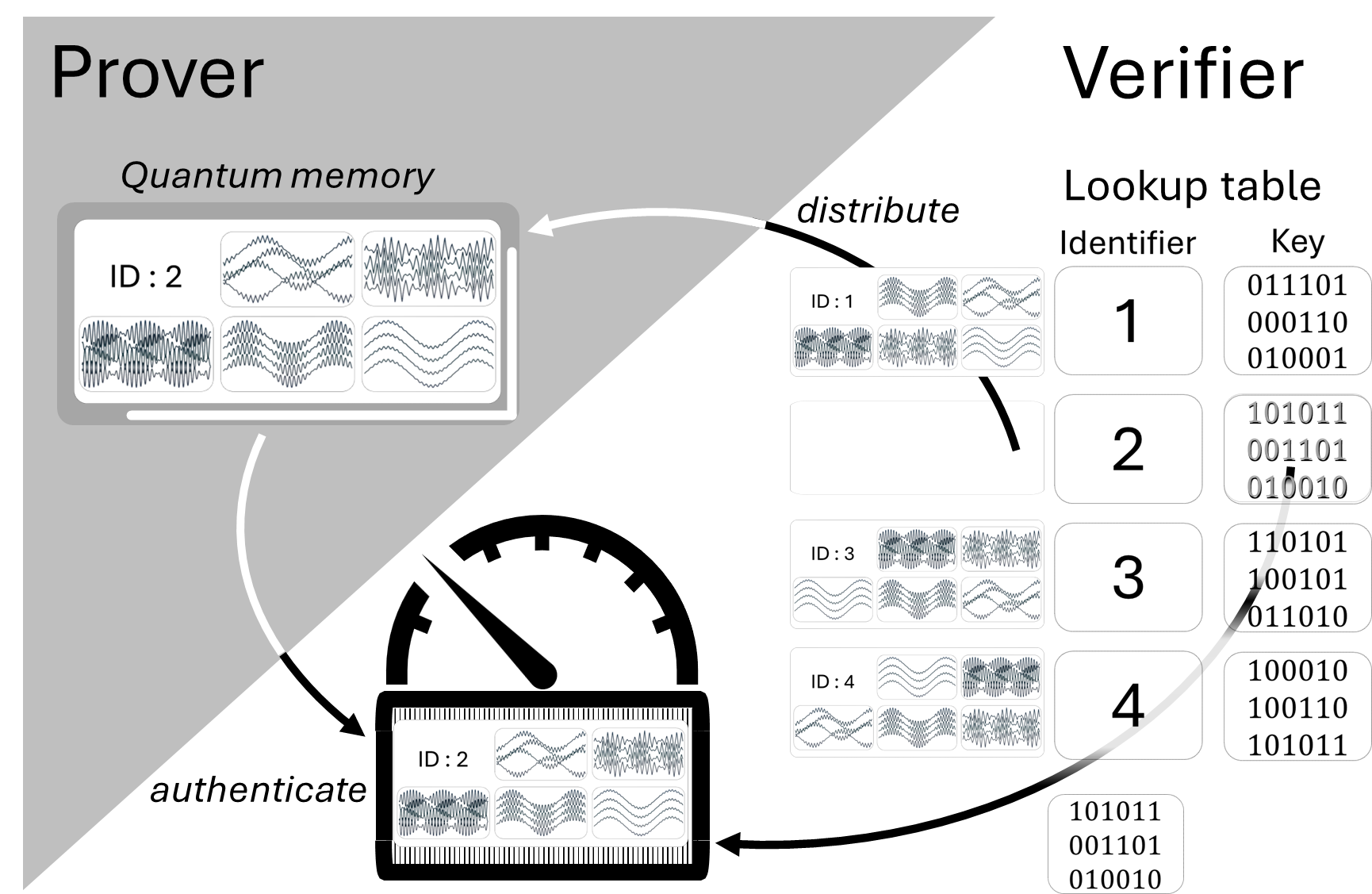}
	\caption{\label{fig:qtok:qr-puf} Authentication scheme based on conjugate coding, following the first research direction.}
\end{figure}

Two types of attacks on qPUFs are considered: A first attack models an adversary trying to get hold of the table of queries and responses. Since the responses of a qPUF may be obtained as quantum states, the no-cloning theorem forbids an attacker to create a copy of the query and response table. Rather, such an attacker must steal the entire table, which may be noticed by the verifier. Further, qPUFs are designed to withstand a specific type of cloning attack where the attacker gets hold of the qPUF and creates their own table of query and response table before returning the qPUF to its legitimate user to avoid detection of the attack: As has been proven in recent work \cite{ghosh2024existential}, any such effort will require a significant amount of resources. 

Quantum Physical Unclonable Functions (qPUFs) have recently gained significant attention as a foundation for secure authentication, storage, and identification in next-generation communication systems. Leveraging the quantum no-cloning principle and the inherent randomness of physical quantum systems, qPUFs offer a promising path toward robust cryptographic primitives that can withstand even quantum adversaries.
One line of research introduces a qPUF-based approach for secure authentication protocols \cite{Kumar2025ICC_Auth}. In these schemes, the unique quantum response of a device to a given input challenge enables reliable and unclonable identity verification. The protocol design ensures that even an adversary with partial knowledge of challenge-response behavior cannot predict new responses, thanks to quantum principles and non-replicability of the physical device.

Extending this idea, another study explores how qPUFs can be applied to the secure storage and identification of data \cite{Kumar2025ICC_Storage}. The work shows how each device’s quantum response can serve as a robust and physically bound identifier or secure storage primitive. These applications are particularly relevant in distributed systems where hardware-level trust anchors are essential.

A further development combines qPUF-based key generation and secure storage in the presence of side information \cite{Kumar2025ISIT_SideInfo}. Here, side information might originate from environmental noise or partial leakage to an adversary. The proposed framework demonstrates how secret keys can still be reliably extracted while ensuring provable security, even in the presence of such information.

These implementations build upon a comprehensive theoretical foundation that quantifies the information-theoretic properties of qPUFs \cite{Kumar2024ISIT_InfoTheory}. This work provides detailed entropy bounds and uniqueness guarantees, showing that qPUFs can generate high-quality randomness suitable for cryptographic use and discusses trade-offs between robustness and key reproducibility.

A complementary approach investigates secret key generation and storage mechanisms purely based on qPUFs \cite{Kumar2025ITW_Storage}. This work considers scenarios with practical constraints such as noisy quantum measurements and device variations. It proposes %coding-based 
strategies to mitigate error effects while maintaining security.% the security benefits inherent to qPUFs.

Potential applications of qPUFs in the Internet of Things (IoT) are examined in another study \cite{Kumar2024WFIoT}. It highlights the suitability of qPUFs for decentralized, low-power devices where conventional cryptographic methods may be infeasible. This line of work opens new avenues for integrating quantum security primitives into embedded systems and edge computing environments.

In \cite{farre2025entanglement} an authenticated QKD protocol using maximally entangled qubit pairs that secures BB84 under minimal assumptions was presented. The protocol enables secret-key expansion through reuse of pre-shared information. In noisy conditions, simulations with cavity-enhanced Atomic Frequency Comb memories show that both statistical methods and trained DNNs can detect adversaries with over 80\% accuracy at 150~$\mu$s memory time and 1~m distance.

Two patents were submitted describing practical qPUF-based security systems. They outline a method for securely storing and retrieving data messages in a memory device using physical identification mechanisms \cite{Boche2025Patent1}, and techniques for securely enrolling data into a device and later identifying whether a specific message was previously stored \cite{Boche2025Patent2}. Both inventions are designed to be resilient against duplication and unauthorized access, supporting future secure hardware development.

These works establish a coherent framework for using quantum randomness as a trust anchor in digital systems, demonstrating the theoretical and practical potential of qPUFs for post-quantum security, secure IoT, and critical infrastructure.

%While QR-PUFs have seen a larger number of successful implementations on the level of laboratory experiment~\cite{}\npk{Janis: a reference is wanted, Can you mention more specific examples of such systems? This would be very relevant, since we talk about physical implementations}, qPUFs have not been implemented yet. A major challenge here is the realization of random unitary transformations over several quantum systems. These random unitaries may be realized in the sense of One-Time Programmable (OTP) memory and may allow the realization of systems such as \npk{???} \cite{Arapinis.2021}. Future research would then need to assess their security properties. To realize more advanced schemes such as \cite{ghosh2024existential}, an additional quantum storage would be necessary. 

While quantum-readout PUFs (QR-PUFs) have already been demonstrated in laboratory experiments, fully quantum PUFs (qPUFs) remain, to the best of our knowledge, without an end-to-end experimental realization that meets the formal qPUF definitions and associated security notions. In particular, quantum-secure authentication of a \emph{classical} multiple-scattering optical key using few-photon (weak coherent) challenges has been experimentally demonstrated, providing security against digital emulation attacks in remote settings~\cite{goorden2014qsa,skoric2009qrpuf}. Related photonic platforms that are compatible with quantum-readout ideas include scalable integrated photonic PUFs (e.g., silicon-photonic interferometric structures) and reconfigurable disordered polymers, which improve stability and facilitate large-scale deployment~\cite{tarik2022sipuf,nocentini2024multilevel}.

In contrast, qPUFs as formalized in \cite{Arapinis.2021} require the device itself to realize an intrinsically quantum primitive (often modeled as an unknown quantum transformation or, more generally, a quantum channel) that is efficiently verifiable and resistant to quantum-polynomial-time forgery. A major obstacle for implementing such constructions is the physical realization of sufficiently high-dimensional \emph{random} quantum transformations with controlled noise, together with a reproducible verification interface \cite{Arapinis.2021}. One pragmatic route is therefore to consider \emph{one-time} (consumable) quantum tokens—analogous in spirit to private-key conjugate coding—where verification necessarily disturbs the stored state and thus each authentication attempt consumes the token, requiring a fresh re-initialization or re-issuance for subsequent use. Such one-time usage can reduce the adversary’s ability to accumulate information across repeated interactions, but shifts the engineering burden to reliable quantum state preparation and (possibly) short-term quantum storage \cite{Arapinis.2021}.
Finally, realizing more advanced qPUF-based authentication schemes that aim at stronger notions of unforgeability via measurement-based or non-unitary constructions~\cite{ghosh2024existential} is expected to require additional quantum resources, in particular quantum storage and coherent control to preserve quantum states across the verification procedure \cite{ghosh2024existential,vernazgris2018memory}.
\newpage
\section{Quantum tokens in the post-quantum world}
\begin{center}
    {J. Krämer}
\end{center}

%\subsection{The Post-Quantum World}

Quantum tokens are related to quantum cryptography, i.e., they use quantum technology to provide cryptographic functionalities. Post-quantum cryptography is conceptually different: Due to Shor's algorithm and its ability to compute discrete logarithms and large prime factors~\cite{shor1994algorithms}, it has been known for many years that the cryptographic public-key algorithms that are currently ubiquitously in use will not provide any security once cryptographically relevant quantum computers exist, i.e., quantum computers with enough stable qubits and low enough error rates to run Shor's algorithm successfully on instances currently used in public-key cryptography. According to the German Federal Office for Information Security, it is likely that such quantum computers will already exist by 2040~\cite{BSI}.

The response of the cryptographic community to this threat is the development of cryptographic schemes which base their security on hard mathematical problems which are assumed to resist attacks with quantum computers, such as Shor's algorithm. This kind of cryptography is called post-quantum cryptography (PQC), or quantum-resistant cryptography. In contrast, the cryptography used today, which will not withstand attacks by quantum computers, is called classical cryptography. Hence, post-quantum cryptography does not use quantum technology (and is not part of quantum cryptography), but promises security against attackers who can use a powerful quantum computer.

For at least two reasons, using post-quantum cryptography in practical applications may not begin only when cryptographically relevant quantum computers exist: First, the migration to post-quantum security, i.e., the exchange of currently used cryptographic schemes with post-quantum schemes, is a very complex and time-consuming undertaking and a field of research in its own right. Second, quantum computers are not only a threat once they exist - from an attacker's perspective it might be a good idea to collect large amounts of classically encrypted data now and store it until cryptographically relevant quantum computers exist. These can then be used to decrypt the data and to learn the information that should actually be confidential. These so-called "store now, decrypt later" attacks can be prevented by using post-quantum cryptography already today.

%\subsection{Differences and Similarities Between Quantum Tokens and PQC}

A major difference between quantum tokens and PQC has already been pointed out: While quantum tokens use quantum technology, post-quantum cryptography is designed to run on classical, non-quantum computing devices.
%, but to provide security against attackers equipped with a quantum computer.
Another difference concerns their security promise: Quantum tokens use quantum technology, which means that they are information-theoretically secure. Post-quantum cryptographic schemes, on the contrary, usually rely on mathematical complexity assumptions, meaning that cryptanalytic attacks might exist, both classical and quantum attacks. The maturity of quantum tokens and post-quantum cryptography is a further difference: As this paper demonstrates, first experimental demonstrations exist, but quantum tokens are still a long way from being widely used in practice. For post-quantum cryptography, however, the US-American National Institute of Standards and Technologies (NIST) initiated a standardization process already in 2016, leading to five post-quantum schemes that have been selected for standardization by now (two key-encapsulation mechanisms and three digital signature schemes). Three of those have already been standardized~\cite{NIST_PQC}. An additional standardization process for digital signature schemes is currently underway~\cite{NIST_PQC_sig}. Consequently, post-quantum cryptography is ready to be used in practice, and many practical applications exist, e.g.,~\cite{OQS,Signal,google,tuta}. 

%similarities: beide werden zukünftig in der IT-Sicherheit verwendet
There are also two relevant similarities between quantum tokens and PQC: The first concerns the purpose of their use: both quantum tokens and post-quantum cryptography are intended to be used to achieve specific security goals. 
%The second concerns their motivation by the rapid development of quantum technologies: for post-quantum cryptography, this rapid development implies that it will become necessary, and for quantum tokens, that they will become possible.
The second similarity is that the rapid development of quantum technologies is greatly accelerating their use in practical applications: for post-quantum cryptography, the rapid development of quantum technologies implies that they are becoming increasingly important, and for quantum tokens, that they can be realized.

%\subsection{Combing Quantum Tokens and PQC}
% the best of both worlds?
%3) for specific solutions they can be combined: auch weil sie gleichzeitig genutzt werden: die einen, weil sie dann nötig sind, die anderen, weil sie dann möglich sind
Since the practical application of both PQC and quantum tokens is driven by the rapid development of quantum technologies, there will come a time when both can be combined in practical use.
Although the migration to post-quantum security is challenging, it is easier than a future migration to security solutions that integrate quantum technology, since PQC principally does not require specialized hardware. Hence, the combination of quantum tokens and PQC is especially interesting for use cases where the necessary effort in quantum infrastructure is justified by the additional security guarantees that quantum tokens offer. For instance, when it comes to putting QKD into practice, proponents emphasize that combining QKD with classical symmetric cryptography, e.g., the AES encryption algorithm, combines unconditionally secure key establishment with efficient and practical encryption.
Opponents, however, stress that QKD requires authentication, which is often realized with public-key cryptography, i.e., complexity-based cryptography. When the authentication of QKD relies on mathematical complexity assumptions, however, the promise of QKD to be unconditionally secure no longer holds true in practice. 
One perspective now could be that investing in QKD technology is not effective, but key establishment should continue to be based on public-key cryptography (and, hence, PQC). 
Another perspective, would be to solve the issue of authentication in certain scenarios differently, e.g., through a physical meeting of the parties involved. While this may sound unnecessary complicated at first glance, there are high-security areas where this kind of authentication is completely adequate. In such high-security use cases, hence, investing in QKD technology and combining QKD with PQC can be useful in order to achieve a very high level of security. 
In the future, similar scenarios will arise where quantum tokens and PQC can be combined and where the advantages of quantum tokens outweigh their supposed investment costs, especially in terms of their unforgeability, the possibility of local validation, and the user privacy that can be achieved with quantum tokens.
%\newpage
%\begin{widetext}
%\hrule
%\hrule
%\end{widetext}

\begin{figure*}[th!]
    \centering
    \includegraphics[width=0.99\textwidth]{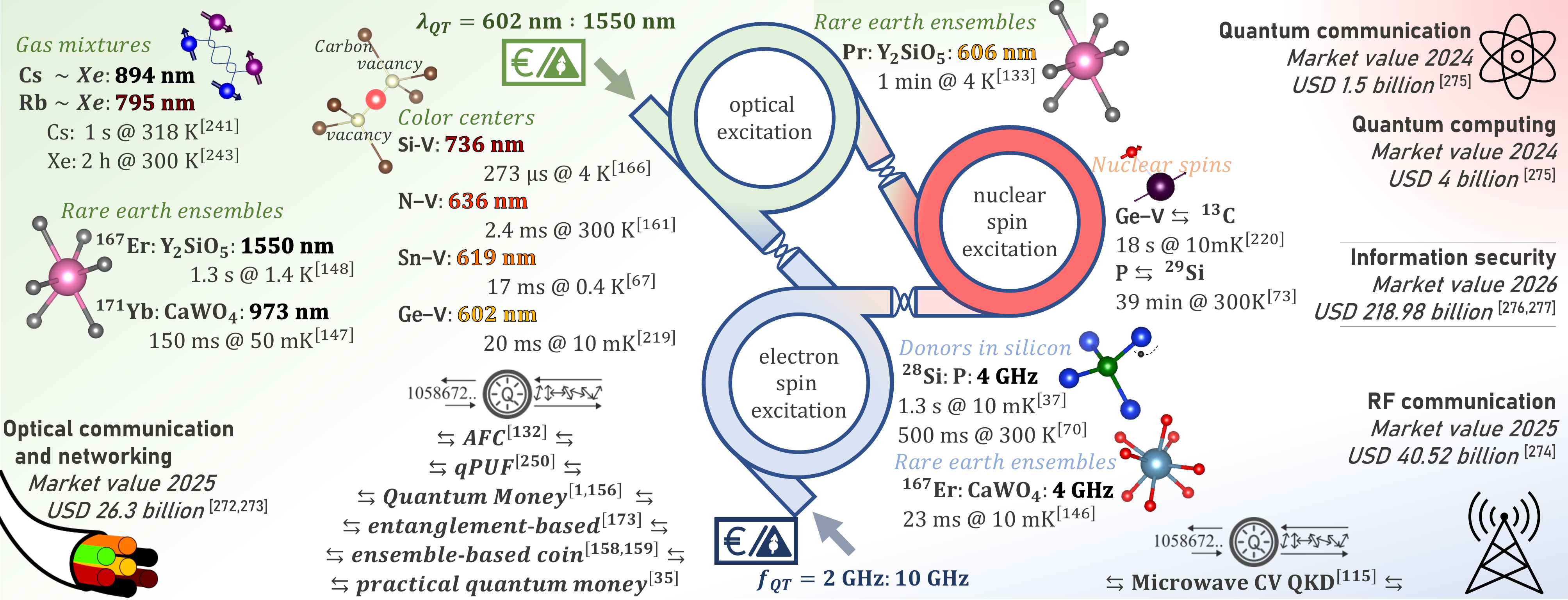}
    \caption{Schematic description of all discussed quantum token realizations. The quantum tokens can be generated in a range of laser optics frequencies (green-colored area), directly absorbed as an excitation on optical states, which can be further converted into electron or nuclear spin transitions, where the latter can be generated either directly (e.g., in Pr:Y\textsubscript{2}SiO\textsubscript{5}) or mediated via electron spin excitations. Alternatively, quantum tokens can be generated using transitions in the RF and microwave frequency regimes (blue-colored area). They can then be directly absorbed into the electron spin states and subsequently transferred into the nuclear spin-state excitations, which offer the longest achievable storage times for quantum states.
    For each memory system, the longest reported storage time and the corresponding thermal conditions are noted. While all listed spin systems aim for storing the quantum token states in nuclear spin excitations, the reported storage times mostly reflect the longest published storage time in electron spin states. Additionally, the suitable QKD protocols are summarized on the laser-optical or microwave side of the figure. The market values for the classical and quantum fields highlight the economic potential of the quantum token research.       }
    \label{fig:sec-conclusion}
\end{figure*}
\newpage

\section{Discussion and conclusion}
Quantum tokens constitute an emerging hardware-assisted security primitive that leverages quantum effects %-- most notably the measurement-induced disturbance in unknown quantum states -- 
to enable information-theoretic security of communication and authentication, which are difficult to achieve in purely classical systems. Quantum token architectures rely largely on optical and solid‑state platforms and and their economical perspective is directly connected to the optical communication market size, which is projected to expand from USD~25.8~billion in 2025 to USD~59.4~billion by 2035~\cite{R1-optical-comm,R5-optical-comm}, driven by demands for high‑capacity networks and advanced photonic infrastructure. The prospective deployment of quantum tokens within wireless and embedded systems similarly aligns with growth in the RF communication market size, forecast to increase from USD~40.5~billion in 2025 to USD~88.0~billion by 2030~\cite{R3-RF-comm}, supported by the proliferation of 5G technologies, IoT devices, and next‑generation RF front‑end technologies. At the same time, the broader quantum technology ecosystem is undergoing rapid maturation, with significant public and private investment advancing quantum communication, computing, and sensing capabilities, as documented in the 2025 Quantum Technology Monitor~\cite{R2-quantum-comm}. Because quantum tokens can provide secure authentication rooted in the no‑cloning theorem rather than in asymmetric computational difficulty, they address security challenges that are becoming increasingly salient within the information security sector, whose global market size is already USD~218.98~billion and expected to reach USD~699.39~billion by 2034~\cite{R4-inf-security,R6-inf-security} in response to escalating cyberthreats and the transition toward quantum‑resilient security frameworks. These converging market trajectories underscore the broader technological and economic importance of research aimed at developing practical, scalable quantum token systems capable of supporting future authentication tasks across emerging communication infrastructures.

While fully deployable, end-to-end quantum token infrastructures (issuance, storage, verification, and lifecycle management under adversarial conditions) have not yet been demonstrated, intensive experimental progress on quantum memories, interfaces, encoding protocols, and photonic and cryogenic integration indicates a credible route toward practical realizations. In this paper, we surveyed and compared selected physical approaches for implementing \emph{state-based quantum tokens} using quantum memory elements, focusing on (i) token-state encoding in the optical and microwave domains, (ii) transmission and access models, and (iii) quantum memory architectures and dominant sources of decoherence, loss, and verification error.

A recurring design principle across several leading platforms is to transfer a token state -- initially prepared in an optical or microwave degree of freedom -- into a \emph{long-lived}  nuclear spin register, thereby decoupling the token lifetime from the comparatively short coherence of excited electronic states and photonic modes. In many prominent proposals and demonstrations, this long-lived register is provided by nuclear-spin or hyperfine degrees of freedom (cf.~\Cref{fig:sec-conclusion}), which often exhibit reduced coupling to environmental noise and can preserve phase coherence substantially longer than electronic excitations. At the same time, alternative long-lived quantum memory registers are also relevant (e.g., engineered bosonic modes in superconducting resonators), emphasizing that nuclear spins are a powerful but not exclusive pathway to long-duration storage.

%\paragraph{Solid-state color centers}
In this paper, we have discussed the potential of \textit{color centers in diamond}, which provide a representative solid-state route to optically addressable token generation and nuclear-spin-assisted storage. In architectures based on \textit{nitrogen-vacancy (NV)} centers, photonic nanostructures such as nanopillars can improve optical access, while nearby \textsuperscript{13}C nuclear spins offer long-lived storage registers. Security mechanisms in this family can exploit limited information extractability from finite quantum resources and the detectability of disturbance under attempted duplication. Where complete hardware token demonstrations are not yet available, protocol-level evaluations using gate-based quantum processors (as experimental emulators) can provide evidence of feasibility under specified noise and adversary models, while leaving open the system-level questions of device-specific imperfections, leakage channels, and physical side channels in a deployable token. Closely related, \textit{silicon-vacancy (SiV)} centers offer attractive optical transitions and compatibility with nanophotonic integration. When coupled to nearby nuclear spins, SiV-based schemes provide a plausible route to compact memories with improved optical interfacing. Integration with platforms such as Si\textsubscript{3}N\textsubscript{4} photonic circuitry is promising for scalability and packaging; however, compatibility with Telecom-band networking generally requires either operation in the appropriate wavelength regime or the inclusion of explicit frequency-conversion interfaces, which should be treated as part of the system design rather than assumed implicitly.

%\paragraph{Rare-earth ion and spin-ensemble memories}
Quantum-token implementations based on \textit{rare-earth materials} are notable for their potential to interface with both optical and microwave fields. In the optical domain, rare-earth memories commonly employ protocols such as atomic frequency comb (AFC) storage or electromagnetically induced transparency (EIT), mapping optical excitations -- directly or via intermediate electronic-spin manifolds -- into longer-lived spin degrees of freedom. Across the literature, these systems have demonstrated long-lived spin coherence and, in selected settings, storage and retrieval for weak coherent optical pulses and single-photon-level inputs, with efficiencies and fidelities that can be substantially enhanced in cavity-assisted or Purcell-enhanced architectures. Certain ions (e.g., praseodymium in suitable host crystals) provide favorable level structures that can facilitate more direct mappings between optical and long-lived spin or hyperfine states, potentially reducing control overhead and limiting error accumulation across multi-step transfer sequences.

%\paragraph{Room-temperature atomic and hybrid-vapor approaches}
Room-temperature \textit{atomic-vapor platforms} offer an alternative route toward deployable token concepts with reduced infrastructure complexity. EIT-based vapor memories have demonstrated storage of photonic encodings, including polarization degrees of freedom, with performance limited by optical depth, magnetic noise, diffusion, and technical stability. Separately, alkali--noble-gas spin-exchange systems provide access to noble-gas nuclear-spin coherence that can be exceptionally long under well-controlled conditions. Combining these ingredients suggests a pathway to ambient-condition tokens, but it also highlights the importance of careful engineering to maintain stability in realistic environments and to quantify performance under adversarial interaction models.

%\paragraph{Microwave-domain token generation and storage}
\textit{Microwave quantum tokens} arise naturally in superconducting circuit platforms, where nonclassical microwave states (e.g., squeezed or displaced states) can be generated and measured with high accuracy in cryogenic environments. Storage can be realized either in long-lived superconducting resonators or via transfer into spin ensembles, such as donors in silicon or rare-earth ions, that act as microwave memories. These architectures are well matched to integration with superconducting quantum processors and can support short-range, high-assurance authentication scenarios. Here, near-field or guided cryogenic interconnects represent realistic near-future targets, while the free-space, open air microwave quantum communication may require more time to be developed and brought to real applications.

%\paragraph{Relation to qPUFs and hybrid security stacks}
Beyond token-based primitives, \textit{Quantum Physical Unclonable Functions (qPUFs)} aim to provide authentication from intrinsically quantum challenge -- response behavior and are studied under formal security definitions against quantum adversaries. qPUFs therefore complement quantum tokens: tokens are often modeled as stored quantum states verified by a measurement, while qPUFs are commonly modeled as devices implementing quantum processes whose behavior should be hard to emulate. Importantly, neither paradigm eliminates the need to address implementation-level security: practical deployments must still account for noise, calibration drift, and classical side channels (timing, power, electromagnetic leakage), and must explicitly specify adversarial access (e.g., single-use versus repeated verification attempts).

Finally, quantum tokens can be naturally integrated with classical cryptographic layers. In particular, combining token-based authentication with post-quantum cryptography (PQC) can yield a robust defense-in-depth architecture: easy to implement PQC can protect classical communication channels, key management, and stored data, while quantum tokens can strengthen authentication and identity verification by providing the information-theoretic security layer. Such hybrid designs are particularly attractive for critical infrastructure and long-term confidentiality, because a possible breakdown of the PQC layer will not imply compromising of the entire system.

%\paragraph{Summary}
In summary, quantum tokens offer a promising direction for quantum-enhanced authentication and secure transactions, and the range of candidate realizations, which span to spanning solid-state defects, rare-earth and atomic ensembles, and superconducting microwave systems, underscores the adaptability of the concept across technological ecosystems. Moving from laboratory prototypes toward deployment will require advances in memory lifetimes and efficiencies under realistic operating conditions, integrated interfaces (including photonics, transduction, and packaging), and end-to-end security analyses that incorporate both device imperfections and explicit adversary models. Continued progress along these lines will clarify the practical roles quantum tokens can play alongside classical and post-quantum security infrastructures in applications ranging from identity management and access control to high-assurance transactions.
\newpage
\section{Acknowledgments}

All authors acknowledge the support of Bundesministerium für Forschung, Technologie und Raumfahrt (BMFTR) via the funding program "Grand challenge of the quantum communication", with projects QuaMToMe (Grant No. 16KISQ036), HybridQToken (Grant No. 16KISQ043K), Q.TOK (Grant numbers 16KISQ039, 16KISQ038 and 16KISQ037K), QPIS (Grant No. 16KISQ032K), DIQTOK (Grant numbers 16KISQ034K and 16KISQ035), NEQSIS (Grant Numbers 16KISQ029K, 16KISQ030, and 16KISQ031), Q-ToRX (Grant Numbers 16KISQ040K, 16KISQ041 and 16KISQ042).
 
NK, KF and HH acknowledge support by the Deutsche Forschungsgemeinschaft via Germany’s Excellence Strategy (Grant No. EXC-2111- 390814868), and JN via the Emmy-Noether program (Grant No. NO 1129/2-1 ).

HB and CD were supported by the German Federal Ministry of Research, Technology and Space (BMFTR) through the projects QuaPhySI (grants 16KISQ1598K and 16KIS2234), QR.N (grants 16KIS2195 and 16KIS2196), QSTARS (grants 16KIS2611 and 16KIS2602), QUIET (grants 16KISQ093 and 16KISQ0170), QD-CamNetz (grants 16KISQ077 and 16KISQ169), and Q-TREX (grants 16KISR027K and 16KISR038). 

TS was supported by the BMFTR through the  project DiNOQuant (Grant No. 13N14921), GP by the EU ERC St. Grant QUREP (851810). Additional support was provided by the 6GQT initiative, funded by the Bavarian State Ministry of Economic Affairs, Regional Development and Energy. HB further acknowledges funding from the German Research Foundation (DFG) under Germany’s Excellence Strategy EXC-2092 – 390781972. 

JK was supported by the German Federal Ministry of Research, Technology and Space (BMFTR) under the project SEQUIN (16KIS2134).

CD additionally received funding from the DFG under grant DE 1915/2-1.

%JW, WK and IG were supported by the German Federal Ministry of Research, Technology and Space (BMFTR) through the  project Q-ToRX (Grant No. 16KISQ040K). 
JW further acknowledges funding from the German Research Foundation (DFG) under SPP1514, Grant. No. 563423354.

%\bibliography{Bibliography, NEQSIS, Q-ToRX, Q-TOK, QuaMToMe, QPIS}
\bibliography{Bibliography_combined}

@online{NIST_PQC,
  author = {NIST},
  title = {Selected Algorithms - Post-Quantum Cryptography},
  year = 2025,
  url = {https://csrc.nist.gov/Projects/post-quantum-cryptography/selected-algorithms},
  urldate = {2025-11-11}
}

@online{NIST_PQC_sig,
  author = {NIST},
  title = {Standardization of Additional Digital Signature Schemes - Post-Quantum Cryptography},
  year = 2025,
  url = {https://csrc.nist.gov/Projects/pqc-dig-sig/standardization},
  urldate = {2025-11-11}
}

@article{Fedorov2016,
    author =  {Fedorov, Kirill G. and Zhong, L. and Pogorzalek, S. and Eder, P. and Fischer, M. and Goetz, J. and Xie, E. and Wulschner, F. and Inomata, K. and Yamamoto, T. and Nakamura, Y. and {Di Candia}, R. and {Las Heras}, U. and Sanz, M. and Solano, E. and Menzel, E. P. and Deppe, F. and Marx, A. and Gross, R.},
    title =   {Displacement of Propagating Squeezed Microwave States},
    journal = {Phys. Rev. Lett.},
    year =    {2016},
    volume =  {117},
    number =  {2},
    pages =   {020502},
    doi =     {10.1103/PhysRevLett.117.020502},
    url =     {https://doi.org/10.1103/PhysRevLett.117.020502}
}

@article{Menzel2010,
  title = {Planck Spectroscopy and Quantum Noise of Microwave Beam Splitters},
  author = {Mariantoni, M. and Menzel, E. P. and Deppe, F. and Araque Caballero, M. \'A. and Baust, A. and Niemczyk, T. and Hoffmann, E. and Solano, E. and Marx, A. and Gross, R.},
  journal = {Phys. Rev. Lett.},
  volume = {105},
  issue = {13},
  pages = {133601},
  numpages = {4},
  year = {2010},
  month = {Sep},
  publisher = {American Physical Society},
  doi = {10.1103/PhysRevLett.105.133601},
  url = {https://link.aps.org/doi/10.1103/PhysRevLett.105.133601}
}

@article{Eichler:2011kt, 
year = {2011}, 
title = {{Experimental State Tomography of Itinerant Single Microwave Photons}}, 
author = {Eichler, C and Bozyigit, D and Lang, C and Steffen, L and Fink, J and Wallraff, A}, 
journal = {Physical Review Letters}, 
doi = {10.1103/physrevlett.106.220503}, 
url = {http://link.aps.org/doi/10.1103/PhysRevLett.106.220503}, 
pages = {220503}, 
number = {22}, 
volume = {106}, 
language = {English}, 
month = {06}
}

@article{Gandorfer2025,
  title = {Two-dimensional {P}lanck spectroscopy for microwave photon calibration},
  author = {Gandorfer, S. and Renger, M. and Yam, W.K. and Fesquet, F. and Marx, A. and Gross, R. and Fedorov, K.G.},
  journal = {Phys. Rev. Appl.},
  volume = {23},
  issue = {2},
  pages = {024064},
  numpages = {10},
  year = {2025},
  month = {Feb},
  publisher = {American Physical Society},
  doi = {10.1103/PhysRevApplied.23.024064}
}

@article{Honasoge2025,
  title = {Fabrication of low-loss Josephson parametric devices},
  author = {Honasoge, K. E. and Handschuh, M. and Yam, W. K. and Gandorfer, S. and Bazulin, D. and Bruckmoser, N. and Koch, L. and Marx, A. and Gross, R. and Fedorov, K. G.},
  journal = {Phys. Rev. B},
  volume = {111},
  issue = {21},
  pages = {214508},
  numpages = {11},
  year = {2025},
  month = {Jun},
  publisher = {American Physical Society},
  doi = {10.1103/PhysRevB.111.214508},
  url = {https://link.aps.org/doi/10.1103/PhysRevB.111.214508}
}

@article{Yam2025,
    title={Quantum teleportation over thermal microwave network}, 
    author={W. K. Yam and S. Gandorfer and F. Fesquet and M. Handschuh and K. E. Honasoge and A. Marx and R. Gross and K. G. Fedorov},
    year={2025},
    journal={arXiv:2508.14691},
    archivePrefix={arXiv},
    primaryClass={quant-ph},
    url={https://arxiv.org/abs/2508.14691},
    doi={10.48550/arXiv.2508.14691}
}

@misc{OQS,
 year = {2025},
 title = {Open Quantum Safe (OQS)},
 urldate = {2025-12-05},
url = {https://openquantumsafe.org/}
}

@misc{tuta,
 year = {2024},
 title = {Tuta Mail jetzt mit postquantensicherer Verschlüsselung},
 urldate = {2025-12-05},
url = {https://tuta.com/de/blog/post-quantum-cryptography}
}

@misc{google,
 year = {2025},
 title = {Google Expands Post-Quantum Cryptography Support with Quantum-Safe Digital Signatures},
 urldate = {2025-12-05},
url = {https://thequantuminsider.com/2025/02/24/google-expands-post-quantum-cryptography-support-with-quantum-safe-digital-signatures/}
}

@misc{Signal,
 author = {Graeme Connell and Rolfe Schmidt},
 year = {2025},
 title = {Signal Protocol and Post-Quantum Ratchets},
 urldate = {2025-12-05},
url = {https://signal.org/blog/spqr/}
}

@misc{BSI,
 author = {Federal Office for Information Security},
 year = {2024},
 title = {Status of quantum computer
development},
 urldate = {2025-12-05},
url = {https://www.bsi.bund.de/SharedDocs/Downloads/DE/BSI/Publikationen/Studien/Quantencomputer/Entwicklungstand_QC_V_2_1.pdf?__blob=publicationFile&v=3}
}

@article{trusheim_lead-related_2019,
	title = {Lead-related quantum emitters in diamond},
	volume = {99},
	url = {https://link.aps.org/doi/10.1103/PhysRevB.99.075430},
	doi = {10.1103/PhysRevB.99.075430},
	number = {7},
	urldate = {2025-10-28},
	journal = {Physical Review B},
	author = {Trusheim, Matthew E. and Wan, Noel H. and Chen, Kevin C. and Ciccarino, Christopher J. and Flick, Johannes and Sundararaman, Ravishankar and Malladi, Girish and Bersin, Eric and Walsh, Michael and Lienhard, Benjamin and Bakhru, Hassaram and Narang, Prineha and Englund, Dirk},
	year = {2019},
	note = {Publisher: American Physical Society},
	pages = {075430},
}

@article{iwasaki_tin-vacancy_2017,
	title = {Tin-{Vacancy} {Quantum} {Emitters} in {Diamond}},
	volume = {119},
	url = {https://link.aps.org/doi/10.1103/PhysRevLett.119.253601},
	doi = {10.1103/PhysRevLett.119.253601},
	number = {25},
	urldate = {2025-10-28},
	journal = {Physical Review Letters},
	author = {Iwasaki, Takayuki and Miyamoto, Yoshiyuki and Taniguchi, Takashi and Siyushev, Petr and Metsch, Mathias H. and Jelezko, Fedor and Hatano, Mutsuko},
	year = {2017},
	pages = {253601},
}

@article{wei_quantum_2024,
	title = {Quantum storage of 1650 modes of single photons at telecom wavelength},
	volume = {10},
	copyright = {2024 The Author(s)},
	issn = {2056-6387},
	url = {https://www.nature.com/articles/s41534-024-00812-1},
	doi = {10.1038/s41534-024-00812-1},
	number = {1},
	urldate = {2025-11-11},
	journal = {npj Quantum Information},
	author = {Wei, Shi-Hai and Jing, Bo and Zhang, Xue-Ying and Liao, Jin-Yu and Li, Hao and You, Li-Xing and Wang, Zhen and Wang, You and Deng, Guang-Wei and Song, Hai-Zhi and Oblak, Daniel and Guo, Guang-Can and Zhou, Qiang},
	month = feb,
	year = {2024},
	note = {Publisher: Nature Publishing Group},
	pages = {19},
}

@article{neu_single_2011,
	title = {Single photon emission from silicon-vacancy colour centres in chemical vapour deposition nano-diamonds on iridium},
	volume = {13},
	issn = {1367-2630},
	doi = {10.1088/1367-2630/13/2/025012},
	number = {2},
	journal = {New Journal of Physics},
	author = {Neu, Elke and Steinmetz, David and Riedrich-Möller, Janine and Gsell, Stefan and Fischer, Martin and Schreck, Matthias and Becher, Christoph},
	year = {2011},
	pages = {025012}
}

@article{tittel_quantum_2025,
	title = {Quantum networks using rare-earth ions},
	volume = {10},
	issn = {2058-9565},
	url = {https://dx.doi.org/10.1088/2058-9565/addd93},
	doi = {10.1088/2058-9565/addd93},
	number = {3},
	urldate = {2025-06-27},
	journal = {Quantum Science and Technology},
	author = {Tittel, Wolfgang and Afzelius, Mikael and Kinos, Adam and Rippe, Lars and Walther, Andreas},
	month = jun,
	year = {2025},
	note = {Publisher: IOP Publishing},
	pages = {033002}
}

@article{saglamyurek_broadband_2011,
	title = {Broadband waveguide quantum memory for entangled photons},
	volume = {469},
	copyright = {2010 Springer Nature Limited},
	issn = {1476-4687},
	url = {https://www.nature.com/articles/nature09719},
	doi = {10.1038/nature09719},
	number = {7331},
	urldate = {2025-11-11},
	journal = {Nature},
	author = {Saglamyurek, Erhan and Sinclair, Neil and Jin, Jeongwan and Slater, Joshua A. and Oblak, Daniel and Bussières, Félix and George, Mathew and Ricken, Raimund and Sohler, Wolfgang and Tittel, Wolfgang},
	month = jan,
	year = {2011},
	pages = {512--515}
}

@article{corrielli_integrated_2016,
	title = {Integrated {Optical} {Memory} {Based} on {Laser}-{Written} {Waveguides}},
	volume = {5},
	url = {https://link.aps.org/doi/10.1103/PhysRevApplied.5.054013},
	doi = {10.1103/PhysRevApplied.5.054013},
	number = {5},
	urldate = {2025-11-11},
	journal = {Physical Review Applied},
	author = {Corrielli, Giacomo and Seri, Alessandro and Mazzera, Margherita and Osellame, Roberto and de Riedmatten, Hugues},
	month = may,
	year = {2016},
	pages = {054013}
}

@article{casabone_cavity-enhanced_2018,
doi = {10.1088/1367-2630/aadf68},
url = {https://doi.org/10.1088/1367-2630/aadf68},
year = {2018},
month = {sep},
publisher = {IOP Publishing},
volume = {20},
number = {9},
pages = {095006},
author = {Casabone, Bernardo and Benedikter, Julia and Hümmer, Thomas and Oehl, Franziska and Lima, Karmel de Oliveira and Hänsch, Theodor W and Ferrier, Alban and Goldner, Philippe and Riedmatten, Hugues de and Hunger, David},
title = {Cavity-enhanced spectroscopy of a few-ion ensemble in Eu3+:Y2O3},
journal = {New Journal of Physics}
}

@article{eichhorn_multimodal_2025,
url = {https://doi.org/10.1515/nanoph-2024-0721},
title = {Multimodal Purcell enhancement and optical coherence of Eu3+ ions in a single nanoparticle coupled to a microcavity},
title = {},
author = {Timon Eichhorn and Nicholas Jobbitt and Sören Bieling and Shuping Liu and Tobias Krom and Diana Serrano and Robert Huber and Ulrich Lemmer and Hugues de Riedmatten and Philippe Goldner and David Hunger},
pages = {1817--1826},
volume = {14},
number = {11},
journal = {Nanophotonics},
doi = {doi:10.1515/nanoph-2024-0721},
year = {2025},
lastchecked = {2026-02-06}
}

@article{MacCabe.2020,
 author = {MacCabe, Gregory S. and Ren, Hengjiang and Luo, Jie and Cohen, Justin D. and Zhou, Hengyun and Sipahigil, Alp and Mirhosseini, Mohammad and Painter, Oskar},
 year = {2020},
 title = {Nano-acoustic resonator with ultralong phonon lifetime},
 pages = {840--843},
 volume = {370},
 number = {6518},
 journal = {Science},
 doi = {10.1126/science.abc7312}
}

@article{Seis.2022,
 author = {Seis, Yannick and Capelle, Thibault and Langman, Eric and Saarinen, Sampo and Planz, Eric and Schliesser, Albert},
 year = {2022},
 title = {Ground state cooling of an ultracoherent electromechanical system},
 pages = {1507},
 volume = {13},
 number = {1},
 issn = {2041-1723},
 journal = {Nat. Commun.},
 doi = {10.1038/s41467-022-29115-9}
}

@article{Dudin.2013,
  title = {Light storage on the time scale of a minute},
  author = {Dudin, Y. O. and Li, L. and Kuzmich, A.},
  journal = {Phys. Rev. A},
  volume = {87},
  issue = {3},
  pages = {031801},
  numpages = {4},
  year = {2013},
  month = {Mar},
  publisher = {American Physical Society},
  doi = {10.1103/PhysRevA.87.031801},
  url = {https://link.aps.org/doi/10.1103/PhysRevA.87.031801}
}

@article{Levine.2019,
 author = {Levine, Harry and Keesling, Alexander and Semeghini, Giulia and Omran, Ahmed and Wang, Tout T. and Ebadi, Sepehr and Bernien, Hannes and Greiner, Markus and Vuleti{\'c}, Vladan and Pichler, Hannes and Lukin, Mikhail D.},
 year = {2019},
 title = {Parallel Implementation of High-Fidelity Multiqubit Gates with Neutral Atoms},
 pages = {170503},
 volume = {123},
 number = {17},
 journal = {Physical Review Letters},
 doi = {10.1103/PhysRevLett.123.170503}
}

@article{Ruster.2016,
 author = {Ruster, T. and Schmiegelow, C. T. and Kaufmann, H. and Warschburger, C. and Schmidt-Kaler, F. and Poschinger, U. G.},
 year = {2016},
 title = {A long-lived Zeeman trapped-ion qubit},
 volume = {122},
 number = {10},
 pages = {254},
 issn = {0946-2171},
 journal = {Applied Physics B},
 doi = {10.1007/s00340-016-6527-4}
}

@article{Wang2021,
 author = {Wang, Pengfei and Luan, Chun Yang and Qiao, Mu and Um, Mark and Zhang, Junhua and Wang, Ye and Yuan, Xiao and Gu, Mile and Zhang, Jingning and Kim, Kihwan},
 year = {2021},
 title = {Single ion qubit with estimated coherence time exceeding one hour},
 volume = {12},
 pages = {233},
 number = {1},
 journal = {Nature Communications},
 doi = {10.1038/s41467-020-20330-w}
}

@article{Yoneda.2018,
 author = {Yoneda, Jun and Takeda, Kenta and Otsuka, Tomohiro and Nakajima, Takashi and Delbecq, Matthieu R. and Allison, Giles and Honda, Takumu and Kodera, Tetsuo and Oda, Shunri and Hoshi, Yusuke and Usami, Noritaka and Itoh, Kohei M. and Tarucha, Seigo},
 year = {2018},
 title = {A quantum-dot spin qubit with coherence limited by charge noise and fidelity higher than 99.9},
 pages = {102--106},
 volume = {13},
 number = {2},
 journal = {Nature nanotechnology},
 doi = {10.1038/s41565-017-0014-x}
}

@article{Place.2021,
 author = {Place, Alexander P. M. and Rodgers, Lila V. H. and Mundada, Pranav and Smitham, Basil M. and Fitzpatrick, Mattias and Leng, Zhaoqi and Premkumar, Anjali and Bryon, Jacob and Vrajitoarea, Andrei and Sussman, Sara and Cheng, Guangming and Madhavan, Trisha and Babla, Harshvardhan K. and Le, Xuan Hoang and Gang, Youqi and J{\"a}ck, Berthold and Gyenis, Andr{\'a}s and Yao, Nan and Cava, Robert J. and de Leon, Nathalie P. and Houck, Andrew A.},
 year = {2021},
 title = {New material platform for superconducting transmon qubits with coherence times exceeding 0.3 milliseconds},
 pages = {1779},
 volume = {12},
 number = {1},
 issn = {2041-1723},
 journal = {Nat. Commun.},
 doi = {10.1038/s41467-021-22030-5}
}

@article{Somoroff.2023,
 author = {Somoroff, Aaron and Ficheux, Quentin and Mencia, Raymond A. and Xiong, Haonan and Kuzmin, Roman and Manucharyan, Vladimir E.},
 year = {2023},
 title = {Millisecond Coherence in a Superconducting Qubit},
 pages = {267001},
 volume = {130},
 number = {26},
 journal = {Physical Review Letters},
 doi = {10.1103/PhysRevLett.130.267001}
}

@article{Wang.2022,
 author = {Wang, Chenlu and Li, Xuegang and Xu, Huikai and Li, Zhiyuan and Wang, Junhua and Yang, Zhen and Mi, Zhenyu and Liang, Xuehui and Su, Tang and Yang, Chuhong and Wang, Guangyue and Wang, Wenyan and Li, Yongchao and Chen, Mo and Li, Chengyao and Linghu, Kehuan and Han, Jiaxiu and Zhang, Yingshan and Feng, Yulong and Song, Yu and Ma, Teng and Zhang, Jingning and Wang, Ruixia and Zhao, Peng and Liu, Weiyang and Xue, Guangming and Jin, Yirong and Yu, Haifeng},
 year = {2022},
 title = {Towards practical quantum computers: transmon qubit with a lifetime approaching 0.5 milliseconds},
 volume = {8},
 number = {1},
 pages = {3},
 journal = {npj Quantum Information},
 doi = {10.1038/s41534-021-00510-2}
}

@article{Romanenko.2020,
  title = {Three-Dimensional Superconducting Resonators at $\mathrm{T}<20 \mathrm{mK}$ with Photon Lifetimes up to $\tau=2$ s},
  author = {Romanenko, A. and Pilipenko, R. and Zorzetti, S. and Frolov, D. and Awida, M. and Belomestnykh, S. and Posen, S. and Grassellino, A.},
  journal = {Phys. Rev. Appl.},
  volume = {13},
  issue = {3},
  pages = {034032},
  numpages = {5},
  year = {2020},
  month = {Mar},
  publisher = {American Physical Society},
  doi = {10.1103/PhysRevApplied.13.034032},
  url = {https://link.aps.org/doi/10.1103/PhysRevApplied.13.034032}
}

@article{Xie.2018,
    author = {Xie, Edwar and Deppe, Frank and Renger, Michael and Repp, Daniel and Eder, Peter and Fischer, Michael and Goetz, Jan and Pogorzalek, Stefan and Fedorov, Kirill G. and Marx, Achim and Gross, Rudolf},
    title = {Compact 3D quantum memory},
    journal = {Applied Physics Letters},
    volume = {112},
    number = {20},
    pages = {202601},
    year = {2018},
    month = {05},
    issn = {0003-6951},
    doi = {10.1063/1.5029514},
    url = {https://doi.org/10.1063/1.5029514}
}

@article{Nojiri.2024,
  title = {Onset of transmon ionization in microwave single-photon detection},
  author = {Nojiri, Yuki and Honasoge, Kedar E. and Marx, Achim and Fedorov, Kirill G. and Gross, Rudolf},
  journal = {Phys. Rev. B},
  volume = {109},
  issue = {17},
  pages = {174312},
  numpages = {14},
  year = {2024},
  month = {May},
  publisher = {American Physical Society},
  doi = {10.1103/PhysRevB.109.174312},
  url = {https://link.aps.org/doi/10.1103/PhysRevB.109.174312}
}

@article{Renger.2022,
  title = {Flow of quantum correlations in noisy two-mode squeezed microwave states},
  author = {Renger, M. and Pogorzalek, S. and Fesquet, F. and Honasoge, K. and Kronowetter, F. and Chen, Q. and Nojiri, Y. and Inomata, K. and Nakamura, Y. and Marx, A. and Deppe, F. and Gross, R. and Fedorov, K. G.},
  journal = {Phys. Rev. A},
  volume = {106},
  issue = {5},
  pages = {052415},
  numpages = {10},
  year = {2022},
  month = {Nov},
  publisher = {American Physical Society},
  doi = {10.1103/PhysRevA.106.052415},
  url = {https://link.aps.org/doi/10.1103/PhysRevA.106.052415}
}

@article{Menzel.2012,
  title = {Path Entanglement of Continuous-Variable Quantum Microwaves},
  author = {Menzel, E. P. and Di Candia, R. and Deppe, F. and Eder, P. and Zhong, L. and Ihmig, M. and Haeberlein, M. and Baust, A. and Hoffmann, E. and Ballester, D. and Inomata, K. and Yamamoto, T. and Nakamura, Y. and Solano, E. and Marx, A. and Gross, R.},
  journal = {Phys. Rev. Lett.},
  volume = {109},
  issue = {25},
  pages = {250502},
  numpages = {4},
  year = {2012},
  month = {Dec},
  publisher = {American Physical Society},
  doi = {10.1103/PhysRevLett.109.250502},
  url = {https://link.aps.org/doi/10.1103/PhysRevLett.109.250502}
}

@inproceedings{Kumar2025ICC_Auth,
  author={Nilesh, Kumar and Deppe, Christian and Boche, Holger},
  booktitle={ICC 2025 - IEEE International Conference on Communications}, 
  title={Authentication Based on Quantum PUF}, 
  year={2025},
  volume={},
  number={},
  pages={6796-6801},
  doi={10.1109/ICC52391.2025.11161044},
  ISSN={1938-1883},
  month={June},}

@inproceedings{Kumar2025ICC_Storage,
  author={Nilesh, Kumar and Deppe, Christian and Boche, Holger},
  booktitle={ICC 2025 - IEEE International Conference on Communications}, 
  title={Secure Storage and Identification Using Quantum PUF}, 
  year={2025},
  volume={},
  number={},
  pages={6802-6807},
  doi={10.1109/ICC52391.2025.11160793},
  ISSN={1938-1883},
  month={June},}

@inproceedings{Kumar2025ISIT_SideInfo,
  author={Nilesh, Kumar and Deppe, Christian and Boche, Holger},
  booktitle={2025 IEEE International Symposium on Information Theory (ISIT)}, 
  title={Quantum PUF Based Secret Key Generation and Secure Storage with Side Information}, 
  year={2025},
  volume={},
  number={},
  pages={1-5},
  doi={10.1109/ISIT63088.2025.11195487},
  ISSN={2157-8117},
  month={June},}

@inproceedings{Kumar2025ITW_Storage,
    author = "Nilesh, Kumar and Deppe, Christian and Boche, Holger",
    title = "{Secret key generation and Storage based on QPUF}",
    doi = "10.1109/ITW62417.2025.11240507",
    month = "9",
    year = "2025"
}

@inproceedings{Kumar2024ISIT_InfoTheory,
  author={Nilesh, Kumar and Deppe, Christian and Boche, Holger},
  booktitle={2024 IEEE International Symposium on Information Theory (ISIT)}, 
  title={Information Theoretic Analysis of a Quantum PUF}, 
  year={2024},
  volume={},
  number={},
  pages={3320-3325},
  doi={10.1109/ISIT57864.2024.10619408}}

@inproceedings{Aaronson.2018,
 author = {Aaronson, Scott},
 title = {Shadow tomography of quantum states},
 pages = {325--338},
 publisher = {{Association for Computing Machinery}},
 isbn = {9781450355599},
 series = {STOC 2018},
 booktitle = {Proceedings of the 50th Annual ACM SIGACT Symposium on Theory of Computing},
 year = {2018},
 address = {New York, NY, USA},
 doi = {10.1145/3188745.3188802}
}

@misc{Afzal2024,
title={Distributed Quantum Computing in Silicon}, 
      author={Photonic Inc and : and Francis Afzal and Mohsen Akhlaghi and Stefanie J. Beale and Olinka Bedroya and Kristin Bell and Laurent Bergeron and Kent Bonsma-Fisher and Polina Bychkova and Zachary M. E. Chaisson and Camille Chartrand and Chloe Clear and Adam Darcie and Adam DeAbreu and Colby DeLisle and Lesley A. Duncan and Chad Dundas Smith and John Dunn and Amir Ebrahimi and Nathan Evetts and Daker Fernandes Pinheiro and Patricio Fuentes and Tristen Georgiou and Biswarup Guha and Rafael Haenel and Daniel Higginbottom and Daniel M. Jackson and Navid Jahed and Amin Khorshidahmad and Prasoon K. Shandilya and Alexander T. K. Kurkjian and Nikolai Lauk and Nicholas R. Lee-Hone and Eric Lin and Rostyslav Litynskyy and Duncan Lock and Lisa Ma and Iain MacGilp and Evan R. MacQuarrie and Aaron Mar and Alireza Marefat Khah and Alex Matiash and Evan Meyer-Scott and Cathryn P. Michaels and Juliana Motira and Narwan Kabir Noori and Egor Ospadov and Ekta Patel and Alexander Patscheider and Danny Paulson and Ariel Petruk and Adarsh L. Ravindranath and Bogdan Reznychenko and Myles Ruether and Jeremy Ruscica and Kunal Saxena and Zachary Schaller and Alex Seidlitz and John Senger and Youn Seok Lee and Orbel Sevoyan and Stephanie Simmons and Oney Soykal and Leea Stott and Quyen Tran and Spyros Tserkis and Ata Ulhaq and Wyatt Vine and Russ Weeks and Gary Wolfowicz and Isao Yoneda},
      year={2024},
      eprint={2406.01704},
      archivePrefix={arXiv},
      primaryClass={quant-ph},
      url={https://arxiv.org/abs/2406.01704}, 
}

@article{Afzelius:2009gc,
 author = {Afzelius, Mikael and Simon, Christoph and de Riedmatten, Hugues and Gisin, Nicolas},
 year = {2009},
 title = {Multimode quantum memory based on atomic frequency combs},
 pages = {052329},
 volume = {79},
 number = {5},
 journal = {Phys Rev A},
 doi = {10.1103/PhysRevA.79.052329}
}

@article{Afzelius2013,
doi = {10.1088/1367-2630/15/6/065008},
url = {https://doi.org/10.1088/1367-2630/15/6/065008},
year = {2013},
month = {jun},
publisher = {IOP Publishing},
volume = {15},
number = {6},
pages = {065008},
author = {Afzelius, M and Sangouard, N and Johansson, G and Staudt, M U and Wilson, C M},
title = {Proposal for a coherent quantum memory for propagating microwave photons},
journal = {New Journal of Physics}
}

@article{altiere_pra_2018,
 author = {Altiere, Emily and Miller, Eric R. and Hayamizu, Tomohiro and Jones, David J. and Madison, Kirk W. and Momose, Takamasa},
 year = {2018},
 title = {High-resolution two-photon spectroscopy of a 5p{\^{}}56p{\l}eftarrow5p{\^{}}6 transition of xenon},
 url = {https://link.aps.org/doi/10.1103/PhysRevA.97.012507},
 pages = {012507},
 volume = {97},
 number = {1},
 journal = {Phys. Rev. A},
 doi = {10.1103/PhysRevA.97.012507}
}

@article{amiri_quantum_2017,
 author = {Amiri, Ryan and Arrazola, Juan Miguel},
 year = {2017},
 title = {Quantum money with nearly optimal error tolerance},
 url = {https://link.aps.org/doi/10.1103/PhysRevA.95.062334},
 urldate = {2025-07-31},
 pages = {062334},
 volume = {95},
 number = {6},
 issn = {2469-9926},
 journal = {Physical Review A},
 doi = {10.1103/PhysRevA.95.062334}
}

@article{Arapinis.2021,
 author = {Arapinis, Myrto and Delavar, Mahshid and Doosti, Mina and Kashefi, Elham},
 year = {2021},
 title = {Quantum Physical Unclonable Functions: Possibilities and Impossibilities},
 pages = {475},
 volume = {5},
 issn = {2521-327X},
 journal = {Quantum},
 doi = {10.22331/q-2021-06-15-475}
}

@article{Arute2019,
 author = {Arute, Frank and Arya, Kunal and Babbush, Ryan and Bacon, Dave and Bardin, Joseph C. and Barends, Rami and Biswas, Rupak and Boixo, Sergio and {Brandao, Fernando G. S. L.} and Buell, David A. and Burkett, Brian and Chen, Yu and Chen, Zijun and Chiaro, Ben and Collins, Roberto and Courtney, William and Dunsworth, Andrew and Farhi, Edward and Foxen, Brooks and Fowler, Austin and Gidney, Craig and Giustina, Marissa and Graff, Rob and Guerin, Keith and Habegger, Steve and Harrigan, Matthew P. and Hartmann, Michael J. and Ho, Alan and Hoffmann, Markus and Huang, Trent and Humble, Travis S. and Isakov, Sergei V. and Jeffrey, Evan and Jiang, Zhang and Kafri, Dvir and Kechedzhi, Kostyantyn and Kelly, Julian and Klimov, Paul V. and Knysh, Sergey and Korotkov, Alexander and Kostritsa, Fedor and Landhuis, David and Lindmark, Mike and Lucero, Erik and Lyakh, Dmitry and Mandr{\`a}, Salvatore and McClean, Jarrod R. and McEwen, Matthew and Megrant, Anthony and Mi, Xiao and Michielsen, Kristel and Mohseni, Masoud and Mutus, Josh and Naaman, Ofer and Neeley, Matthew and Neill, Charles and Niu, Murphy Yuezhen and Ostby, Eric and Petukhov, Andre and Platt, John C. and Quintana, Chris and Rieffel, Eleanor G. and Roushan, Pedram and Rubin, Nicholas C. and Sank, Daniel and Satzinger, Kevin J. and Smelyanskiy, Vadim and Sung, Kevin J. and Trevithick, Matthew D. and Vainsencher, Amit and Villalonga, Benjamin and White, Theodore and Yao, Z. Jamie and Yeh, Ping and Zalcman, Adam and Neven, Hartmut and Martinis, John M.},
 year = {2019},
 title = {Quantum supremacy using a programmable superconducting processor},
 pages = {505--510},
 volume = {574},
 number = {7779},
 issn = {1476-4687},
 journal = {Nature},
 doi = {10.1038/s41586-019-1666-5}
}

@article{Axline2018,
 author = {Axline, Christopher J. and Burkhart, Luke D. and Pfaff, Wolfgang and Zhang, Mengzhen and Chou, Kevin and Campagne-Ibarcq, Philippe and Reinhold, Philip and Frunzio, Luigi and Girvin, S. M. and Jiang, Liang and Devoret, M. H. and Schoelkopf, R. J.},
 year = {2018},
 title = {On-demand quantum state transfer and entanglement between remote microwave cavity memories},
 pages = {705--710},
 volume = {14},
 number = {7},
 issn = {1745-2481},
 journal = {Nature Physics},
 doi = {10.1038/s41567-018-0115-y}
}

@article{Bartkiewicz.2017,
 author = {Bartkiewicz, Karol and {\v{C}}ernoch, Anton{\'i}n and Chimczak, Grzegorz and Lemr, Karel and Miranowicz, Adam and Nori, Franco},
 year = {2017},
 title = {Experimental quantum forgery of quantum optical money},
 volume = {3},
 pages = {7},
 number = {1},
 journal = {npj Quantum Information},
 doi = {10.1038/s41534-017-0010-x}
}

@article{BenDavid.2023,
 author = {Ben-David, Shalev and Sattath, Or},
 year = {2023},
 title = {Quantum Tokens for Digital Signatures},
 pages = {901},
 volume = {7},
 issn = {2521-327X},
 journal = {Quantum},
 doi = {10.22331/q-2023-01-19-901}
}

@article{Bennett.1984,
 author = {Bennett, C. H. and Brassard, G.},
 year = {1984},
 title = {Quantum Cryptography: Public Key Distribution and Coin Tossing},
 pages = {175--179},
 volume = {10},
 doi = {10.1016/j.tcs.2011.08.039},
 journal = {Proceedings of the IEEE International Conference on Computers, Systems and Signal Processing, Bangalore}
}

@incollection{Bennett.1983,
 author = {Bennett, Charles H. and Brassard, Gilles and Breidbart, Seth and Wiesner, Stephen},
 title = {Quantum Cryptography, or Unforgeable Subway Tokens},
 pages = {267--275},
 publisher = {{Springer US}},
 isbn = {978-1-4757-0604-8},
 editor = {Chaum, David and Rivest, Ronald L. and Sherman, Alan T.},
 booktitle = {Advances in Cryptology},
 year = {1983},
 address = {Boston, MA},
 doi = {10.1007/978-1-4757-0602-4}
}

@article{Bennett1992,
 author = {Bennett, Charles H. and Bessette, Fran{\c{c}}ois and Brassard, Gilles and Salvail, Louis and Smolin, John},
 year = {1992},
 title = {Experimental quantum cryptography},
 pages = {3--28},
 volume = {5},
 number = {1},
 journal = {Journal of Cryptology},
 doi = {10.1007/BF00191318}
}

@article{Beukers2025,
 author = {{Beukers,  Hans K. C.} and Waas, Christopher and Pasini, Matteo and {van Ommen}, Hendrik B. and Ademi, Zarije and Iuliano, Mariagrazia and Codreanu, Nina and Brevoord, Julia M. and Turan, Tim and Taminiau, Tim H. and Hanson, Ronald},
 year = {2025},
 title = {Control of Solid-State Nuclear Spin Qubits Using an Electron Spin},
 volume = {15},
 number = {2},
 page = {021011},
 journal = {Physical Review X},
 doi = {10.1103/physrevx.15.021011}
}

@article{Bhaskar2020,
 author = {Bhaskar, M. K. and Riedinger, R. and Machielse, B. and Levonian, D. S. and Nguyen, C. T. and Knall, E. N. and Park, H. and Englund, D. and Lon{\v{c}}ar, M. and Sukachev, D. D. and Lukin, M. D.},
 year = {2020},
 title = {Experimental demonstration of memory-enhanced quantum communication},
 pages = {60--64},
 volume = {580},
 number = {7801},
 issn = {1476-4687},
 journal = {Nature},
 doi = {10.1038/s41586-020-2103-5}
}

@article{birowosuto_fast_2012,
 author = {Birowosuto, Muhammad Danang and Sumikura, Hisashi and Matsuo, Shinji and Taniyama, Hideaki and {van Veldhoven}, Peter J. and N{\"o}tzel, Richard and Notomi, Masaya},
 year = {2012},
 title = {Fast Purcell-enhanced single photon source in 1,550-nm telecom band from a resonant quantum dot-cavity coupling},
 url = {https://www.nature.com/articles/srep00321},
 urldate = {2025-07-31},
 pages = {321},
 volume = {2},
 number = {1},
 journal = {Scientific reports},
 doi = {10.1038/srep00321}
}

@article{Blais.2004,
  title = {Cavity quantum electrodynamics for superconducting electrical circuits: An architecture for quantum computation},
  author = {Blais, Alexandre and Huang, Ren-Shou and Wallraff, Andreas and Girvin, S. M. and Schoelkopf, R. J.},
  journal = {Phys. Rev. A},
  volume = {69},
  issue = {6},
  pages = {062320},
  numpages = {14},
  year = {2004},
  month = {Jun},
  publisher = {American Physical Society},
  doi = {10.1103/PhysRevA.69.062320},
  url = {https://link.aps.org/doi/10.1103/PhysRevA.69.062320}
}

@misc{Boche2025Patent1,
 author = {{Holger Boche} and {Christian Deppe} and {Nilesh Kumar} and {Marc Geitz}},
 year = {2025},
 title = {Method and apparatus for storing a data message in a data storage, and method and apparatus for retrieving a data message from a data storage},
 type = {patent},
 number ={Submitted to patent attorney},
}

@misc{Boche2025Patent2,
 author = {{Holger Boche} and {Christian Deppe} and {Nilesh Kumar} and {Marc Geitz}},
 year = {2025},
 title = {Method and apparatus for enrolling a data message for identification in a data storage, and method and apparatus for identifying the presence of one or more query data messages},
 type = {patent},
 number ={Submitted to patent attorney},
}

@article{bopp_sawfish_2024,
 author = {Bopp, Julian M. and Plock, Matthias and Turan, Tim and Pieplow, Gregor and Burger, Sven and Schr{\"o}der, Tim},
 year = {2024},
 title = {`Sawfish' Photonic Crystal Cavity for Near-Unity Emitter-to-Fiber Interfacing in Quantum Network Applications},
 urldate = {2025-02-18},
 pages = {2301286},
 volume = {12},
 number = {13},
 issn = {2195-1071},
 journal = {Adv. Opt. Mater.},
 doi = {10.1002/adom.202301286}
}

@article{borregaard_one-way_2020,
 author = {Borregaard, Johannes and Pichler, Hannes and Schr{\"o}der, Tim and Lukin, Mikhail D. and Lodahl, Peter and S{\o}rensen, Anders S.},
 year = {2020},
 title = {One-Way Quantum Repeater Based on Near-Deterministic Photon-Emitter Interfaces},
 url = {https://link.aps.org/doi/10.1103/PhysRevX.10.021071},
 urldate = {2025-07-31},
 pages = {021071},
 volume = {10},
 number = {2},
 journal = {Physical Review X},
 doi = {10.1103/PhysRevX.10.021071}
}

@article{Bouchard2022,
 author = {Bouchard, Fr{\'e}d{\'e}ric and England, Duncan and Bustard, Philip J. and Heshami, Khabat and Sussman, Benjamin},
 year = {2022},
 title = {Quantum Communication with Ultrafast Time-Bin Qubits},
 volume = {3},
 number = {1},
 issn = {2691-3399},
 journal = {Phys. Rev. X Quantum}
}

@article{Bouwmeester.1997,
 author = {Bouwmeester, Dik and Pan, Jian-Wei and Mattle, Klaus and Eibl, Manfred and Weinfurter, Harald and Zeilinger, Anton},
 year = {1997},
 title = {Experimental quantum teleportation},
 pages = {575--579},
 volume = {390},
 number = {6660},
 issn = {1476-4687},
 journal = {Nature},
 doi = {10.1038/37539}
}

@article{Bozzio.2018,
 author = {Bozzio, Mathieu and Orieux, Adeline and {Trigo Vidarte}, Luis and Zaquine, Isabelle and Kerenidis, Iordanis and Diamanti, Eleni},
 year = {2018},
 title = {Experimental investigation of practical unforgeable quantum money},
 volume = {4},
 pages = {5},
 number = {1},
 journal = {npj Quantum Information},
 doi = {10.1038/s41534-018-0058-2}
}

@inproceedings{Brassard.2005,
 author = {Brassard, G.},
 title = {Brief history of quantum cryptography: a personal perspective},
 pages = {19--23},
 publisher = {IEEE},
 isbn = {0-7803-9491-7},
 booktitle = {2005 IEEE information theory workshop on theory and practice in information-theoretic security},
 year = {2005},
 address = {New York},
 doi = {10.1109/ITWTPI.2005.1543949}
}

@article{Braunstein2000,
 author = {Braunstein, Samuel L. and Kimble, H. J.},
 year = {2000},
 title = {Dense coding for continuous variables},
 pages = {042302},
 volume = {61},
 number = {4},
 journal = {Phys. Rev. A},
 doi = {10.1103/PhysRevA.61.042302}
}

@article{bus22,
 author = {Buser, Gianni and Mottola, Roberto and Cotting, Bj{\"o}rn and Wolters, Janik and Treutlein, Philipp},
 year = {2022},
 title = {Single-Photon Storage in a Ground-State Vapor Cell Quantum Memory},
 pages = {020349},
 volume = {3},
 journal = {PRX Quantum}
}

@article{casabone_dynamic_2021,
 author = {Casabone, Bernardo and Deshmukh, Chetan and Liu, Shuping and Serrano, Diana and Ferrier, Alban and H{\"u}mmer, Thomas and Goldner, Philippe and Hunger, David and de Riedmatten, Hugues},
 title = {Dynamic control of Purcell enhanced emission of erbium ions in nanoparticles},
 url = {https://www.nature.com/articles/s41467-021-23632-9},
 urldate = {2022-10-13},
 pages = {3570},
 volume = {12},
year = {2021},
 number = {1},
 issn = {2041-1723},
 journal = {Nat. Commun.},
 doi = {10.1038/s41467-021-23632-9}
}

@article{Cerf2001,
 author = {Cerf, N. J. and L{\'e}vy, M. and {van Assche}, G.},
 year = {2001},
 title = {Quantum distribution of Gaussian keys using squeezed states},
 pages = {052311},
 volume = {63},
 number = {5},
 journal = {Phys. Rev. A},
 doi = {10.1103/PhysRevA.63.052311}
}

@misc{chamier_low-noise_2025,
      title={Low-Noise Cascaded Frequency Conversion of $637.2$ nm Light to the Telecommunication C-Band in a Single-Waveguide Device}, 
      author={Fabrice von Chamier and Joscha Hanel and Chris Müller and Wanrong Li and Roger Alfredo Kögler and Oliver Benson},
      year={2025},
      eprint={2502.14557},
      archivePrefix={arXiv},
      primaryClass={quant-ph},
      url={https://arxiv.org/abs/2502.14557}, 
}

@inproceedings{che2015puf,
 author={Che, Wenjie and Saqib, Fareena and Plusquellic, Jim},
  booktitle={2015 IEEE/ACM International Conference on Computer-Aided Design (ICCAD)}, 
  title={PUF-based authentication}, 
  year={2015},
  volume={},
  number={},
  pages={337-344},
  doi={10.1109/ICCAD.2015.7372589},
  ISSN={},
  month={Nov},}

@inproceedings{Chen.9520219102021,
 author = {Chen, Huimin and Jia, Hengyue and Wu, Xia and Wang, Xiuli and Wang, Maoning},
 title = {Quantum Token for Network Authentication},
 pages = {688--692},
 publisher = {IEEE},
 isbn = {978-1-6654-1681-8},
 booktitle = {2021 IEEE International Conference on Web Services (ICWS)},
 year = {2021},
 doi = {10.1109/ICWS53863.2021.00095}
}

@article{chen_parallel_2020,
 author = {Chen, Songtao and Raha, Mouktik and Phenicie, Christopher M. and Ourari, Salim and Thompson, Jeff D.},
 year = {2020},
 title = {Parallel single-shot measurement and coherent control of solid-state spins below the diffraction limit},
 url = {https://www.science.org/doi/full/10.1126/science.abc7821    ,},
 urldate = {2025-06-26},
 pages = {592--595},
 volume = {370},
 number = {6516},
 journal = {Science},
 doi = {10.1126/science.abc7821}
}

@article{Kun.2025,
  title = {Direct and efficient detection of quantum superposition},
  author = {Kun, Daniel and Str\"omberg, Teodor and Spagnolo, Michele and Daki\ifmmode \acute{c}\else \'{c}\fi{}, Borivoje and Rozema, Lee A. and Walther, Philip},
  journal = {Phys. Rev. A},
  volume = {111},
  issue = {5},
  pages = {L050402},
  numpages = {6},
  year = {2025},
  month = {May},
  publisher = {American Physical Society},
  doi = {10.1103/PhysRevA.111.L050402},
  url = {https://link.aps.org/doi/10.1103/PhysRevA.111.L050402}
}

@article{Terhal.2025,
 author = {Terhal, Barbara M. and Wolf, Michael M. and Doherty, Andrew C.},
 year = {2025},
 title = {Quantum entanglement: A modern perspective},
 pages = {40--46},
 volume = {78},
 number = {1},
 issn = {0031-9228},
 journal = {Physics Today},
 doi = {10.1063/pt.rzbf.mjwy}
}

@article{Chen2021,
 author = {Chen, Yu-Ao and Zhang, Qiang and Chen, Teng-Yun and Cai, Wen-Qi and Liao, Sheng-Kai and Zhang, Jun and Chen, Kai and Yin, Juan and Ren, Ji-Gang and Chen, Zhu and Han, Sheng-Long and Yu, Qing and Liang, Ken and Zhou, Fei and Yuan, Xiao and Zhao, Mei-Sheng and Wang, Tian-Yin and Jiang, Xiao and Zhang, Liang and Liu, Wei-Yue and Li, Yang and Shen, Qi and Cao, Yuan and Lu, Chao-Yang and Shu, Rong and Wang, Jian-Yu and Li, Li and Liu, Nai-Le and Xu, Feihu and Wang, Xiang-Bin and Peng, Cheng-Zhi and Pan, Jian-Wei},
 year = {2021},
 title = {An integrated space-to-ground quantum communication network over 4,600 kilometres},
 pages = {214--219},
 volume = {589},
 number = {7841},
 issn = {1476-4687},
 journal = {Nature},
 doi = {10.1038/s41586-020-03093-8}
}

@article{Chiorescu.2004,
 author = {Chiorescu, I. and Bertet, P. and Semba, K. and Nakamura, Y. and {Harmans, C. J. P. M.} and Mooij, J. E.},
 year = {2004},
 title = {Coherent dynamics of a flux qubit coupled to a harmonic oscillator},
 pages = {159--162},
 volume = {431},
 number = {7005},
 issn = {1476-4687},
 journal = {Nature},
 doi = {10.1038/nature02831}
}

@misc{costa_telecom_2025,
      title={Telecom quantum dots on GaAs substrates as integration-ready high performance single-photon sources}, 
      author={Beatrice Costa and Bianca Scaparra and Xiao Wei and Hubert Riedl and Gregor Koblmüller and Eugenio Zallo and Jonathan Finley and Lukas Hanschke and Kai Müller},
      year={2025},
      eprint={2505.22886},
      archivePrefix={arXiv},
      primaryClass={cond-mat.mes-hall},
      url={https://arxiv.org/abs/2505.22886}, 
}

@misc{cra23,
  title = {High-rate subgigahertz-linewidth bichromatic entanglement source for quantum networking},
  author = {Craddock, Alexander N. and Wang, Yang and Giraldo, Felipe and Sekelsky, Rourke and Flament, Mael and Namazi, Mehdi},
  journal = {Phys. Rev. Appl.},
  volume = {21},
  issue = {3},
  pages = {034012},
  numpages = {6},
  year = {2024},
  month = {Mar},
  publisher = {American Physical Society},
  doi = {10.1103/PhysRevApplied.21.034012},
  url = {https://link.aps.org/doi/10.1103/PhysRevApplied.21.034012}
}

@article{Davies:1974gu,
 author = {Davies, E. R.},
 year = {1974},
 title = {A new pulse endor technique},
 pages = {1--2},
 volume = {47},
 journal = {Phys. Lett. A},
 doi = {10.1016/0375-9601(74)90078-4}
}

@article{Debroux2021,
 author = {Debroux, Romain and Michaels, Cathryn P. and Purser, Carola M. and Wan, Noel and Trusheim, Matthew E. and {Arjona Mart{\'i}nez}, Jes{\'u}s and Parker, Ryan A. and Stramma, Alexander M. and Chen, Kevin C. and de Santis, Lorenzo and Alexeev, Evgeny M. and Ferrari, Andrea C. and Englund, Dirk and Gangloff, Dorian A. and Atat{\"u}re, Mete},
 year = {2021},
 title = {Quantum Control of the Tin-Vacancy Spin Qubit in Diamond},
 volume = {11},
 number = {4},
 pages ={041041},
 journal = {Physical Review X},
 doi = {10.1103/physrevx.11.041041}
}

@article{delvaux2014helper,
  author={Delvaux, Jeroen and Gu, Dawu and Schellekens, Dries and Verbauwhede, Ingrid},
  journal={IEEE Transactions on Computer-Aided Design of Integrated Circuits and Systems}, 
  title={Helper Data Algorithms for PUF-Based Key Generation: Overview and Analysis}, 
  year={2015},
  volume={34},
  number={6},
  pages={889-902},
  doi={10.1109/TCAD.2014.2370531},
  ISSN={1937-4151},
  month={June},}

@article{DeSantis2021,
 author = {de Santis, Lorenzo and Trusheim, Matthew E. and Chen, Kevin C. and Englund, Dirk R.},
 year = {2021},
 title = {Investigation of the Stark Effect on a Centrosymmetric Quantum Emitter in Diamond},
 volume = {127},
 number = {14},
 journal = {Physical Review Letters},
 doi = {10.1103/physrevlett.127.147402}
}

@article{deshmukh_detection_2023,
 author = {Deshmukh, Chetan and Beattie, Eduardo and Casabone, Bernardo and Grandi, Samuele and Serrano, Diana and Ferrier, Alban and Goldner, Philippe and Hunger, David and de Riedmatten, Hugues},
 year = {2023},
 title = {Detection of single ions in a nanoparticle coupled to~a fiber cavity},
 urldate = {2024-02-08},
 pages = {1339--1344},
 volume = {10},
 number = {10},
 issn = {2334-2536},
 journal = {Optica},
 doi = {10.1364/OPTICA.491692}
}

@article{dibos_atomic_2018,
  title = {Atomic Source of Single Photons in the Telecom Band},
  author = {Dibos, A. M. and Raha, M. and Phenicie, C. M. and Thompson, J. D.},
  journal = {Phys. Rev. Lett.},
  volume = {120},
  issue = {24},
  pages = {243601},
  numpages = {6},
  year = {2018},
  month = {Jun},
  publisher = {American Physical Society},
  doi = {10.1103/PhysRevLett.120.243601},
  url = {https://link.aps.org/doi/10.1103/PhysRevLett.120.243601}
}

@article{Dicke.1954,
 author = {Dicke, R. H.},
 year = {1954},
 title = {Coherence in Spontaneous Radiation Processes},
 pages = {99--110},
 volume = {93},
 number = {1},
 journal = {Physical Review},
 doi = {10.1103/PhysRev.93.99}
}

@article{Dieks.1982,
 author = {Dieks, D.},
 year = {1982},
 title = {Communication by EPR devices},
 pages = {271--272},
 volume = {92},
 number = {6},
 issn = {03759601},
 journal = {Physics Letters A},
 doi = {10.1016/0375-9601(82)90084-6}
}

@article{doo21,
 author = {Doosti, Mina and Kumar, Niraj and Delavar, Mahshid and Kashefi, Elham},
 year = {2021},
 title = {Client-server Identification Protocols with Quantum PUF},
 volume = {2},
 number = {3},
page = {197},
 journal = {ACM Transactions on Quantum Computing},
 doi = {10.1145/3484197}
}

@article{dreau_quantum_2018,
 author = {Dr{\'e}au, Ana{\"i}s and Tchebotareva, Anna and Mahdaoui, Aboubakr El and Bonato, Cristian and Hanson, Ronald},
 year = {2018},
 title = {Quantum Frequency Conversion of Single Photons from a Nitrogen-Vacancy Center in Diamond to Telecommunication Wavelengths},
 url = {https://link.aps.org/doi/10.1103/PhysRevApplied.9.064031},
 urldate = {2025-08-12},
 pages = {064031},
 volume = {9},
 number = {6},
 journal = {Physical Review Applied},
 doi = {10.1103/PhysRevApplied.9.064031}
}

@article{esg23,
 title = {Optimization and readout-noise analysis of a warm-vapor electromagnetically-induced-transparency memory on the Cs ${D}_{1}$ line},
  author = {Esguerra, Luisa and Me\ss{}ner, Leon and Robertson, Elizabeth and Ewald, Norman Vincenz and G\"undo\ifmmode \breve{g}\else \u{g}\fi{}an, Mustafa and Wolters, Janik},
  journal = {Phys. Rev. A},
  volume = {107},
  issue = {4},
  pages = {042607},
  numpages = {6},
  year = {2023},
  month = {Apr},
  publisher = {American Physical Society},
  doi = {10.1103/PhysRevA.107.042607},
  url = {https://link.aps.org/doi/10.1103/PhysRevA.107.042607}
}

@article{farre2025entanglement,
doi = {10.1088/1402-4896/ae0838},
url = {https://doi.org/10.1088/1402-4896/ae0838},
year = {2025},
month = {oct},
publisher = {IOP Publishing},
volume = {100},
number = {10},
pages = {105113},
author = {Farré, Pol Julià and Galetsky, Vladlen and Ghosh, Soham and Nötzel, Janis and Deppe, Christian},
title = {Entanglement-assisted authenticated BB84 protocol},
journal = {Physica Scripta}
}

@article{Fedorov.2021,
 author = {{Kirill G. Fedorov} and {Michael Renger} and {Stefan Pogorzalek} and {Roberto Di Candia} and {Qiming Chen} and {Yuki Nojiri} and {Kunihiro Inomata} and {Yasunobu Nakamura} and {Matti Partanen} and {Achim Marx} and {Rudolf Gross} and {Frank Deppe}},
 year = {2021},
 title = {Experimental quantum teleportation of propagating microwaves},
 pages = {eabk0891},
 volume = {7},
 number = {52},
 journal = {Sci. Adv.},
 doi = {10.1126/sciadv.abk0891}
}

@article{Goorden2014qsa,
author = {Sebastianus A. Goorden and Marcel Horstmann and Allard P. Mosk and Boris \v{S}kori\'{c} and Pepijn W. H. Pinkse},
journal = {Optica},
number = {6},
pages = {421--424},
publisher = {Optica Publishing Group},
title = {Quantum-secure authentication of a physical unclonable key},
volume = {1},
month = {Dec},
year = {2014},
url = {https://opg.optica.org/optica/abstract.cfm?URI=optica-1-6-421},
doi = {10.1364/OPTICA.1.000421},
}

@article{tarik2022sipuf,
 author = {Tarik, Farhan Bin and Famili, Azadeh and Lao, Yingjie and Ryckman, Judson D.},
 year = {2022},
 title = {Scalable and CMOS compatible silicon photonic physical unclonable functions for supply chain assurance},
 pages = {15653},
 volume = {12},
 number = {1},
 journal = {Scientific reports},
 doi = {10.1038/s41598-022-19796-z}
}

@article{vernazgris2018memory,
 author = {Vernaz-Gris, Pierre and Huang, Kun and Cao, Mingtao and Sheremet, Alexandra S. and Laurat, Julien},
 year = {2018},
 title = {Highly-efficient quantum memory for polarization qubits in a spatially-multiplexed cold atomic ensemble},
 pages = {363},
 volume = {9},
 number = {1},
 issn = {2041-1723},
 journal = {Nat. Commun.},
 doi = {10.1038/s41467-017-02775-8}
}

@article{nocentini2024multilevel,
 author = {Nocentini, Sara and R{\"u}hrmair, Ulrich and Barni, Mauro and Wiersma, Diederik S. and Riboli, Francesco},
 year = {2024},
 title = {All-optical multilevel physical unclonable functions},
 pages = {369--376},
 volume = {23},
 number = {3},
 journal = {Nature Materials},
 doi = {10.1038/s41563-023-01734-7}
}

@misc{skoric2009qrpuf,
      author = {Boris Skoric},
      title = {Quantum readout of Physical Unclonable Functions: Remote authentication without trusted readers and authenticated Quantum Key Exchange without initial shared secrets},
      howpublished = {Cryptology {ePrint} Archive, Paper 2009/369},
      year = {2009},
      url = {https://eprint.iacr.org/2009/369}
}

@article{Fedorov2018,
 author = {Fedorov, K. G. and Pogorzalek, S. and {Las Heras}, U. and Sanz, M. and Yard, P. and Eder, P. and Fischer, M. and Goetz, J. and Xie, E. and Inomata, K. and Nakamura, Y. and {Di Candia}, R. and Solano, E. and Marx, A. and Deppe, F. and Gross, R.},
 year = {2018},
 title = {Finite-time quantum entanglement in propagating squeezed microwaves},
 pages = {6416},
 volume = {8},
 number = {1},
 journal = {Sci. Rep.},
 doi = {10.1038/s41598-018-24742-z}
}

@article{feher59b,
 author = {Feher, G. and Gere, E. A.},
 year = {1959},
 title = {Electron Spin Resonance Experiments on Donors in Silicon. $\mathrm{II}$. Electron Spin Relaxation Effects},
 url = {http://link.aps.org/doi/10.1103/PhysRev.114.1245},
 pages = {1245},
 volume = {114},
 number = {5},
 journal = {Physical Review},
 doi = {10.1103/physrev.114.1245}
}

@article{Fesquet2024,
 author = {Fesquet, Florian and Kronowetter, Fabian and Renger, Michael and Yam, Wun Kwan and Gandorfer, Simon and Inomata, Kunihiro and Nakamura, Yasunobu and Marx, Achim and Gross, Rudolf and Fedorov, Kirill G.},
 year = {2024},
 title = {Demonstration of microwave single-shot quantum key distribution},
 pages = {7544},
 volume = {15},
 number = {1},
 journal = {Nature Communications},
 doi = {10.1038/s41467-024-51421-7}
}

@article{Fesquet.2023,
 author = {Fesquet, F. and Kronowetter, F. and Renger, M. and Chen, Q. and Honasoge, K. and Gargiulo, O. and Nojiri, Y. and Marx, A. and Deppe, F. and Gross, R. and Fedorov, K. G.},
 year = {2023},
 title = {Perspectives of microwave quantum key distribution in the open air},
 volume = {108},
 pages = {032607},
 number = {3},
 issn = {2469-9926},
 journal = {Physical Review A},
 doi = {10.1103/PhysRevA.108.032607}
}

@article{GonzalezRaya.2022,
  title = {Open-Air Microwave Entanglement Distribution for Quantum Teleportation},
  author = {Gonzalez-Raya, Tasio and Casariego, Mateo and Fesquet, Florian and Renger, Michael and Salari, Vahid and M\"ott\"onen, Mikko and Omar, Yasser and Deppe, Frank and Fedorov, Kirill G. and Sanz, Mikel},
  journal = {Phys. Rev. Appl.},
  volume = {18},
  issue = {4},
  pages = {044002},
  numpages = {30},
  year = {2022},
  month = {Oct},
  publisher = {American Physical Society},
  doi = {10.1103/PhysRevApplied.18.044002},
  url = {https://link.aps.org/doi/10.1103/PhysRevApplied.18.044002}
}

@article{fin18,
 author = {{Ran Finkelstein} and {Eilon Poem} and {Ohad Michel} and {Ohr Lahad} and {Ofer Firstenberg}},
 year = {2018},
 title = {Fast, noise-free memory for photon synchronization at room temperature},
 pages = {eaap8598},
 volume = {4},
 number = {1},
 journal = {Science Advances},
 doi = {10.1126/sciadv.aap8598}
}

@article{flaschmann_optimizing_2023,
 author = {Flaschmann, Rasmus and Schmid, Christian and Zugliani, Lucio and Strohauer, Stefan and Wietschorke, Fabian and Grotowski, Stefanie and Jonas, Bj{\"o}rn and M{\"u}ller, Manuel and Althammer, Matthias and Gross, Rudolf and Finley, Jonathan J. and M{\"u}ller, Kai},
 year = {2023},
 title = {Optimizing the growth conditions of Al mirrors for superconducting nanowire single-photon detectors},
 urldate = {2025-08-12},
 pages = {035002},
 volume = {3},
 number = {3},
 issn = {2633-4356},
 journal = {Materials for Quantum Technology},
 doi = {10.1088/2633-4356/ace490}
}

@inproceedings{Gavinsky.2012,
 author = {Gavinsky, Dmitry},
 title = {Quantum Money with Classical Verification},
 pages = {42--52},
 publisher = {IEEE},
 isbn = {978-0-7695-4708-4},
 booktitle = {2012 IEEE 27th Conference on Computational Complexity},
 year = {6/26/2012 - 6/29/2012},
 doi = {10.1109/CCC.2012.10}
}

@article{genov_arbitrarily_2017,
 author = {Genov, Genko T. and Schraft, Daniel and Vitanov, Nikolay V. and Halfmann, Thomas},
 title = {Arbitrarily Accurate Pulse Sequences for Robust Dynamical Decoupling},
 url = {https://link.aps.org/doi/10.1103/PhysRevLett.118.133202    ,},
 urldate = {2025-06-27},
 pages = {133202},
 volume = {118},
year = {2017},
 number = {13},
 journal = {Physical Review Letters},
 doi = {10.1103/PhysRevLett.118.133202}
}

@article{Gentile2017_RMP_45004,
  title = {Optically polarized $^{3}\mathrm{He}$},
  author = {Gentile, T. R. and Nacher, P. J. and Saam, B. and Walker, T. G.},
  journal = {Rev. Mod. Phys.},
  volume = {89},
  issue = {4},
  pages = {045004},
  numpages = {59},
  year = {2017},
  month = {Dec},
  publisher = {American Physical Society},
  doi = {10.1103/RevModPhys.89.045004},
  url = {https://link.aps.org/doi/10.1103/RevModPhys.89.045004}
}

@article{ghosh2024existential,
  title = {Existential unforgeability in quantum authentication from quantum physical unclonable functions based on random von Neumann measurement},
  author = {Ghosh, Soham and Galetsky, Vladlen and Juli\`a Farr\'e, Pol and Deppe, Christian and Ferrara, Roberto and Boche, Holger},
  journal = {Phys. Rev. Res.},
  volume = {6},
  issue = {4},
  pages = {043306},
  numpages = {18},
  year = {2024},
  month = {Dec},
  doi = {https://link.aps.org/doi/10.1103/PhysRevResearch.6.043306},
  url = {https://link.aps.org/doi/10.1103/PhysRevResearch.6.043306}
}

@incollection{goldner_chapter_2015,
 author = {Goldner, Philippe and Ferrier, Alban and Guillot-No{\"e}l, Olivier},
 title = {Chapter 267 - Rare Earth-Doped Crystals for Quantum Information Processing},
 url = {https://www.sciencedirect.com/science/article/pii/B9780444632609002674},
 urldate = {2025-06-27},
 pages = {1--78},
 volume = {46},
year = {2015},
 publisher = {Elsevier},
 editor = {B{\"u}nzli, Jean-Claude G. and Pecharsky, Vitalij K.},
 booktitle = {Handbook on the Physics and Chemistry of Rare Earths},
 doi = {10.1016/B978-0-444-63260-9.00267-4}
}

@article{grimm_coherent_2025,
 author = {Grimm, Nick and Senkalla, Katharina and Vetter, Philipp J. and Frey, Jurek and Gundlapalli, Prithvi and Calarco, Tommaso and Genov, Genko and M{\"u}ller, Matthias M. and Jelezko, Fedor},
 year = {2025},
 title = {Coherent Control of a Long-Lived Nuclear Memory Spin in a Germanium-Vacancy Multi-Qubit Node},
 urldate = {2025-08-01},
 pages = {043603},
 volume = {134},
 number = {4},
 journal = {Physical Review Letters},
 doi = {10.1103/PhysRevLett.134.043603}
}

@article{gritsch_optical_2025,
 author = {Gritsch, Andreas and Ulanowski, Alexander and Pforr, Jakob and Reiserer, Andreas},
 title = {Optical single-shot readout of spin qubits in silicon},
 url = {https://www.nature.com/articles/s41467-024-55552-9},
 urldate = {2025-02-24},
 pages = {64},
 volume = {16},
year = {2025},
 number = {1},
 issn = {2041-1723},
 journal = {Nat. Commun.},
 doi = {10.1038/s41467-024-55552-9}
}

@article{grotowski_optimizing_2025,
 author = {Grotowski, Stefanie and Zugliani, Lucio and Jonas, Bj{\"o}rn and Flaschmann, Rasmus and Schmid, Christian and Strohauer, Stefan and Wietschorke, Fabian and Bruckmoser, Niklas and M{\"u}ller, Manuel and Althammer, Matthias and Gross, Rudolf and M{\"u}ller, Kai and Finley, Jonathan},
 year = {2025},
 title = {Optimizing the growth conditions of superconducting MoSi thin films for single photon detection},
 url = {https://www.nature.com/articles/s41598-025-86303-5},
 urldate = {2025-07-31},
 pages = {2438},
 volume = {15},
 number = {1},
 journal = {Scientific reports},
 doi = {10.1038/s41598-025-86303-5}
}

@article{Guan.2018,
 author = {Guan, Jian-Yu and Arrazola, Juan Miguel and Amiri, Ryan and Zhang, Weijun and Li, Hao and You, Lixing and Wang, Zhen and Zhang, Qiang and Pan, Jian-Wei},
 year = {2018},
 title = {Experimental preparation and verification of quantum money},
 volume = {97},
 number = {3},
 pages = {032338},
 issn = {2469-9926},
 journal = {Physical Review A},
 doi = {10.1103/PhysRevA.97.032338},
 url = {https://link.aps.org/doi/10.1103/PhysRevA.97.032338}
}

@article{gundogdu_alganaln_2025,
 author = {G{\"u}ndo{\u{g}}du, Sinan and Pazzagli, Sofia and Pregnolato, Tommaso and Kolbe, Tim and Hagedorn, Sylvia and Weyers, Markus and Schr{\"o}der, Tim},
 year = {2025},
 title = {AlGaN/AlN heterostructures: an emerging platform for integrated photonics},
 url = {https://www.nature.com/articles/s44310-024-00048-z},
 urldate = {2025-08-01},
 pages = {2},
 volume = {2},
 number = {1},
 issn = {2948-216X},
 journal = {npj Nanophotonics},
 doi = {10.1038/s44310-024-00048-z}
}

@article{guo19,
 author = {Guo, Jinxian and Feng, Xiaotian and Yang, Peiyu and Yu, Zhifei and Chen, L. Q. and Yuan, Chun-Hua and Zhang, Weiping},
 year = {2019},
 title = {High-performance Raman quantum memory with optimal control in room temperature atoms},
 pages = {148},
 volume = {10},
 doi = {https://doi.org/10.1038/s41467-018-08118-5},
 number = {1},
 journal = {Nature Communications}
}

@misc{gupta_dual_2023,
      title={Dual epitaxial telecom spin-photon interfaces with correlated long-lived coherence}, 
      author={Shobhit Gupta and Yizhong Huang and Shihan Liu and Yuxiang Pei and Natasha Tomm and Richard J. Warburton and Tian Zhong},
      year={2023},
      eprint={2310.07120},
      archivePrefix={arXiv},
      primaryClass={quant-ph},
      url={https://arxiv.org/abs/2310.07120}, 
}

@article{hain_light_2025,
 author = {Hain, Marcel and Stewen, Niklas and Halfmann, Thomas},
 year = {2025},
 title = {Light storage by electromagnetically induced transparency for one second at the level of a single photon in \textrm{Pr}:\textrm{Y}$_2$\textrm{SiO}$_5$ prepared with multiple frequency ensembles},
 urldate = {2025-02-28},
 pages = {023029},
 volume = {27},
 number = {2},
 journal = {New Journal of Physics},
 doi = {10.1088/1367-2630/adb510}
}

@article{Harris2024,
  title = {Coherence of group-IV color centers},
  author = {Harris, Isaac B. W. and Englund, Dirk},
  journal = {Phys. Rev. B},
  volume = {109},
  issue = {8},
  pages = {085414},
  numpages = {10},
  year = {2024},
  month = {Feb},
  publisher = {American Physical Society},
  doi = {10.1103/PhysRevB.109.085414},
  url = {https://link.aps.org/doi/10.1103/PhysRevB.109.085414}
}

@article{heinze_stopped_2013,
 author = {Heinze, Georg and Hubrich, Christian and Halfmann, Thomas},
 title = {Stopped Light and Image Storage by Electromagnetically Induced Transparency up to the Regime of One Minute},
 url = {https://link.aps.org/doi/10.1103/PhysRevLett.111.033601   ,},
 urldate = {2025-06-27},
 pages = {033601},
year = {2023},
 volume = {111},
 number = {3},
 journal = {Physical Review Letters},
 doi = {10.1103/PhysRevLett.111.033601}
}

@article{hes16,
 author = {{Khabat Heshami} and {Duncan G. England} and {Peter C. Humphreys} and {Philip J. Bustard} and {Victor M. Acosta} and {Joshua Nunn} and {Benjamin J. Sussman and}},
 year = {2016},
 title = {Quantum memories: emerging applications and recent advances},
 pages = {2005--2028},
 volume = {63},
 number = {20},
 journal = {Journal of Modern Optics},
 doi = {10.1080/09500340.2016.1148212}
}

@article{Hoehne:2011ii,
 author = {Hoehne, Felix and Dreher, Lukas and Huebl, Hans and Stutzmann, Martin and Brandt, Martin S.},
 year = {2011},
 title = {Electrical Detection of Coherent Nuclear Spin Oscillations in Phosphorus-Doped Silicon using Pulsed ENDOR},
 pages = {187601},
 volume = {106},
 number = {18},
 journal = {Physical Review Letters},
 doi = {10.1103/PhysRevLett.106.187601}
}

@article{Jensen2011,
 author = {Jensen, K. and Wasilewski, W. and Krauter, H. and Fernholz, T. and Nielsen, B. M. and Owari, M. and Plenio, M. B. and Serafini, A. and Wolf, M. M. and Polzik, E. S.},
 year = {2011},
 title = {Quantum memory for entangled continuous-variable states},
 pages = {13--16},
 volume = {7},
 number = {1},
 issn = {1745-2481},
 journal = {Nature Physics},
 doi = {10.1038/nphys1819}
}

@misc{Jiang2024,
      title={Experimental practical quantum tokens with transaction time advantage}, 
      author={Yang-Fan Jiang and Adrian Kent and Damián Pitalúa-García and Xiaochen Yao and Xiaohan Chen and Jia Huang and George Cowperthwaite and Qibin Zheng and Hao Li and Lixing You and Yang Liu and Qiang Zhang and Jian-Wei Pan},
      year={2024},
      eprint={2408.13063},
      archivePrefix={arXiv},
      primaryClass={quant-ph},
      url={https://arxiv.org/abs/2408.13063}
}

@article{Julsgaard2013,
 author = {Julsgaard, Brian and Grezes, C{\'e}cile and Bertet, Patrice and M{\o}lmer, Klaus},
 year = {2013},
 title = {Quantum memory for microwave photons in an inhomogeneously broadened spin ensemble},
 pages = {250503},
 volume = {110},
 number = {25},
 journal = {Physical Review Letters},
 doi = {10.1103/PhysRevLett.110.250503}
}

@article{jut24,
 author = {Jutisz, Martin and Erl, Alexander and Wolters, Janik and G{\"u}ndo{\u{g}}an, Mustafa and Krutzik, Markus},
 year = {2024},
 title = {Standalone mobile quantum memory system},
 journal = {arXiv preprint arXiv:2410.21209}
}

@article{Karapatzakis2024,
 author = {Karapatzakis, Ioannis and Resch, Jeremias and Schrodin, Marcel and Fuchs, Philipp and Kieschnick, Michael and Heupel, Julia and Kussi, Luis and S{\"u}rgers, Christoph and Popov, Cyril and Meijer, Jan and Becher, Christoph and Wernsdorfer, Wolfgang and Hunger, David},
 year = {2024},
 title = {Microwave Control of the Tin-Vacancy Spin Qubit in Diamond with a Superconducting Waveguide},
 volume = {14},
 pages = {031036},
 number = {3},
 journal = {Physical Review X},
 doi = {10.1103/physrevx.14.031036}
}

@article{kat18,
 author = {Katz, Or and Firstenberg, Ofer},
 year = {2018},
 title = {Light storage for one second in room-temperature alkali vapor},
 pages = {2074},
 volume = {9},
 pages = {https://doi.org/10.1038/s41467-018-04458-4},
 number = {1},
 journal = {Nature Communications}
}

@article{kat22,
 author = {Katz, Or and Shaham, Roy and Reches, Eran and Gorshkov, Alexey V. and Firstenberg, Ofer},
 year = {2022},
 title = {Optical quantum memory for noble-gas spins based on spin-exchange collisions},
 pages = {042606},
 volume = {105},
 number = {4},
 doi = {https://doi.org/10.1103/PhysRevA.105.042606},
 issn = {2469-9926},
 journal = {Physical Review A}
}

@article{kaupp_purcell-enhanced_2023,
 author = {Kaupp, Jochen and Reum, Yorick and Kohr, Felix and Michl, Johannes and Buchinger, Quirin and Wolf, Adriana and Peniakov, Giora and Huber-Loyola, Tobias and Pfenning, Andreas and H{\"o}fling, Sven},
 year = {2023},
 title = {Purcell-Enhanced Single-Photon Emission in the Telecom C-Band},
 urldate = {2025-07-31},
 pages = {2300242},
 volume = {6},
 number = {12},
 issn = {2511-9044},
 journal = {Advanced Quantum Technologies},
 doi = {10.1002/qute.202300242}
}

@article{Kent.2019,
 author = {Kent, Adrian},
 year = {2019},
 title = {S-money: virtual tokens for a relativistic economy},
 pages = {20190170},
 volume = {475},
 number = {2225},
 issn = {1364-5021},
 journal = {Proceedings. Mathematical, physical, and engineering sciences},
 doi = {10.1098/rspa.2019.0170}
}

@article{Kent.2022,
 author = {Kent, Adrian and Lowndes, David and Pital{\'u}a-Garc{\'i}a, Dami{\'a}n and Rarity, John},
 year = {2022},
 title = {Practical quantum tokens without quantum memories and experimental tests},
 volume = {8},
 pages = {28},
 number = {1},
 journal = {npj Quantum Information},
 doi = {10.1038/s41534-022-00524-4}
}

@article{kindem_control_2020,
 author = {Kindem, Jonathan M. and Ruskuc, Andrei and Bartholomew, John G. and Rochman, Jake and Huan, Yan Qi and Faraon, Andrei},
 year = {2020},
 title = {Control and single-shot readout of an ion embedded in a nanophotonic cavity},
 url = {https://www.nature.com/articles/s41586-020-2160-9},
 urldate = {2025-01-26},
 pages = {201--204},
 volume = {580},
 number = {7802},
 issn = {1476-4687},
 journal = {Nature},
 doi = {10.1038/s41586-020-2160-9}
}

@article{Kjaergaard2020,
 author = {Kjaergaard, Morten and Schwartz, Mollie E. and Braum{\"u}ller, Jochen and Krantz, Philip and Wang, Joel I.-J. and Gustavsson, Simon and Oliver, William D.},
 year = {2020},
 title = {Superconducting Qubits: Current State of Play},
 pages = {369--395},
 volume = {11},
 number = {Volume 11, 2020},
 journal = {Annu. Rev. Condens. Matter Phys.},
 doi = {10.1146/annurev-conmatphys-031119-050605}
}

@article{Knaut2024,
 author = {{C. M. Knaut} and {A. Suleymanzade} and {Y.-C. Wei} and {D. R. Assumpcao} and {P.-J. Stas} and {Y. Q. Huan} and {B. Machielse} and others},
 year = {2024},
 title = {Entanglement of Nanophotonic Quantum Memory Nodes in a Telecom Network},
 pages = {573--578},
 volume = {629},
 number = {8012},
 issn = {1476-4687},
 journal = {Nature},
 doi = {10.1038/s41586-024-07252-z}
}

@article{Kornack2002_PRL_253002,
 author = {Kornack, T. W. and Romalis, M. V.},
 year = {2002},
 title = {Dynamics of Two Overlapping Spin Ensembles Interacting by Spin Exchange},
 pages = {253002},
 volume = {89},
 journal = {Physical Review Letters},
 doi = {10.1103/physrevlett.89.253002}
}

@article{Kukharchyk2018coh,
 author = {Kukharchyk, N. and Sholokhov, D. and Morozov, O. and Korableva, S. L. and Kalachev, A. A. and Bushev, P. A.},
 year = {2018},
 title = {Optical coherence of $^{166}\mathrm{Er}:^7\mathrm{LiYF}_4$ crystal below 1 $\mathrm{K}$},
 volume = {20},
 number = {2},
 pages = {023044},
 journal = {New Journal of Physics},
 doi = {10.1088/1367-2630/aaa7e4}
}

@article{Kukharchyk2020,
 author = {Kukharchyk, Nadezhda and Sholokhov, Dmitriy and Morozov, Oleg and Korableva, Stella L. and Kalachev, Alexey A. and Bushev, Pavel A.},
 year = {2020},
 title = {Electromagnetically induced transparency in a mono-isotopic $^{167}\mathrm{Er}:^7\mathrm{LiYF}_4$ crystal below 1 Kelvin: microwave photonics approach},
 pages = {29166},
 volume = {28},
 number = {20},
 journal = {Optics Express},
 doi = {10.1364/oe.400222}
}

@inproceedings{Kumar2024WFIoT,
  author={Nilesh, Kumar and Deppe, Christian and Boche, Holger},
  booktitle={2024 IEEE 10th World Forum on Internet of Things (WF-IoT)}, 
  title={Quantum PUF and its Applications with Information Theoretic Analysis}, 
  year={2024},
  volume={},
  number={},
  pages={1-6},
  doi={10.1109/WF-IoT62078.2024.10811185}}

@article{le_dantec_twenty-threemillisecond_2021,
 author = {{Le Dantec}, Marianne and Ran{\v{c}}i{\'c}, Milo{\v{s}} and Lin, Sen and Billaud, Eric and Ranjan, Vishal and Flanigan, Daniel and Bertaina, Sylvain and Chaneli{\`e}re, Thierry and Goldner, Philippe and Erb, Andreas and Liu, Ren Bao and Est{\`e}ve, Daniel and Vion, Denis and Flurin, Emmanuel and Bertet, Patrice},
 title = {Twenty-three--millisecond electron spin coherence of erbium ions in a natural-abundance crystal},
 url = {https://www.science.org/doi/10.1126/sciadv.abj9786        ,},
 urldate = {2025-06-27},
year = {2021},
 pages = {eabj9786},
 volume = {7},
 number = {51},
 journal = {Science Advances},
 doi = {10.1126/sciadv.abj9786}
}

@article{Lee2019,
 author = {{J P Lee} and {B Villa} and {A J Bennett} and {R M Stevenson} and {D J P Ellis} and {I Farrer} and {D A Ritchie} and {A J Shields}},
 year = {2019},
 title = {A quantum dot as a source of time-bin entangled multi-photon states},
 pages = {025011},
 volume = {4},
 number = {2},
 journal = {Quantum Science and Technology},
 doi = {10.1088/2058-9565/ab0a9b}
}

@article{Leung2019,
 author = {Leung, N. and Lu, Y. and Chakram, S. and Naik, R. K. and Earnest, N. and Ma, R. and Jacobs, K. and Cleland, A. N. and Schuster, D. I.},
 year = {2019},
 title = {Deterministic bidirectional communication and remote entanglement generation between superconducting qubits},
 pages = {18},
 volume = {5},
 number = {1},
 journal = {npj Quantum Information},
 doi = {10.1038/s41534-019-0128-0}
}

@misc{Liu2025,
      title={Robust quantum computational advantage with programmable 3050-photon Gaussian boson sampling}, 
      author={Hua-Liang Liu and Hao Su and Si-Qiu Gong and Yi-Chao Gu and Hao-Yang Tang and Meng-Hao Jia and Qian Wei and Yukun Song and Dongzhou Wang and Mingyang Zheng and Faxi Chen and Libo Li and Siyu Ren and Xuezhi Zhu and Meihong Wang and Yaojian Chen and Yanfei Liu and Longsheng Song and Pengyu Yang and Junshi Chen and Hong An and Lei Zhang and Lin Gan and Guangwen Yang and Jia-Min Xu and Yu-Ming He and Hui Wang and Han-Sen Zhong and Ming-Cheng Chen and Xiao Jiang and Li Li and Nai-Le Liu and Yu-Hao Deng and Xiao-Long Su and Qiang Zhang and Chao-Yang Lu and Jian-Wei Pan},
      year={2025},
      eprint={2508.09092},
      archivePrefix={arXiv},
      primaryClass={quant-ph},
      url={https://arxiv.org/abs/2508.09092}, 
}

@article{ma_one-hour_2021,
 author = {Ma, Yu and Ma, You-Zhi and Zhou, Zong-Quan and Li, Chuan-Feng and Guo, Guang-Can},
 year = {2021},
 title = {One-hour coherent optical storage in an atomic frequency comb memory},
 url = {https://www.nature.com/articles/s41467-021-22706-y},
 urldate = {2025-06-26},
 pages = {2381},
 volume = {12},
 number = {1},
 issn = {2041-1723},
 journal = {Nat. Commun.},
 doi = {10.1038/s41467-021-22706-y}
}

@article{maa24,
 author = {Maa{\ss}, Benjamin and Ewald, Norman Vincenz and Barua, Avijit and Reitzenstein, Stephan and Wolters, Janik},
 year = {2024},
 title = {Room-temperature ladder-type optical memory compatible with single photons from semiconductor quantum dots},
 pages = {044050},
 volume = {22},
 number = {4},
 journal = {Physical Review Applied}
}

@misc{maa25,
      title={On-demand storage and retrieval of single photons from a semiconductor quantum dot in a room-temperature atomic vapor memory}, 
      author={Benjamin Maaß and Avijit Barua and Norman Vincenz Ewald and Elizabeth Robertson and Kartik Gaur and Suk In Park and Sven Rodt and Jin-Dong Song and Stephan Reitzenstein and Janik Wolters},
      year={2025},
      eprint={2501.15663},
      archivePrefix={arXiv},
      primaryClass={quant-ph},
      url={https://arxiv.org/abs/2501.15663}, 
}

@article{Magnard2020,
 author = {Magnard, P. and Storz, S. and Kurpiers, P. and Sch{\"a}r, J. and Marxer, F. and L{\"u}tolf, J. and Walter, T. and Besse, J.-C. and Gabureac, M. and Reuer, K. and Akin, A. and Royer, B. and Blais, A. and Wallraff, A.},
 year = {2020},
 title = {Microwave Quantum Link between Superconducting Circuits Housed in Spatially Separated Cryogenic Systems},
 pages = {260502},
 volume = {125},
 number = {26},
 journal = {Phys. Rev. Lett.},
 doi = {10.1103/PhysRevLett.125.260502}
}

@article{Majety2023,
 author = {Majety, Sridhar and Strohauer, Stefan and Saha, Pranta and Wietschorke, Fabian and Finley, Jonathan J. and M{\"u}ller, Kai and Radulaski, Marina},
 year = {2023},
 title = {Triangular quantum photonic devices with integrated detectors in silicon carbide},
 pages = {015004},
 volume = {3},
 number = {1},
 issn = {2633-4356},
 journal = {Materials for Quantum Technology},
 doi = {10.1088/2633-4356/acc302}
}

@misc{Mamann.31.03.2025,
 author = {Mamann, H. and Nieddu, T. and Hoffet, F. and Bozzio, M. and de Loubresse, F. Garreau and Kerenidis, I. and Diamanti, E. and Urvoy, A. and Laurat, J.},
 title = {Quantum cryptography integrating an optical quantum memory},
 url = {http://arxiv.org/pdf/2504.00094v1},
year = {2025}
}

@article{mann_low-noise_2023,
 author = {Mann, Felix and Chrzanowski, Helen M. and Gewers, Felipe and Placke, Marlon and Ramelow, Sven},
 year = {2023},
 title = {Low-noise quantum frequency conversion in a monolithic cavity with bulk periodically poled potassium titanyl phosphate},
 url = {https://link.aps.org/doi/10.1103/PhysRevApplied.20.054010},
 urldate = {2025-08-12},
 pages = {054010},
 volume = {20},
 number = {5},
 journal = {Physical Review Applied},
 doi = {10.1103/PhysRevApplied.20.054010}
}

@article{mes23,
 author = {Leon Me{\ss}ner and Elizabeth Robertson and Luisa Esguerra and Kathy L\"{u}dge and Janik Wolters},
journal = {Opt. Express},
number = {6},
pages = {10150--10158},
publisher = {Optica Publishing Group},
title = {Multiplexed random-access optical memory in warm cesium vapor},
volume = {31},
month = {Mar},
year = {2023},
url = {https://opg.optica.org/oe/abstract.cfm?URI=oe-31-6-10150},
doi = {10.1364/OE.483642},
}

@article{morello2010,
 author = {Morello, Andrea and Pla, Jarryd J. and Zwanenburg, Floris A. and Chan, Kok W. and Tan, Kuan Y. and Huebl, Hans and Mottonen, Mikko and Nugroho, Christopher D. and Yang, Changyi and {van Donkelaar}, Jessica A. and Alves, Andrew D. C. and Jamieson, David N. and Escott, Christopher C. and Hollenberg, Lloyd C. L. and Clark, Robert G. and Dzurak, Andrew S.},
 year = {2010},
 title = {Single-shot readout of an electron spin in silicon},
 pages = {687},
 volume = {467},
 issn = {1476-4687},
 journal = {Nature},
 doi = {10.1038/nature09392}
}

@article{mot20,
author = {Roberto Mottola and Gianni Buser and Chris M\"{u}ller and Tim Kroh and Andreas Ahlrichs and Sven Ramelow and Oliver Benson and Philipp Treutlein and Janik Wolters},
journal = {Opt. Express},
number = {3},
pages = {3159--3170},
publisher = {Optica Publishing Group},
title = {An efficient, tunable, and robust source of narrow-band photon pairs at the 87Rb D1 line},
volume = {28},
month = {Feb},
year = {2020},
url = {https://opg.optica.org/oe/abstract.cfm?URI=oe-28-3-3159},
doi = {10.1364/OE.384081}
}

@incollection{Murray.2021,
 author = {Murray, Hazel and Malone, David},
 title = {Quantum Multi-factor Authentication},
 pages = {50--67},
 volume = {13136},
 publisher = {{Springer International Publishing}},
 isbn = {978-3-030-93746-1},
 series = {Lecture Notes in Computer Science},
 editor = {Saracino, Andrea and Mori, Paolo},
 booktitle = {Emerging Technologies for Authorization and Authentication},
 year = {2021},
 address = {Cham},
 doi = {10.1007/978-3-030-93747-8$\backslash$textunderscore}
}

@article{nam17,
 author = {Namazi, Mehdi and Kupchak, Connor and Jordaan, Bertus and Shahrokhshahi, Reihaneh and Figueroa, Eden},
 year = {2017},
 title = {Ultralow-Noise Room-Temperature Quantum Memory for Polarization Qubits},
 url = {https://link.aps.org/doi/10.1103/PhysRevApplied.8.034023},
 pages = {034023},
 volume = {8},
 number = {3},
 journal = {Phys. Rev. Appl.},
 doi = {10.1103/PhysRevApplied.8.034023}
}

@article{Nikolopoulos.2017,
 author = {Nikolopoulos, Georgios M. and Diamanti, Eleni},
 year = {2017},
 title = {Continuous-variable quantum authentication of physical unclonable keys},
 pages = {46047},
 volume = {7},
 journal = {Scientific reports},
 doi = {10.1038/srep46047}
}

@article{Oehrl.2025,
 author = {P. Oehrl and F. Fesquet and K. E. Honasoge and M. Handschuh and A. Marx and R. Gross and K. G. Fedorov, H. Huebl},
 year = {2025},
 title = {Probing an electron spin ensemble with squeezed microwave signals},
 journal = {arXiv: 2512.17490},
archivePrefix={arXiv},
 doi = {10.48550/arXiv.2512.17490},
 url = {https://doi.org/10.48550/arXiv.2512.17490}
}

@article{Omlor2025,
 author = {Omlor, Ferdinand and Tissot, Benedikt and Burkard, Guido},
 year = {2025},
 title = {Entanglement generation using single-photon pulse reflection in realistic networks},
 volume = {111},
 number = {1},
 pages = {012612},
 journal = {Phys. Rev. A},
 doi = {10.1103/PhysRevA.111.012612}
}

@article{OrphalKobin2023,
  title = {Optically Coherent Nitrogen-Vacancy Defect Centers in Diamond Nanostructures},
  author = {Orphal-Kobin, Laura and Unterguggenberger, Kilian and Pregnolato, Tommaso and Kemf, Natalia and Matalla, Mathias and Unger, Ralph-Stephan and Ostermay, Ina and Pieplow, Gregor and Schr\"oder, Tim},
  journal = {Phys. Rev. X},
  volume = {13},
  issue = {1},
  pages = {011042},
  numpages = {22},
  year = {2023},
  month = {Mar},
  publisher = {American Physical Society},
  doi = {10.1103/PhysRevX.13.011042},
  url = {https://link.aps.org/doi/10.1103/PhysRevX.13.011042}
}

@article{OSullivan:2022,
 author = {O'Sullivan, James and Kennedy, Oscar W. and Debnath, Kamanasish and Alexander, Joseph and Zollitsch, Christoph W. and {\v{S}}im{\.{e}}nas, Mantas and Hashim, Akel and Thomas, Christopher N. and Withington, Stafford and Siddiqi, Irfan and M{\o}lmer, Klaus and Morton, John J. L.},
 year = {2022},
 title = {Random-Access Quantum Memory Using Chirped Pulse Phase Encoding},
 pages = {041014},
 volume = {12},
 number = {4},
 journal = {Physical Review X},
 doi = {10.1103/physrevx.12.041014}
}

@article{ourari_indistinguishable_2023,
 author = {Ourari, Salim and Dusanowski, ukasz and Horvath, Sebastian P. and Uysal, Mehmet T. and Phenicie, Christopher M. and Stevenson, Paul and Raha, Mouktik and Chen, Songtao and Cava, Robert J. and de Leon, Nathalie P. and Thompson, Jeff D.},
 title = {Indistinguishable telecom band photons from a single Er ion in the solid state},
 url = {https://www.nature.com/articles/s41586-023-06281-4},
 urldate = {2023-11-03},
 pages = {977--981},
year = {2023},
 volume = {620},
 number = {7976},
 issn = {1476-4687},
 journal = {Nature},
 doi = {10.1038/s41586-023-06281-4}
}

@article{Owari2008,
 author = {{M Owari} and {M B Plenio} and {E S Polzik} and {A Serafini} and {M M Wolf}},
 year = {2008},
 title = {Squeezing the limit: quantum benchmarks for the teleportation and storage of squeezed states},
 pages = {113014},
 volume = {10},
 number = {11},
 journal = {New Journal of Physics},
 doi = {10.1088/1367-2630/10/11/113014}
}

@article{Pappu.2002,
 author = {Pappu, Ravikanth and Recht, Ben and Taylor, Jason and Gershenfeld, Neil},
 year = {2002},
 title = {Physical one-way functions},
 pages = {2026--2030},
 volume = {297},
 number = {5589},
 journal = {Science (New York, N.Y.)},
 doi = {10.1126/science.1074376}
}

@article{Pastawski.2012,
 author = {Pastawski, Fernando and Yao, Norman Y. and Jiang, Liang and Lukin, Mikhail D. and Cirac, J. Ignacio},
 year = {2012},
 title = {Unforgeable noise-tolerant quantum tokens},
 pages = {16079--16082},
 volume = {109},
 number = {40},
 issn = {0027-8424},
 journal = {Proceedings of the National Academy of Sciences},
 doi = {10.1073/pnas.1203552109}
}

@article{Phillips2001,
 author = {Phillips, D. F. and Fleischhauer, A. and Mair, A. and Walsworth, R. L. and Lukin, M. D.},
 year = {2001},
 title = {Storage of light in atomic vapor},
 pages = {783--786},
 volume = {86},
 number = {5},
 journal = {Physical Review Letters},
 doi = {10.1103/PhysRevLett.86.783}
}

@misc{pieplow_deterministic_2023,
      title={Deterministic Creation of Large Photonic Multipartite Entangled States with Group-IV Color Centers in Diamond}, 
      author={Gregor Pieplow and Yannick Strocka and Mariano Isaza-Monsalve and Joseph H. D. Munns and Tim Schröder},
      year={2023},
      eprint={2312.03952},
      archivePrefix={arXiv},
      primaryClass={quant-ph},
      url={https://arxiv.org/abs/2312.03952}, 
}

@article{pieplow_quantum_2025,
 author = {Pieplow, Gregor and Torun, Cem G{\"u}ney and Gurr, Charlotta and Munns, Joseph H. D. and Herrmann, Franziska Marie and Thies, Andreas and Pregnolato, Tommaso and Schr{\"o}der, Tim},
 year = {2025},
 title = {Quantum electrometer for time-resolved material science at the atomic lattice scale},
 url = {https://www.nature.com/articles/s41467-025-61839-2},
 urldate = {2025-08-12},
 pages = {6435},
 volume = {16},
 number = {1},
 issn = {2041-1723},
 journal = {Nat. Commun.},
 doi = {10.1038/s41467-025-61839-2}
}

@article{Pieplow2024,
  title = {Efficient microwave spin control of negatively charged group-IV color centers in diamond},
  author = {Pieplow, Gregor and Belhassen, Mohamed and Schr\"oder, Tim},
  journal = {Phys. Rev. B},
  volume = {109},
  issue = {11},
  pages = {115409},
  numpages = {12},
  year = {2024},
  month = {Mar},
  publisher = {American Physical Society},
  doi = {10.1103/PhysRevB.109.115409},
  url = {https://link.aps.org/doi/10.1103/PhysRevB.109.115409}
}

@article{Pirandola2015,
 author = {Pirandola, S. and Eisert, J. and Weedbrook, C. and Furusawa, A. and Braunstein, S. L.},
 year = {2015},
 title = {Advances in quantum teleportation},
 pages = {641--652},
 volume = {9},
 number = {10},
 journal = {Nature Photonics},
 doi = {10.1038/nphoton.2015.154}
}

@article{Pogorzalek2019,
 author = {Pogorzalek, S. and Fedorov, K. G. and Xu, M. and Parra-Rodriguez, A. and Sanz, M. and Fischer, M. and Xie, E. and Inomata, K. and Nakamura, Y. and Solano, E. and Marx, A. and Deppe, F. and Gross, R.},
 year = {2019},
 title = {Secure quantum remote state preparation of squeezed microwave states},
 pages = {2604},
 volume = {10},
 journal = {Nat. Comm.},
 doi = {10.1038/s41467-019-10727-7}
}

@article{pregnolato_fabrication_2024,
 author = {Pregnolato, Tommaso and Stucki, Marco E. and Bopp, Julian M. and {v. d. Hoeven}, Maarten H. and Gokhale, Alok and Kr{\"u}ger, Olaf and Schr{\"o}der, Tim},
 year = {2024},
 title = {Fabrication of Sawfish photonic crystal cavities in bulk diamond},
 url = {https://pubs.aip.org/app/article/9/3/036105/3269905/Fabrication-of-Sawfish-photonic-crystal-cavities},
 urldate = {2025-02-28},
 pages = {036105},
 volume = {9},
 number = {3},
 issn = {2378-0967},
 journal = {APL Photonics},
 doi = {10.1063/5.0186509}
}

@article{Rancic2018,
 author = {Ran{\v{c}}i{\'c}, Milo{\v{s}} and Hedges, Morgan P. and Ahlefeldt, Rose L. and Sellars, Matthew J.},
 year = {2018},
 title = {Coherence time of over a second in a telecom-compatible quantum memory storage material},
 pages = {50--54},
 volume = {14},
 number = {1},
 issn = {1745-2481},
 journal = {Nature Physics},
 doi = {10.1038/nphys4254}
}

@article{rob24,
  title = {Machine-learning optimal control pulses in an optical quantum memory experiment},
  author = {Robertson, Elizabeth and Esguerra, Luisa and Me\ss{}ner, Leon and Gallego, Guillermo and Wolters, Janik},
  journal = {Phys. Rev. Appl.},
  volume = {22},
  issue = {2},
  pages = {024026},
  numpages = {10},
  year = {2024},
  month = {Aug},
  publisher = {American Physical Society},
  doi = {10.1103/PhysRevApplied.22.024026},
  url = {https://link.aps.org/doi/10.1103/PhysRevApplied.22.024026}
}

@misc{rob25,
      title={A digital twin of atomic ensemble quantum memories}, 
      author={Elizabeth Robertson and Benjamin Maaß and Konrad Tschernig and Janik Wolters},
      year={2025},
      eprint={2506.20403},
      archivePrefix={arXiv},
      primaryClass={quant-ph},
      url={https://arxiv.org/abs/2506.20403}, 
}

@article{Saeedi.2013,
author = {Kamyar Saeedi  and Stephanie Simmons  and Jeff Z. Salvail  and Phillip Dluhy  and Helge Riemann  and Nikolai V. Abrosimov  and Peter Becker  and Hans-Joachim Pohl  and John J. L. Morton  and Mike L. W. Thewalt },
title = {Room-Temperature Quantum Bit Storage Exceeding 39 Minutes Using Ionized Donors in Silicon-28},
journal = {Science},
volume = {342},
number = {6160},
pages = {830-833},
year = {2013},
doi = {10.1126/science.1239584},
URL = {https://www.science.org/doi/abs/10.1126/science.1239584},
eprint = {https://www.science.org/doi/pdf/10.1126/science.1239584}
}

@article{san11,
 author = {Sangouard, Nicolas and Simon, Christoph and de Riedmatten, Hugues and Gisin, Nicolas},
 year = {2011},
 title = {Quantum repeaters based on atomic ensembles and linear optics},
 url = {https://link.aps.org/doi/10.1103/RevModPhys.83.33},
 pages = {33--80},
 volume = {83},
 number = {1},
 journal = {Rev. Mod. Phys.},
 doi = {10.1103/RevModPhys.83.33}
}

@article{Scarani2009,
 author = {Scarani, Valerio and Bechmann-Pasquinucci, Helle and Cerf, Nicolas J. and Dusek, Miloslav and L{\"u}tkenhaus, Norbert and Peev, Momtchil},
 year = {2009},
 title = {The security of practical quantum key distribution},
 pages = {1301--1350},
 volume = {81},
 number = {3},
 journal = {Rev. Mod. Phys.},
 doi = {10.1103/RevModPhys.81.1301}
}

@article{schafer_two-stage_2025,
 author = {Sch{\"a}fer, Marlon and Kambs, Benjamin and Herrmann, Dennis and Bauer, Tobias and Becher, Christoph},
 year = {2025},
 title = {Two-Stage, Low Noise Quantum Frequency Conversion of Single Photons from Silicon-Vacancy Centers in Diamond to the Telecom C-Band},
 urldate = {2025-08-12},
 pages = {2300228},
 volume = {8},
 number = {2},
 issn = {2511-9044},
 journal = {Advanced Quantum Technologies},
 doi = {10.1002/qute.202300228}
}

@article{Schiansky.2023,
 author = {Schiansky, Peter and Kalb, Julia and Sztatecsny, Esther and Roehsner, Marie-Christine and Guggemos, Tobias and Trenti, Alessandro and Bozzio, Mathieu and Walther, Philip},
 year = {2023},
 title = {Demonstration of quantum-digital payments},
 pages = {3849},
 volume = {14},
 number = {1},
 journal = {Nature Communications},
 doi = {10.1038/s41467-023-39519-w}
}

@article{senkalla_germanium_2024,
 author = {Senkalla, Katharina and Genov, Genko and Metsch, Mathias H. and Siyushev, Petr and Jelezko, Fedor},
 year = {2024},
 title = {Germanium Vacancy in Diamond Quantum Memory Exceeding 20 ms},
 url = {https://link.aps.org/doi/10.1103/PhysRevLett.132.026901},
 urldate = {2025-08-01},
 pages = {026901},
 volume = {132},
 number = {2},
 journal = {Physical Review Letters},
 doi = {10.1103/PhysRevLett.132.026901}
}

@article{Shaham2022_NP,
 author = {{R. Shaham} and {O. Katz} and {O. Firstenberg}},
 year = {2022},
 title = {Strong coupling of alkali-metal spins to noble-gas spins with an hour-long coherence time},
 pages = {506--510},
 volume = {18},
 issn = {1745-2481},
 journal = {Nature Physics},
 doi = {10.1038/s41567-022-01535-w}
}

@article{shamsoshoara2020survey,
title = {A survey on physical unclonable function (PUF)-based security solutions for Internet of Things},
journal = {Computer Networks},
volume = {183},
pages = {107593},
year = {2020},
issn = {1389-1286},
doi = {https://doi.org/10.1016/j.comnet.2020.107593},
url = {https://www.sciencedirect.com/science/article/pii/S1389128620312275},
author = {Alireza Shamsoshoara and Ashwija Korenda and Fatemeh Afghah and Sherali Zeadally}
}

@incollection{shi23,
 author = {{Kai Shinbrough} and {Donny R. Pearson} and {Bin Fang} and {Elizabeth A. Goldschmidt} and {Virginia O. Lorenz}},
 title = {Chapter Five - Broadband quantum memory in atomic ensembles},
 url = {https://www.sciencedirect.com/science/article/pii/S1049250X23000010},
 pages = {297--360},
 volume = {72},
 publisher = {{Academic Press}},
 series = {Advances In Atomic, Molecular, and Optical Physics},
 editor = {Louis F. DiMauro and Hélène Perrin and Susanne F. Yelin},
 booktitle = {Advances in Atomic, Molecular, and Optical Physics},
 year = {2023},
 doi = {10.1016/bs.aamop.2023.04.001}
}

@misc{Skoric.2010,
 author = {{\v{S}}kori{\'c}, Boris},
 year = {2010},
 title = {Quantum Readout of Physical Unclonable Functions},
 publisher = {{Springer Berlin Heidelberg}},
 isbn = {0302-9743},
 series = {Lecture Notes in Computer Science},
 doi = {10.1007/978-3-642-12678-9$\backslash$textunderscore}
}

@article{Skoric.2012,
 author = {{\v{S}}kori{\'c}, Boris},
 year = {2012},
 title = {Quantum readout of Physical Unclonable Functions},
 pages = {1250001},
 volume = {10},
 number = {01},
 issn = {1793-6918},
 journal = {International Journal of Quantum Information},
 doi = {10.1142/S0219749912500013}
}

@article{Steger:2012ev,
 author = {Steger, M. and Saeedi, K. and Thewalt, M. L. W. and Morton, J. J. L. and Riemann, H. and Abrosimov, N. V. and Becker, P. and Pohl, H. J.},
 year = {2012},
 title = {Quantum Information Storage for over 180 s Using Donor Spins in a $^{28}\mathrm{Si}$},
 pages = {1280},
 volume = {336},
 number = {6086},
 journal = {Science},
 doi = {10.1126/science.1217635}
}

@article{stolk_metropolitan-scale_2024,
 author = {Stolk, Arian J. and {van der Enden}, Kian L. and Slater, Marie-Christine and {te Raa-Derckx}, Ingmar and Botma, Pieter and {van Rantwijk}, Joris and Biemond, J. J. Benjamin and Hagen, Ronald A. J. and Herfst, Rodolf W. and Koek, Wouter D. and Meskers, Adrianus J. H. and Vollmer, Ren{\'e} and {van Zwet}, Erwin J. and Markham, Matthew and Edmonds, Andrew M. and Geus, J. Fabian and Elsen, Florian and Jungbluth, Bernd and Haefner, Constantin and Tresp, Christoph and Stuhler, J{\"u}rgen and Ritter, Stephan and Hanson, Ronald},
 year = {2024},
 title = {Metropolitan-scale heralded entanglement of solid-state qubits},
 url = {https://www.science.org/doi/10.1126/sciadv.adp6442},
 urldate = {2025-08-12},
 pages = {eadp6442},
 volume = {10},
 number = {44},
 journal = {Science Advances},
 doi = {10.1126/sciadv.adp6442}
}

@article{Strinic2025,
   title = {Broadband electron paramagnetic resonance spectroscopy of $^{167}\mathrm{Er}$:$^{7}\mathrm{LiYF}_{4}$ at millikelvin temperatures},
  author = {Strini\ifmmode \acute{c}\else \'{c}\fi{}, Ana and Oehrl, Patricia and Marx, Achim and Bushev, Pavel A. and Huebl, Hans and Gross, Rudolf and Kukharchyk, Nadezhda},
  journal = {Phys. Rev. B},
  volume = {111},
  issue = {21},
  pages = {214430},
  numpages = {12},
  year = {2025},
  month = {Jun},
  publisher = {American Physical Society},
  doi = {10.1103/5j7f-wfyl},
  url = {https://link.aps.org/doi/10.1103/5j7f-wfyl}
}

@misc{strocka_secure_2025,
      title={Secure Quantum Token Processing with Color Centers in Diamond}, 
      author={Yannick Strocka and Mohamed Belhassen and Tim Schröder and Gregor Pieplow},
      year={2025},
      eprint={2503.04985},
      archivePrefix={arXiv},
      primaryClass={quant-ph},
      url={https://arxiv.org/abs/2503.04985}, 
}

@article{strohauer_current_2025,
 author = {Strohauer, Stefan and Wietschorke, Fabian and Schmid, Christian and Grotowski, Stefanie and Zugliani, Lucio and Jonas, Bj{\"o}rn and M{\"u}ller, Kai and Finley, Jonathan J.},
 year = {2025},
 title = {Current crowding--free superconducting nanowire single-photon detectors},
 url = {https://www.science.org/doi/full/10.1126/sciadv.adt0502},
 urldate = {2025-07-31},
 pages = {eadt0502},
 volume = {11},
 number = {13},
 journal = {Science Advances},
 doi = {10.1126/sciadv.adt0502}
}

@article{strohauer_site-selective_2023,
 author = {Strohauer, Stefan and Wietschorke, Fabian and Zugliani, Lucio and Flaschmann, Rasmus and Schmid, Christian and Grotowski, Stefanie and M{\"u}ller, Manuel and Jonas, Bj{\"o}rn and Althammer, Matthias and Gross, Rudolf and M{\"u}ller, Kai and Finley, Jonathan J.},
 year = {2023},
 title = {Site-Selective Enhancement of Superconducting Nanowire Single-Photon Detectors via Local Helium Ion Irradiation},
 urldate = {2025-08-12},
 pages = {2300139},
 volume = {6},
 number = {12},
 issn = {2511-9044},
 journal = {Advanced Quantum Technologies},
 doi = {10.1002/qute.202300139}
}

@article{teller_solid-state_2025,
 author = {Teller, Markus and Plascencia, Susana and {Sastre Jachimska}, Cristina and Grandi, Samuele and de Riedmatten, Hugues},
 year = {2025},
 title = {A solid-state temporally multiplexed quantum memory array at the single-photon level},
 volume = {11},
 number = {1},
 pages = {92},
 journal = {npj Quantum Information},
 doi = {10.1038/s41534-025-01042-9}
}

@article{Thiering2018,
 author = {Thiering, Gergő and Gali, Adam},
 year = {2018},
 title = {Ab Initio  Magneto-Optical Spectrum of Group-IV Vacancy Color Centers in Diamond},
 volume = {8},
 pages = {021063},
 number = {2},
 journal = {Physical Review X},
 doi = {10.1103/physrevx.8.021063}
}

@article{tho24,
 author = {{Sarah E. Thomas} and {Lukas Wagner} and {Raphael Joos} and {Robert Sittig} and {Cornelius Nawrath} and {Paul Burdekin} and {Ilse Maillette de Buy Wenniger} and {Mikhael J. Rasiah} and {Tobias Huber-Loyola} and {Steven Sagona-Stophel} and {Sven Hafling} and {Michael Jetter} and {Peter Michler} and {Ian A. Walmsley} and {Simone L. Portalupi} and {Patrick M. Ledingham}},
 year = {2024},
 title = {Deterministic storage and retrieval of telecom light from a quantum dot single-photon source interfaced with an atomic quantum memory},
 pages = {eadi7346},
 volume = {10},
 number = {15},
 journal = {Science Advances},
 doi = {10.1126/sciadv.adi7346}
}

@misc{tiranov_sub-second_2025,
      title={Sub-second spin and lifetime-limited optical coherences in $^{171}$Yb$^{3+}$:CaWO$_4$}, 
      author={Alexey Tiranov and Emanuel Green and Sophie Hermans and Erin Liu and Federico Chiossi and Diana Serrano and Pascal Loiseau and Achuthan Manoj Kumar and Sylvain Bertaina and Andrei Faraon and Philippe Goldner},
      year={2025},
      eprint={2504.01592},
      archivePrefix={arXiv},
      primaryClass={quant-ph},
      url={https://arxiv.org/abs/2504.01592}
}

@misc{torun_optical_2023,
 author = {Torun, Cem G{\"u}ney and Munns, Joseph H. D. and Herrmann, Franziska Marie and Villafane, Viviana and M{\"u}ller, Kai and Thies, Andreas and Pregnolato, Tommaso and Pieplow, Gregor and Schr{\"o}der, Tim},
 title = {Optical probing of phononic properties of a tin-vacancy color center in diamond},
 year={2023},
      eprint={2312.05335},
      archivePrefix={arXiv},
 doi = {10.48550/arXiv.2312.05335}
}

@misc{torun2023,
 author = {Torun, Cem G{\"u}ney and G{\"o}k{\c{c}}e, Mustafa and Bracht, Thomas K. and Monsalve, Mariano Isaza and Benbouabdellah, Sarah and Nacitarhan, {\"O}zg{\"u}n Ozan and Stucki, Marco E. and Markham, Matthew L. and Pieplow, Gregor and Pregnolato, Tommaso and {Munns,  Joseph H. D.} and Reiter, Doris E. and Schr{\"o}der, Tim},
 year = {2023},
 title = {SUPER and subpicosecond coherent control of an optical qubit in a tin-vacancy color center in diamond},
 publisher = {arXiv},
 doi = {10.48550/ARXIV.2312.05246}
}

@article{Trusheim2020,
 author = {Trusheim, Matthew E. and Pingault, Benjamin and Wan, Noel H. and G{\"u}ndo{\u{g}}an, Mustafa and de Santis, Lorenzo and Debroux, Romain and Gangloff, Dorian and Purser, Carola and Chen, Kevin C. and Walsh, Michael and Rose, Joshua J. and Becker, Jonas N. and Lienhard, Benjamin and Bersin, Eric and Paradeisanos, Ioannis and Wang, Gang and Lyzwa, Dominika and Montblanch, Alejandro R-P. and Malladi, Girish and Bakhru, Hassaram and Ferrari, Andrea C. and Walmsley, Ian A. and Atat{\"u}re, Mete and Englund, Dirk},
 year = {2020},
 title = {Transform-Limited Photons From a Coherent Tin-Vacancy Spin in Diamond},
 volume = {124},
 pages = {023602},
 number = {2},
 journal = {Physical Review Letters},
 doi = {10.1103/physrevlett.124.023602}
}

@article{Tyryshkin:2011fi,
 author = {Tyryshkin, Alexei M. and Tojo, Shinichi and Morton, John J. L. and Riemann, Helge and Abrosimov, Nikolai V. and Becker, Peter and Pohl, Hans-Joachim and Schenkel, Thomas and Thewalt, Michael L. W. and Itoh, Kohei M. and Lyon, S. A.},
 year = {2011},
 title = {Electron spin coherence exceeding seconds in high-purity silicon},
 url = {http://www.nature.com/doifinder/10.1038/nmat3182},
 pages = {143},
 volume = {11},
 number = {2},
 journal = {Nature Materials},
 doi = {10.1038/nmat3182}
}

@article{ulanowski_spectral_2022,
 author = {Ulanowski, Alexander and Merkel, Benjamin and Reiserer, Andreas},
 title = {Spectral multiplexing of telecom emitters with stable transition frequency},
 url = {https://www.science.org/doi/full/10.1126/sciadv.abo4538   ,},
 urldate = {2025-06-27},
 pages = {eabo4538},
year = {2022},
 volume = {8},
 number = {43},
 journal = {Science Advances},
 doi = {10.1126/sciadv.abo4538}
}

@article{Wallraff.2004,
 author = {Wallraff, A. and Schuster, D. I. and Blais, A. and Frunzio, L. and Huang, R-S and Majer, J. and Kumar, S. and Girvin, S. M. and Schoelkopf, R. J.},
 year = {2004},
 title = {Strong coupling of a single photon to a superconducting qubit using circuit quantum electrodynamics},
 pages = {162--167},
 volume = {431},
 number = {7005},
 issn = {1476-4687},
 journal = {Nature},
 doi = {10.1038/nature02851}
}

@article{wan22,
 author = {Wang, Yang and Craddock, Alexander N. and Sekelsky, Rourke and Flament, Mael and Namazi, Mehdi},
 year = {2022},
 title = {Field-Deployable Quantum Memory for Quantum Networking},
 url = {https://link.aps.org/doi/10.1103/PhysRevApplied.18.044058},
 pages = {044058},
 volume = {18},
 number = {4},
 journal = {Phys. Rev. Appl.},
 doi = {10.1103/PhysRevApplied.18.044058}
}

@misc{wan25,
      title={High-fidelity entanglement between a telecom photon and a room-temperature quantum memory}, 
      author={Yang Wang and Alexander N. Craddock and Jaeda M. Mendoza and Rourke Sekelsky and Mael Flament and Mehdi Namazi},
      year={2025},
      eprint={2503.11564},
      archivePrefix={arXiv},
      primaryClass={quant-ph},
      url={https://arxiv.org/abs/2503.11564}, 
}

@article{wang_nuclear_2025,
 author = {Wang, Fudong and Ren, Miaomiao and Sun, Weiye and Guo, Mucheng and Sellars, Matthew J. and Ahlefeldt, Rose L. and Bartholomew, John G. and Yao, Juan and Liu, Shuping and Zhong, Manjin},
 year = {2025},
 title = {Nuclear Spins in a Solid Exceeding 10-Hour Coherence Times for Ultra-Long-Term Quantum Storage},
 url = {https://link.aps.org/doi/10.1103/PRXQuantum.6.010302   ,},
 urldate = {2025-02-13},
 pages = {010302},
 volume = {6},
 number = {1},
 journal = {PRX Quantum},
 doi = {10.1103/PRXQuantum.6.010302}
}

@article{Wei2025,
 author = {Wei, Y.-C. and Stas, P.-J. and Suleymanzade, A. and Baranes, G. and Machado, F. and Huan, Y. Q. and Knaut, C. M. and Ding, S. W. and Merz, M. and Knall, E. N. and Yazlar, U. and Sirotin, M. and Wang, I. W. and Machielse, B. and Yelin, S. F. and Borregaard, J. and Park, H. and Lon{\v{c}}ar, M. and Lukin, M. D.},
 year = {2025},
 title = {Universal distributed blind quantum computing with solid-state qubits},
 pages = {509--513},
 volume = {388},
 number = {6746},
 journal = {Science},
 doi = {10.1126/science.adu6894}
}

@article{Weichselbaumer:2019hi,
 author = {Weichselbaumer, Stefan and Natzkin, Petio and Zollitsch, Christoph W. and Weiler, Mathias and Gross, Rudolf and Huebl, Hans},
 year = {2019},
 title = {Quantitative Modeling of Superconducting Planar Resonators for Electron Spin Resonance},
 pages = {024021},
 volume = {12},
 number = {2},
 journal = {Physical Review Applied},
 doi = {10.1103/PhysRevApplied.12.024021}
}

@article{Weichselbaumer:2020hn,
 author = {Weichselbaumer, Stefan and Zens, Matthias and Zollitsch, Christoph W. and Brandt, Martin S. and Rotter, Stefan and Gross, Rudolf and Huebl, Hans},
 year = {2020},
 title = {Echo Trains in Pulsed Electron Spin Resonance of a Strongly Coupled Spin Ensemble},
 pages = {137701},
 volume = {125},
 number = {13},
 journal = {Physical Review Letters},
 doi = {10.1103/PhysRevLett.125.137701}
}

@article{Wesenberg:2009es,
 author = {Wesenberg, J. H. and Ardavan, A. and Briggs, G. A. D. and Morton, J. J. L. and Schoelkopf, R. J. and Schuster, D. I. and Molmer, K.},
 year = {2009},
 title = {Quantum Computing with an Electron Spin Ensemble},
 url = {http://link.aps.org/doi/10.1103/PhysRevLett.103.070502},
 pages = {070502},
 volume = {103},
 number = {7},
 journal = {Physical Review Letters},
 doi = {10.1103/physrevlett.103.070502}
}

@article{Wiesner.1983,
 author = {Wiesner, Stephen},
 year = {1983},
 title = {Conjugate coding},
 pages = {78--88},
 volume = {15},
 number = {1},
 issn = {0163-5700},
 journal = {ACM SIGACT News},
 doi = {10.1145/1008908.1008920}}

@article{wol17,
   title = {Simple Atomic Quantum Memory Suitable for Semiconductor Quantum Dot Single Photons},
  author = {Wolters, Janik and Buser, Gianni and Horsley, Andrew and B\'eguin, Lucas and J\"ockel, Andreas and Jahn, Jan-Philipp and Warburton, Richard J. and Treutlein, Philipp},
  journal = {Phys. Rev. Lett.},
  volume = {119},
  issue = {6},
  pages = {060502},
  numpages = {5},
  year = {2017},
  month = {Aug},
  publisher = {American Physical Society},
  doi = {10.1103/PhysRevLett.119.060502},
  url = {https://link.aps.org/doi/10.1103/PhysRevLett.119.060502}
}

@article{Wootters.1982,
 author = {Wootters, W. K. and Zurek, W. H.},
 year = {1982},
 title = {A single quantum cannot be cloned},
 pages = {802--803},
 volume = {299},
 number = {5886},
 issn = {1476-4687},
 journal = {Nature},
 doi = {10.1038/299802a0}
}

@article{xia_tunable_2022,
 author = {Xia, Kangwei and Sardi, Fiammetta and Sauerzapf, Colin and Kornher, Thomas and Becker, Hans-Werner and Kis, Zsolt and Kovacs, Laszlo and Dertli, Denis and Foglszinger, Jonas and Kolesov, Roman and Wrachtrup, J{\"o}rg},
 title = {Tunable microcavities coupled to rare-earth quantum emitters},
 urldate = {2025-06-27},
 pages = {445--450},
year = {2022},
 volume = {9},
 number = {4},
 issn = {2334-2536},
 journal = {Optica},
 doi = {10.1364/OPTICA.453527}
}

@article{ruskuc_multiplexed_2025,
	title = {Multiplexed entanglement of multi-emitter quantum network nodes},
	volume = {639},
	copyright = {2025 The Author(s), under exclusive licence to Springer Nature Limited},
	issn = {1476-4687},
	url = {https://www.nature.com/articles/s41586-024-08537-z},
	doi = {10.1038/s41586-024-08537-z},
	number = {8053},
	urldate = {2025-08-08},
	journal = {Nature},
	author = {Ruskuc, A. and Wu, C.-J. and Green, E. and Hermans, S. L. N. and Pajak, W. and Choi, J. and Faraon, A.},
	month = mar,
	year = {2025},
	pages = {54--59},
}

@article{yang_toward_2023,
 author = {Yang, Likai and Wang, Sihao and Tang, Hong X.},
 title = {Toward radiative-limited coherence of erbium dopants in a nanophotonic resonator},
 urldate = {2023-10-23},
 pages = {124001},
 volume = {123},
 number = {12},
year = {2023},
 issn = {0003-6951},
 journal = {Applied Physics Letters},
 doi = {10.1063/5.0165971}
}

@article{Yu2025,
 author = {Yu, Hao and Sciara, Stefania and Chemnitz, Mario and Montaut, Nicola and Crockett, Benjamin and Fischer, Bennet and Helsten, Robin and Wetzel, Benjamin and Goebel, Thorsten A. and Kr{\"a}mer, Ria G. and Little, Brent E. and Chu, Sai T. and Nolte, Stefan and Wang, Zhiming and Aza{\~n}a, Jos{\'e} and Munro, William J. and Moss, David J. and Morandotti, Roberto},
 year = {2025},
 title = {Quantum key distribution implemented with d-level time-bin entangled photons},
 volume = {16},
 number = {1},
 issn = {2041-1723},
 pages = {171},
 journal = {Nat. Commun.},
 doi = {10.1038/s41467-024-55345-0}
}

@article{zaske_visible-telecom_2012,
 author = {Zaske, Sebastian and Lenhard, Andreas and Ke{\ss}ler, Christian A. and Kettler, Jan and Hepp, Christian and Arend, Carsten and Albrecht, Roland and Schulz, Wolfgang-Michael and Jetter, Michael and Michler, Peter and Becher, Christoph},
 year = {2012},
 title = {Visible-to-Telecom Quantum Frequency Conversion of Light from a Single Quantum Emitter},
 url = {https://link.aps.org/doi/10.1103/PhysRevLett.109.147404},
 urldate = {2025-08-12},
 pages = {147404},
 volume = {109},
 number = {14},
 journal = {Physical Review Letters},
 doi = {10.1103/PhysRevLett.109.147404}
}

@article{zhong_optically_2018,
 author = {Zhong, Tian and Kindem, Jonathan M. and Bartholomew, John G. and Rochman, Jake and Craiciu, Ioana and Verma, Varun and Nam, Sae Woo and Marsili, Francesco and Shaw, Matthew D. and Beyer, Andrew D. and Faraon, Andrei},
 title = {Optically Addressing Single Rare-Earth Ions in a Nanophotonic Cavity},
 url = {https://link.aps.org/doi/10.1103/PhysRevLett.121.183603   ,},
 urldate = {2024-01-23},
 pages = {183603},
 volume = {121},
year = {2018},
 number = {18},
 journal = {Physical Review Letters},
 doi = {10.1103/PhysRevLett.121.183603}
}

@inproceedings{sardi_photonic_2024,
	title = {Photonic integration of 171ytterbium single photon sources into an {LiNbO3}-based photonic platform},
	volume = {12911},
	url = {https://www.spiedigitallibrary.org/conference-proceedings-of-spie/12911/129110X/Photonic-integration-of-171ytterbium-single-photon-sources-into-an-LiNbO3/10.1117/12.3001722.full},
	doi = {10.1117/12.3001722},
	urldate = {2025-12-15},
	booktitle = {Quantum {Computing}, {Communication}, and {Simulation} {IV}},
	publisher = {SPIE},
	author = {Sardi, Fiammetta and Foteinou, Varvara and Stöhr, Rainer and Kolesov, Roman and Wrachtrup, Jörg},
	month = mar,
	year = {2024},
	pages = {301--305}
	}

@article{Zollitsch:2015ut,
    author = {Zollitsch, Christoph W. and Mueller, Kai and Franke, David P. and Goennenwein, Sebastian T. B. and Brandt, Martin S. and Gross, Rudolf and Huebl, Hans},
    title = {High cooperativity coupling between a phosphorus donor spin ensemble and a superconducting microwave resonator},
    journal = {Applied Physics Letters},
    volume = {107},
    number = {14},
    pages = {142105},
    year = {2015},
    month = {10},
    issn = {0003-6951},
    doi = {10.1063/1.4932658},
    url = {https://doi.org/10.1063/1.4932658}
}

@article{afzelius_multimode_2009,
	title = {Multimode quantum memory based on atomic frequency combs},
	volume = {79},
	url = {https://link.aps.org/doi/10.1103/PhysRevA.79.052329},
	doi = {10.1103/PhysRevA.79.052329},
	number = {5},
	urldate = {2025-06-27},
	journal = {Physical Review A},
	author = {Afzelius, Mikael and Simon, Christoph and de Riedmatten, Hugues and Gisin, Nicolas},
	month = may,
	year = {2009},
	note = {Publisher: American Physical Society},
	pages = {052329},
}

@inproceedings{zugliani_tailoring_2023,
 author = {Zugliani, Lucio and Schmid, Christian and Wietschorke, Fabian and Jonas, Bj{\"o}rn and Spedicato, Simone and Strohauer, Stefan and Grotowski, Stefanie and Flaschmann, Rasmus and M{\"u}ller, Manuel and Althammer, Matthias and Gross, Rudolf and Finley, Jonathan and M{\"u}ller, Kai},
 title = {Tailoring Superconducting Nanowire Single-Photon Detectors for Quantum Technologies},
 url = {https://ieeexplore.ieee.org/document/10465075},
 urldate = {2025-08-12},
 pages = {1051--1056},
 booktitle = {2023 IEEE Globecom Workshops (GC Wkshps)},
 year = {2023},
 doi = {10.1109/GCWkshps58843.2023.10465075}
}

@article{Businger.2020,
 author = {Businger, M. and Tiranov, A. and Kaczmarek, K. T. and Welinski, S. and Zhang, Z. and Ferrier, A. and Goldner, P. and Afzelius, M.},
 year = {2020},
 title = {Optical Spin-Wave Storage in a Solid-State Hybridized Electron-Nuclear Spin Ensemble},
 pages = {053606},
 volume = {124},
 number = {5},
 journal = {Physical Review Letters},
 doi = {10.1103/PhysRevLett.124.053606}
}

@misc{Kunz.2019,
      title={Spatial Multiplexing in a Cavity-Enhanced Quantum Memory}, 
      author={Paul D. Kunz and Siddhartha Santra and David H. Meyer and Zachary A. Castillo and Vladimir S. Malinovsky and Kevin C. Cox},
      year={2019},
      eprint={1903.02625},
      archivePrefix={arXiv},
      primaryClass={physics.atom-ph},
      url={https://arxiv.org/abs/1903.02625}, 
}

@article{EPR.1935,
 author = {Einstein, A. and Podolsky, B. and Rosen, N.},
 year = {1935},
 title = {Can Quantum-Mechanical Description of Physical Reality Be Considered Complete?},
 pages = {777--780},
 volume = {47},
 number = {10},
 issn = {0031-899X},
 journal = {Physical Review},
 doi = {10.1103/PhysRev.47.777}
}

@article{Horodecki.2009,
 author = {Horodecki, Ryszard and Horodecki, Pawe{\l} and Horodecki, Micha{\l} and Horodecki, Karol},
 year = {2009},
 title = {Quantum entanglement},
 pages = {865--942},
 volume = {81},
 number = {2},
 issn = {0034-6861},
 journal = {Reviews of Modern Physics},
 doi = {10.1103/RevModPhys.81.865}
}

@article{Lan.2009,
 author = {Lan, S-Y and Radnaev, A. G. and Collins, O. A. and Matsukevich, D. N. and Kennedy, T. A. and Kuzmich, A.},
 year = {2009},
 title = {A multiplexed quantum memory},
 pages = {13639--13645},
 volume = {17},
 number = {16},
 journal = {Optics Express},
 doi = {10.1364/oe.17.013639}
}

@article{Ma.2021,
 author = {Ma, Yu and Ma, You-Zhi and Zhou, Zong-Quan and Li, Chuan-Feng and Guo, Guang-Can},
 year = {2021},
 title = {One-hour coherent optical storage in an atomic frequency comb memory},
 pages = {2381},
 volume = {12},
 number = {1},
 issn = {2041-1723},
 journal = {Nat. Commun.},
 doi = {10.1038/s41467-021-22706-y}
}

@article{Saunders.2016,
 author = {Saunders, D. J. and Munns, J. H. D. and Champion, T. F. M. and Qiu, C. and Kaczmarek, K. T. and Poem, E. and Ledingham, P. M. and Walmsley, I. A. and Nunn, J.},
 year = {2016},
 title = {Cavity-Enhanced Room-Temperature Broadband Raman Memory},
 pages = {090501},
 volume = {116},
 number = {9},
 journal = {Physical Review Letters},
 doi = {10.1103/PhysRevLett.116.090501}
}

@article{Seri.2017,
 author = {Seri, Alessandro and Lenhard, Andreas and Riel{\"a}nder, Daniel and G{\"u}ndo{\u{g}}an, Mustafa and Ledingham, Patrick M. and Mazzera, Margherita and de Riedmatten, Hugues},
 year = {2017},
 title = {Quantum Correlations between Single Telecom Photons and a Multimode On-Demand Solid-State Quantum Memory},
 volume = {7},
 pages = {021028},
 number = {2},
 journal = {Phys. Rev. X},
 doi = {10.1103/PhysRevX.7.021028}
}

@article{VernazGris.2018,
 author = {Vernaz-Gris, Pierre and Huang, Kun and Cao, Mingtao and Sheremet, Alexandra S. and Laurat, Julien},
 year = {2018},
 title = {Highly-efficient quantum memory for polarization qubits in a spatially-multiplexed cold atomic ensemble},
 pages = {363},
 volume = {9},
 number = {1},
 issn = {2041-1723},
 journal = {Nat. Commun.},
 doi = {10.1038/s41467-017-02775-8}
}

@article{Kalachev.2019,
doi = {10.1088/1555-6611/ab4049},
url = {https://doi.org/10.1088/1555-6611/ab4049},
year = {2019},
month = {sep},
publisher = {IOP Publishing},
volume = {29},
number = {10},
pages = {104001},
author = {Kalachev, A and Berezhnoi, A and Hemmer, P and Kocharovskaya, O},
title = {Raman quantum memory based on an ensemble of silicon-vacancy centers in diamond},
journal = {Laser Physics}
}

@inproceedings{shor1994algorithms,
  author={Shor, P.W.},
  booktitle={Proceedings 35th Annual Symposium on Foundations of Computer Science}, 
  title={Algorithms for quantum computation: discrete logarithms and factoring}, 
  year={1994},
  volume={},
  number={},
  pages={124-134},
  doi={10.1109/SFCS.1994.365700}}

@article{Klotz.2025,
 author = {Klotz, Marco and Tangemann, Andreas and Kubanek, Alexander},
 year = {2025},
 title = {Ultra-high strained diamond spin register with coherent optical control},
 volume = {11},
 number = {1},
 pages = {91},
 journal = {npj Quantum Information},
 doi = {10.1038/s41534-025-01049-2}
}

@article{Maurer.2012,
 author = {Maurer, P. C. and Kucsko, G. and Latta, C. and Jiang, L. and Yao, N. Y. and Bennett, S. D. and Pastawski, F. and Hunger, D. and Chisholm, N. and Markham, M. and Twitchen, D. J. and Cirac, J. I. and Lukin, M. D.},
 year = {2012},
 title = {Room-temperature quantum bit memory exceeding one second},
 pages = {1283--1286},
 volume = {336},
 number = {6086},
 journal = {Science},
 doi = {10.1126/science.1220513}
}

@article{Ekert.1991,
 author = {Ekert, A. K.},
 year = {1991},
 title = {Quantum cryptography based on Bell's theorem},
 pages = {661--663},
 volume = {67},
 number = {6},
 journal = {Physical Review Letters},
 doi = {10.1103/PhysRevLett.67.661}
}

@article{Scarani.2004,
 author = {Scarani, Valerio and Ac{\'i}n, Antonio and Ribordy, Gr{\'e}goire and Gisin, Nicolas},
 year = {2004},
 title = {Quantum cryptography protocols robust against photon number splitting attacks for weak laser pulse implementations},
 pages = {057901},
 volume = {92},
 number = {5},
 journal = {Physical Review Letters},
 doi = {10.1103/PhysRevLett.92.057901}
}

@article{Hajomer.2024,
 author = {Hajomer, Adnan A. E. and Derkach, Ivan and Jain, Nitin and Chin, Hou-Man and Andersen, Ulrik L. and Gehring, Tobias},
 year = {2024},
 title = {Long-distance continuous-variable quantum key distribution over 100-km fiber with local local oscillator},
 pages = {eadi9474},
 volume = {10},
 number = {1},
 journal = {Science Advances},
 doi = {10.1126/sciadv.adi9474}
}

@article{Singh.2025,
 author = {Singh, Manmohan and Sood, Sandeep Kumar and Bhatia, Munish},
 year = {2025},
 title = {Post-quantum Cryptography: A Review on Cryptographic Solutions for the Era of Quantum Computing},
 pages = {1134-3060},
 journal = {Archives of Computational Methods in Engineering},
 doi = {10.1007/s11831-025-10412-7}
}

@article{Jain.2022,
 author = {Jain, Nitin and Chin, Hou-Man and Mani, Hossein and Lupo, Cosmo and Nikolic, Dino Solar and Kordts, Arne and Pirandola, Stefano and Pedersen, Thomas Brochmann and Kolb, Matthias and {\"O}mer, Bernhard and Pacher, Christoph and Gehring, Tobias and Andersen, Ulrik L.},
 year = {2022},
 title = {Practical continuous-variable quantum key distribution with composable security},
 pages = {4740},
 volume = {13},
 number = {1},
 issn = {2041-1723},
 journal = {Nat. Commun.},
 doi = {10.1038/s41467-022-32161-y}
}

@article{Zhang.2024b,
 author = {Zhang, Yichen and Bian, Yiming and Li, Zhengyu and Yu, Song and Guo, Hong},
 year = {2024},
 title = {Continuous-variable quantum key distribution system: Past, present, and future},
 url = {https://pubs.aip.org/aip/apr/article/11/1/011318/3279669/Continuous-variable-quantum-key-distribution},
 volume = {11},
 pages = {011318},
 number = {1},
 journal = {Applied Physics Reviews},
 doi = {10.1063/5.0179566}
}

@article{Hajomer.2025,
  title = {Coexistence of Continuous-Variable Quantum Key Distribution and Classical Data over 120 km Fiber},
  author = {Hajomer, Adnan A. E. and Derkach, Ivan and Usenko, Vladyslav C. and Andersen, Ulrik L. and Gehring, Tobias},
  journal = {Phys. Rev. Lett.},
  volume = {135},
  issue = {17},
  pages = {170804},
  numpages = {6},
  year = {2025},
  month = {Oct},
  publisher = {American Physical Society},
  doi = {10.1103/zy2d-m3ch},
  url = {https://link.aps.org/doi/10.1103/zy2d-m3ch}
}

@misc{Liu.2023,
 author = {Liu, Chenxu and Wang, Meng and Stein, Samuel A. and Ding, Yufei and Li, Ang},
 date = {25.09.2023},
 title = {Quantum Memory: A Missing Piece in Quantum Computing Units},
 year={2023},
      eprint={2309.14432},
      archivePrefix={arXiv},
      primaryClass={quant-ph},
      url={https://arxiv.org/abs/2309.14432}
}

@article{klotz_2025,
 author = {Klotz, Marco and Tangemann, Andreas and Kubanek, Alexander},
 year = {2025},
 title = {Ultra-high strained diamond spin register with coherent optical control},
 url = {https://doi.org/10.1038/s41534-025-01049-2},
 pages = {91},
 volume = {11},
 number = {91},
 journal = {npj Quantum Inf},
 doi = {10.1038/s41534-025-01049-2}
}

@article{klotz_2024,
 author = {Klotz, Marco and Waltrich, Richard and Lettner, Niklas and Agafonov, Viatcheslav N and Kubanek, Alexander},
 year = {2024},
 title = {Strongly coupled spins of silicon-vacancy centers inside a nanodiamond with sub-megahertz linewidth},
 url = {https://doi.org/10.1515/nanoph-2023-0927},
 pages = {2361-2366},
 volume = {13},
 number = {13},
 journal = {Nanophotonics},
 doi = {10.1515/nanoph-2023-0927}
}

@article{klotz_2022,
 author = {Klotz, Marco and Fehler, Konstantin G and Waltrich, Richard and Steiger, Elena S and Häußler, Stefan and Reddy, Prithvi and Kulikova, Liudmila F and Davydov, Valery A and Agafonov, Viatcheslav N and Doherty, Marcus W and Kubanek, Alexander},
 year = {2022},
 title = {Prolonged Orbital Relaxation by Locally Modified Phonon Density of States for the SiV- Center in Nanodiamonds},
 url = {https://doi.org/10.1103/PhysRevLett.128.153602},
 pages = {153602},
 volume = {128},
 number = {15},
 journal = {Physical Review Letters},
 doi = {10.1103/PhysRevLett.128.153602}
}

@article{kubanek_2022,
 author = {Kubanek, Alexander and Ovvyan, Anna P. and Antoniuk, Lukas and Lettner, Niklas and Pernice, Wolfram HP},
 year = {2022},
 title = {Hybrid Quantum Nanophotonics—Interfacing Color Center in Nanodiamonds with-Photonics},
 url = {https://doi.org/10.1007/978-3-031-16518-4_5},
 pages = {123-174},
 volume = {147},
 number = {7},
 journal = {Progress in Nanophotonics},
 doi = {10.1007/978-3-031-16518-4_5}
}

@article{Fehler_2020,
 author = {Fehler, Konstantin G and Ovvyan, Anna P and Antoniuk, Lukas and Lettner, Niklas and Gruhler, Nico and  Davydov, Valery A and Agafonov, Viatcheslav N and Pernice, Wolfram HP and Kubanek, Alexander},
 year = {2020},
 title = {Purcell-enhanced emission from individual SiV-center in nanodiamonds coupled to a Si3N4-based, photonic crystal cavity},
 url = {https://doi.org/10.1515/nanoph-2020-0257},
 pages = {3655-3662},
 volume = {9},
 number = {11},
 journal = {Nanophotonics},
 doi = {10.1515/nanoph-2020-0257}
}

@article{Fehler_2021,
 author = {Fehler, Konstantin G and Antoniuk, Lukas and Lettner, Niklas and Ovvyan, Anna P and Waltrich, Richard and Gruhler, Nico and  Davydov, Valery A and Agafonov, Viatcheslav N and Pernice, Wolfram HP and Kubanek, Alexander},
 year = {2021},
 title = {Hybrid quantum photonics based on artificial atoms placed inside one hole of a photonic crystal cavity},
 url = {https://doi.org/10.1021/acsphotonics.1c00530},
 pages = {2635-2641},
 volume = {8},
 number = {9},
 journal = {ACS photonics},
 doi = {10.1021/acsphotonics.1c00530}
}

@article{Lettner_2024,
 author = {Lettner, Niklas and Antoniuk, Lukas and Ovvyan, Anna P and Gehring, Helge and Wendland, Daniel and Agafonov, Viatcheslav N and Pernice, Wolfram HP and Kubanek, Alexander},
 year = {2024},
 title = {Controlling all degrees of freedom of the optical coupling in hybrid quantum photonics},
 url = {https://doi.org/10.1021/acsphotonics.3c01559},
 pages = {696-702},
 volume = {11},
 number = {2},
 journal = {ACS photonics},
 doi = {10.1021/acsphotonics.3c01559}
}

@article{Antoniuk_2024,
 author = {Antoniuk, Lukas and Lettner, Niklas and Ovvyan, Anna P and Haugg, Simon and Klotz, Marco and Gehring, Helge and Wendland, Daniel and Agafonov, Viatcheslav N and Pernice, Wolfram HP and Kubanek, Alexander},
 year = {2024},
 title = {All-optical spin access via a cavity-broadened optical transition in on-chip hybrid quantum photonics},
 url = {https://doi.org/10.1103/PhysRevApplied.21.054032},
 pages = {054032},
 volume = {21},
 number = {5},
 journal = {Physical Review Applied},
 doi = {10.1103/PhysRevApplied.21.054032}
}

@article{Waltrich_2023,
 author = {Waltrich, Richard and Klotz, Marco and Agafonov, Viatcheslav N and Kubanek, Alexander},
 year = {2023},
 title = {Two-photon interference from silicon-vacancy centers in remote nanodiamonds},
 url = { https://doi.org/10.1515/nanoph-2023-0379},
 pages = {3663-3669},
 volume = {12},
 number = {18},
 journal = {Nanophotonics},
 doi = {10.1515/nanoph-2023-0379}
}

@article{Schäfer_2025,
 author = {Schäfer, Marlon and Kambs, Benjamin and Herrmann, Dennis and Bauer, Tobias and Becher, Christoph},
 year = {2025},
 title = {Two-Stage, Low Noise Quantum Frequency Conversion of Single Photons from Silicon-Vacancy Centers in Diamond to the Telecom C-Band},
 url = {https://doi.org/10.1002/qute.202300228},
 pages = {2300228},
 volume = {8},
 number = {2},
 journal = {Advanced Quantum Technologies},
 doi = {10.1002/qute.202300228}
}

@article{Maurer_2012,
 author = {Maurer, P. C. and Kucsko, G. and Latta, C. and Jiang, L. and Yao, N. Y. and Bennett, S. D. and Pastawski, F. and Hunger, D. and Chisholm, N. and Markham, M. and Twitchen, D. J. and Cirac, J. I. and Lukin, M. D.},
 year = {2012},
 title = {Room-Temperature Quantum Bit Memory Exceeding One Second},
 url = {https://doi.org/10.1126/science.1220513},
 pages = {1283-1286},
 volume = {336},
 number = {6086},
 journal = {Science},
 doi = {10.1126/science.1220513}
}

@article{Bernien_2013,
 author = {Bernien, H. and Hensen, B. and Pfaff; W. and Koolstra, G. and Blok, M. S. and Robledo, L. and Taminiau, T. H. Markham, M. and Twitchen, D. J. and Childress, L. and Hanson, R.},
 year = {2013},
 title = {Heralded entanglement between solid-state qubits separated by three metres},
 url = {https://doi.org/10.1038/nature12016},
 pages = {86-90},
 volume = {497},
 number = {},
 journal = {Nature},
 doi = {10.1038/nature12016}
}

@article{Bersin_2024,
 author = {Bersin, Eric and Sutula, Madison and Huan, Yan Qi and Suleymanzade, Aziza and Assumpcao, Daniel R. and Wei, Yan-Cheng and  Stas, Pieter-Jan and Knaut, Can M. and Knall, Erik N. and Langrock, Carsten and Sinclair, Neil and Murphy, Ryan and Riedinger, Ralf and  Yeh, Matthew and Xin, C.J. and Bandyopadhyay, Saumil and Sukachev, Denis D. and Machielse, Bartholomeus and Levonian, David S. and  Bhaskar, Mihir K. and Hamilton, Scott and Park, Hongkun and Lončar, Marko and Fejer, Martin M. and Dixon, P. Benjamin and Englund, Dirk R. and Lukin, Mikhail D.},
 year = {2024},
 title = {Telecom Networking with a Diamond Quantum Memory},
 url = {https://doi.org/10.1103/PRXQuantum.5.010303},
 pages = {010303 },
 volume = {5},
 number = {},
 journal = {PRX Quantum},
 doi = {10.1103/PRXQuantum.5.010303}
}

@article{bhaskar_2020,
 author = {Bhaskar, M. K. and Riedinger, R. and Machielse, B. and Levonian, D. S. and Nguyen, C. T. and Knall, E. N. and Park, H. and Englund, D and Lončar, M and Sukachev, D. D. and Lukin, Mikhail D.},
 year = {2020},
 title = {Experimental demonstration of memory-enhanced quantum communication},
 url = {https://doi.org/10.1038/s41586-020-2103-5},
 pages = {60–64},
 volume = {580},
 number = {},
 journal = {Nature},
 doi = {10.1038/s41586-020-2103-5}
}

@article{bradley_2019,
  title = {A Ten-Qubit Solid-State Spin Register with Quantum Memory up to One Minute},
  author = {Bradley, C. E. and Randall, J. and Abobeih, M. H. and Berrevoets, R. C. and Degen, M. J. and Bakker, M. A. and Markham, M. and Twitchen, D. J. and Taminiau, T. H.},
  journal = {Phys. Rev. X},
  volume = {9},
  issue = {3},
  pages = {031045},
  numpages = {12},
  year = {2019},
  month = {Sep},
  doi = {10.1103/PhysRevX.9.031045},
  url = {https://link.aps.org/doi/10.1103/PhysRevX.9.031045}
}

@article{Horodecki.2021,
 author = {Horodecki, R.},
 year = {2021},
 title = {Quantum Information},
 pages = {197--2018},
 volume = {139},
 number = {3},
 issn = {1898-794X},
 journal = {Acta Physica Polonica A},
 doi = {10.12693/APhysPolA.139.197}
}

@article{Naydenov_2011,
  title = {Dynamical decoupling of a single-electron spin at room temperature},
  author = {Naydenov, Boris and Dolde, Florian and Hall, Liam T. and Shin, Chang and Fedder, Helmut and Hollenberg, Lloyd C. L. and Jelezko, Fedor and Wrachtrup, J\"org},
  journal = {Physical Review B},
  volume = {83},
  issue = {8},
  pages = {081201},
  numpages = {4},
  year = {2011},
  month = {Feb},
  publisher = {American Physical Society},
  doi = {10.1103/PhysRevB.83.081201},
  url = {https://link.aps.org/doi/10.1103/PhysRevB.83.081201}
}

@article{Munuera-Javaloy_2021,
doi = {10.1209/0295-5075/ac0ed1},
url = {https://doi.org/10.1209/0295-5075/ac0ed1},
year = {2021},
month = {jul},
publisher = {EDP Sciences, IOP Publishing and Società Italiana di Fisica},
volume = {134},
number = {3},
pages = {30001},
author = {Munuera-Javaloy, C. and Puebla, R. and Casanova, J.},
title = {Dynamical decoupling methods in nanoscale NMR},
journal = {Europhysics Letters}
}

@article{Delgado_2025a,
  title           = {Impact of annealing and nanostructuring on properties of $\mathrm{NV}$ centers created by different techniques},
  journal         = {Diamond and Related Materials},
  volume          = {154},
  pages           = {112126},
  year            = {2025},
  issn            = {0925-9635},
  doi             = {10.1016/j.diamond.2025.112126},
  author          = {Mendoza Delgado, Miriam and Lucas Tsunaki and Shaul Michaelson and Mohan K. Kuntumalla and Johann P. Reithmaier and Alon Hoffman and Boris Naydenov and Cyril Popov}
}

@inproceedings{Delgado_2025b,
  title={Technological Steps for Realization of Diamond-Based Quantum Tokens},
  author={Mendoza Delgado, Miriam and Tsunaki, Lucas and Michaelson, Shaul and Thieme, Jan and Kuntumalla, Mohan K and Trofimov, Sergei and Reithmaier, Johann Peter and Singer, Kilian and Hoffman, Alon and Naydenov, Boris and Popov, Cyril},
  booktitle={Nanotechnological Advances in Environmental, Cyber and CBRN Security},
  series =   {{NATO Science for Peace and Security Series B: Physics and Biophysics}},
  pages={29--45},
  year={2025},
  editor = { {P. Petkov et al.}},
  organization={Springer Nature B. V.},
url = {https://www.researchgate.net/publication/394012951_Technological_Steps_for_Realization_of_Diamond-Based_Quantum_Tokens.},
 doi = {10.1007/978-94-024-2316-7_2}
}

@misc{Diqtok_patent,
 title={Verfahren zum Erstellen eines Quanten-Datentokens, {Patent} {DE} 10 2022 107 528 {A1} 2023.10.05},
 author={Kilian Singer and Cyril Popov and Boris Naydenov},
 type = {patent},
 number = {DE 10 2022 107 528 A1 2023.10.05},
 month = {October},
 year={2022}
 }

@article{Tsunaki_2025,
doi = {10.1088/2058-9565/ae03e6},
url = {https://doi.org/10.1088/2058-9565/ae03e6},
year = {2025},
month = {sep},
publisher = {IOP Publishing},
volume = {10},
number = {4},
pages = {045042},
author = {Tsunaki, Lucas and Bauerhenne, Bernd and Xibraku, Malwin and Garcia, Martin E and Singer, Kilian and Naydenov, Boris},
title = {Ensemble-based quantum token protocol benchmarked on IBM quantum processors},
journal = {Quantum Science and Technology}
}

@article{Bauerhenne_2025,
doi = {10.1088/2058-9565/ae03e7},
url = {https://doi.org/10.1088/2058-9565/ae03e7},
year = {2025},
month = {sep},
publisher = {IOP Publishing},
volume = {10},
number = {4},
pages = {045043},
author = {Bauerhenne, Bernd and Tsunaki, Lucas and Thieme, Jan and Naydenov, Boris and Singer, Kilian},
title = {Security analysis of ensemble-based quantum token protocol under advanced attacks},
journal = {Quantum Science and Technology}
}

@article{Herbschleb_2019,
author = {Herbschleb, E. D. and Kato, H. and Maruyama, Y. and Danjo, T. and Makino, T. and Yamasaki, S. and Ohki, I. and Hayashi, K. and Morishita, H. and Fujiwara, M. and Mizuochi, N.},
title = {Ultra-long coherence times amongst room-temperature solid-state spins},
journal = {Nat Commun},
year = {2019},
volume = {10},
pages = {3766},
doi = {10.1038/s41467-019-11776-8},
url = {https://doi.org/10.1038/s41467-019-11776-8}
}

@online{R1-optical-comm,
  publisher = {The Business Research Company},
  title = {Optical Communication and Networking Global Market Report 2025},
  year = {2025},
  url = {https://www.thebusinessresearchcompany.com/report/optical-communication-and-networking-global-market-report},
  addendum = "(accessed: 21.01.2026)",
}

@online{R5-optical-comm,
  publisher = {Market Growth Reports},
  title = {Optical Communication and Networking Market Size and Share Forecast Outlook 2025 to 2035},
  year = {2025},
  url = {https://www.futuremarketinsights.com/reports/optical-communication-and-networking-market},
  addendum = "(accessed: 21.01.2026)",
}

@online{R3-RF-comm,
  publisher = {The Business Research Company},
  title = {RF Components Market Report 2026},
  year = {2026},
  url = {https://www.thebusinessresearchcompany.com/report/optical-communication-and-networking-global-market-report},
  addendum = "(accessed: 21.01.2026)",
}

@online{R2-quantum-comm,
  publisher = {McKinsey $\&$ Company},
  title = {The Year of Quantum: From concept to reality in 2025},
  year = {2025},
  url = {https://www.mckinsey.com/capabilities/tech-and-ai/our-insights/the-year-of-quantum-from-concept-to-reality-in-2025/},
  addendum = "(accessed: 21.01.2026)",
}

@online{R4-inf-security,
  publisher = {Market Growth Reports},
  title = {Information Security Market Size, Share, Growth, and Industry Analysis, By Type, By Application, Regional Insights and Forecast to 2035},
  year = {2026},
  url = {https://www.marketgrowthreports.com/market-reports/information-security-market-108695},
  addendum = "(accessed: 21.01.2026)",
}

@online{R6-inf-security,
  publisher = {Fortune Business Insights},
  title = {Cybersecurity Market 2026-2034},
  year = {2025},
  url = {https://www.fortunebusinessinsights.com/industry-reports/cyber-security-market-101165},
  addendum = "(accessed: 21.01.2026)",
}

@article{Kilian2007_EPJD_197,
  author    = {Kilian, W. and Haller, A. and Seifert, F. and Grosenick, D. and Rinneberg, H.},
  doi       = {10.1140/epjd/e2007-00026-8},
  journal  = {Eur. Phys. J. D},
  pages     = {197-202},
  title     = {Free precession and transverse relaxation of hyperpolarized $^{129}$Xe gas detected by SQUIDs in ultra-low magnetic fields},
  volume    = {42},
  year      = {2007},
}
	
\end{document}